%% file: paper_ttbb.tex
\documentclass[a4paper,12pt]{article}
\pdfoutput=1
\usepackage{a4wide}
\usepackage[fleqn]{amsmath}
\usepackage{amsfonts,amssymb,cite}
\usepackage{pifont}
\usepackage{bm}
\usepackage{subfigure}
\usepackage{graphicx}
\usepackage{enumerate}
\usepackage{epsfig}
\usepackage{soul}
\usepackage{verbatim,setspace}
\usepackage{tikz}
\usetikzlibrary{calc}
\usepackage{array,multirow,graphicx}
\usepackage{makecell}
\usepackage{lineno}

\newcommand{\be}{\begin{equation}}
\newcommand{\ee}{\end{equation}}
\newcommand{\bea}{\begin{eqnarray}}
\newcommand{\eea}{\end{eqnarray}}

\newcommand{\MET}{E\llap{/\kern1.5pt}_T}

\newcommand{\ttbb}{t\bar{t}b\bar{b}}
\newcommand{\tttt}{t\bar{t}t\bar{t}}

\newcommand{\lepone}{$\ell_{1}$}
\newcommand{\leptwo}{$\ell_{2}$}
\newcommand{\bone}{$b_{1}$}
\newcommand{\btwo}{$b_{2}$}
\newcommand{\adone}{$add_{1}$}
\newcommand{\adtwo}{$add_{2}$}

\newcommand{\cmark}{\text{\ding{51}}}
\newcommand{\xmark}{ }
\newcommand{\sss}{\scriptscriptstyle}

\def\L{\mathcal{L}}

\def\beq{\begin{equation}}
\def\eeq{\end{equation}}

\def\psqr#1#2{{\vcenter{\vbox{\hrule height.#2pt
        \hbox{\vrule width.#2pt height#1pt \kern#1pt
        \vrule width.#2pt}
        \hrule height.#2pt \hrule height.#2pt
        \hbox{\vrule width.#2pt height#1pt \kern#1pt
        \vrule width.#2pt}
        \hrule height.#2pt}}}}
\def\sqr#1#2{{\vcenter{\vbox{\hrule height.#2pt
        \hbox{\vrule width.#2pt height#1pt \kern#1pt
        \vrule width.#2pt}
        \hrule height.#2pt}}}}

\newcounter{oldcounter}
\addtocounter{equation}{1}
\begin{document}

\begin{center}
{\Large \bf Learning to pinpoint effective operators at the LHC: \\[0.7ex]
a study of the $\boldsymbol{\ttbb}$ signature}\\
\vspace{0.5cm}
{\bf Jorgen D'Hondt$^a$, Alberto Mariotti$^{a,b}$,\\
 Ken Mimasu$^c$, Seth Moortgat$^a$, Cen Zhang$^d$}
\vspace{0.3cm}

\noindent \textit{{\small 
$^a$\,Inter-University Institute for High Energies (IIHE), Vrije Universiteit Brussel\\
Pleinlaan 2, B-1050 Brussels, Belgium \\[2mm]
$^b$\,Theoretische Natuurkunde, Vrije Universiteit Brussel\\
Pleinlaan 2, B-1050 Brussels, Belgium \\[2mm]
$^{c}$\,Centre for Cosmology, Particle Physics and Phenomenology (CP3) \\
Universit\'e catholique de Louvain \\
Chemin du Cyclotron 2, B-1348 Louvain-la-Neuve,Belgium\\
$^{d}$\,Institute of High Energy Physics, Chinese Academy of Sciences\\ 
Beijing 100049, China
}}
\end{center}

\begin{abstract}
\noindent \normalsize
In the context of the 
Standard Model effective field theory (SMEFT),
we study the LHC sensitivity to four fermion operators involving heavy quarks by employing cross section measurements in the $\ttbb$ final state. Starting 
from the measurement of total rates, we progressively exploit kinematical 
information and machine learning techniques to optimize the projected 
sensitivity at the end of Run III. Indeed, in final states with high 
multiplicity containing inter-correlated kinematical information, multi-variate 
methods provide a robust way of isolating the regions of phase space where the 
SMEFT contribution is enhanced. We also show that training for multiple output 
classes allows for the discrimination between operators mediating the 
production of tops in different helicity states. Our projected sensitivities 
not only constrain a host of new directions in the SMEFT parameter space but 
also improve on existing limits demonstrating that, on one hand, $\ttbb$ 
production is an indispensable component in a future global fit for top quark 
interactions in the SMEFT, and on the other, multi-class machine learning 
algorithms can be a valuable tool for interpreting LHC data in this framework.

\end{abstract}

\newpage

\tableofcontents

\input{intro.tex}

\input{EFT.tex}

\input{simulation.tex}

\input{sensitivity.tex}

\input{MachineLearning.tex}

\input{conclusion.tex}

\section*{Acknowledgements}

The authors would like to thank fruitful discussions with F. Maltoni, D. Pagani and G. Durieux. SM is an Aspirant van het Fonds Wetenschappelijk Onderzoek - Vlaanderen. CZ is supported by IHEP under Contract No. Y7515540U1. KM is supported by a Marie Sk\l{}odowska-Curie Individual Fellowship of the European Commission's Horizon 2020 Programme under contract number 707983.
AM is supported by the Strategic Research Program High-Energy Physics and the Research Council of the Vrije Universiteit Brussel, and by FWO under the ``Excellence of Science - EOS'' - be.h project n.30820817.

\appendix
\section*{Appendices}
\addcontentsline{toc}{section}{Appendices}
\renewcommand{\thesubsection}{\Alph{subsection}}
\input{AxigluonAppendix.tex}

\input{NeuralNetworkAppendix.tex}
\newpage
\bibliography{bib_ttbb}
\bibliographystyle{JHEP}

\end{document}

%% file: intro.tex
\section{Introduction}
\label{sec:intro}

The lack of evidence for signatures of new physics at the Large Hadron Collider (LHC) has led to an increased interest in the Standard Model Effective Field Theory (SMEFT)~\cite{Burges:1983zg,Leung:1984ni,Buchmuller:1985jz,Hagiwara:1986vm,Grzadkowski:2010es}
as a model-independent approach to interpret experimental measurements in the context of physics Beyond the Standard Model (BSM).
The main phenomenological consequences of the presence of SMEFT operators involve
heightened energy dependence and modified kinematics in Standard Model (SM) processes. It is therefore important to go beyond inclusive measurements and access the full kinematical information available in a given final state. 
Machine learning classifiers are well suited to the task of discriminating between SM and SMEFT effects, particularly with increasing final state multiplicity and complexity in which a great deal of inter-correlated kinematical information is present. In this work we explore the power of these methods, introducing a novel application of multi-class discriminants trained to distinguish among different classes of operators. We quantify the potential to optimally constrain operators and also more accurately pinpoint the origin of an observed deviation in the parameter space. 
%

We focus our investigation on SMEFT operators that contribute to top pair production in association with two $b$-jets.
The top-quark sector provides an interesting place to search for deviations from the SM, given the relatively large production rate of tops at the LHC. Additionally, the large mass of the top is often considered as a motivation to expect BSM physics to be connected to the top quark itself. The large Yukawa coupling of the top quark makes it an ideal probe of the Higgs sector and therefore the mechanism behind electroweak symmetry breaking. Consequently, the study of SMEFT effects in top quark processes has been a subject of intense study in recent years~\cite{Degrande:2010kt,AguilarSaavedra:2010zi,Zhang:2010dr,Greiner:2011tt,Degrande:2014tta,Jung:2014kxa,Rosello:2015sck,Franzosi:2015osa,Zhang:2016omx,Bylund:2016phk,Maltoni:2016yxb,Degrande:2018fog,deBeurs:2018pvs}.

Moreover, the production of top quark pairs with additional heavy-flavour jet activity is an active field of research for the CMS and ATLAS experiments and forms part of a rich top physics programme at the LHC. More precisely the production of two top quarks in association with two bottom quarks is an important background for $t\bar{t}H \ (H \rightarrow b\bar{b})$ analyses which have recently contributed to the discovery of this particular Higgs boson production mode~\cite{Sirunyan:2018hoz,Aaboud:2018urx}. $\ttbb$ production has therefore long been investigated by CMS at 8 TeV \cite{CMS:2014yxa} and 13 TeV \cite{Sirunyan:2017snr} as well as by the ATLAS experiment at 7 TeV \cite{Aad:2013tua} and 8 TeV \cite{Aad:2015yja}. These analyses have however not yet received a lot of attention in terms of BSM interpretations, even though this process has previously been discussed as a probe of new physics~\cite{Degrande:2010kt,Hajer:2015gka,Alvarez:2017wwr}. 

In this work we will present the possible reach of the $\ttbb$ process at 13 TeV centre-of-mass energy to a set of four-heavy-quark EFT operators of dimension six. The process provides sensitivity to previously unconstrained directions in the SMEFT parameter space. 
We begin with a discussion on the sensitivity that can be achieved with the current inclusive 13 TeV CMS measurement \cite{Sirunyan:2017snr}
with 2.3 fb$^{-1}$, as well as a projection to 300 fb$^{-1}$.
Afterwards, we 
estimate the improvement in sensitivity obtained when
exploiting the kinematical information contained  
in this multi-body final state. 
%
In particular,
we explore the use of machine learning methods to 
improve the LHC reach.
We show that these techniques optimally combine kinematical properties to select the region in phase space that is enriched in SMEFT contributions. Moreover, by exploiting a multi-class shallow neural network we can additionally distinguish amongst different classes of SMEFT operators. This leads to improved performance in the presence of more than one SMEFT 
contribution, by focusing on phase space regions preferred by each class of operators.
Our results suggest that
the methods we explore
could be beneficial for generic SMEFT interpretations beyond the final state that we consider.

The paper is structured as follows:
In Section~\ref{sec:EFT}, we identify the relevant four-fermion operators involving heavy quarks that contribute to $\ttbb$ and discuss the complementarity provided with respect to four top production. We further discuss the validity/perturbativity of the EFT expansion concerning this process by considering some explicit power-counting schemes and ultra-violet completions. In Section~\ref{sec:strategy} we outline the analysis strategy, including details on the sample generation, detector simulation, event selection and the statistical procedure used when deriving limits on the Wilson coefficients. Section~\ref{sec:sensitivity} presents a selection of sensitivity studies on individual operators for this process that exploit inclusive, differential and machine-learning-based observables.
Section~\ref{sec:MultipleOperators} describes our novel neural-network discriminant based on a multi-class output. 
We summarize and conclude in Section~\ref{sec:summary}.

%% file: EFT.tex
\section{\boldsymbol{$\ttbb$} in the SMEFT and its virtues
\label{sec:EFT}}
\subsection{Four-fermion operators for $\ttbb$
\label{subsec:operators}}
In the construction of an EFT one extends the SM Lagrangian with operators of dimension larger than four \cite{Weinberg:1979sa,Leung:1984ni,Buchmuller:1985jz}. Since dimension five operators only generate baryon or lepton number violating couplings, the first extension happens with the addition of dimension six effective operators that are suppressed by the square of an energy scale $\Lambda$, as expressed in Eq. \eqref{eq:lagrangian} where $C_{i}$ is the Wilson coefficient corresponding to the EFT operator $O_{i}$.

\begin{align}
\L = \L_{SM} + \sum_{i} \frac{C_{i}}{\Lambda^{2}} \ O_{i} \label{eq:lagrangian}
\end{align}
Assuming flavor universality, a total of 59 independent operators are present~\cite{Grzadkowski:2010es}. 

The most important feature that makes the $\ttbb$ process different than many
others, is its capability of exploring new contact interactions among the
third-generation quarks.  The study of the corresponding operators is well
motivated in non-flavour-universal scenarios, where couplings to the third
generation could be enhanced.  Randall-Sundrum models of a warped extra
dimension could be one example, see e.g.~Refs.~\cite{Agashe:2003zs,Agashe:2007ki}.  It is also natural in
models addressing naturalness of the electroweak symmetry breaking scale 
where the third generation typically plays a special role (see e.g. \cite{Panico:2015jxa} for composite Higgs models).
In the SMEFT approach, to focus on this class of operators
in a model-independent way, we introduce some flavor assumptions such that we can single out the coefficients of operators involving the $t$- and $b$-quarks.

Inspired by Minimal Flavor Violation~\cite{DAmbrosio:2002ex}, we impose a 
$U(2)_q\times U(2)_u\times U(2)_d$ flavor symmetry in the light quark sector.  
Accordingly, four-quark operators composed of vector currents break down into 
three sub-classes involving either four-light, two-heavy two-light or four-heavy 
quarks each with an independent Wilson coefficient and a $U(2)$ flavour 
symmetry among the first two generations where present 
(see~\cite{AguilarSaavedra:2018nen} for a comprehensive review). This also 
permits scalar current operators only among the third generation quarks. In 
this work we focus on the operators in the four-heavy class. The primary reason 
for this is that the two-heavy two-light operators contribute to $t\bar{t}$ and 
$b\bar{b}$ production via the $q\bar{q}$ initial state. Precise measurements of 
top pair production already constrain the $q\bar{q}t\bar{t}$ operators quite 
well~\cite{Buckley:2015lku} and some additional sensitivity is also gained by 
including four-top measurements~\cite{Zhang:2017mls}. Consequently, we do not 
expect to gain further information from $\ttbb$. The $q\bar{q}b\bar{b}$ 
operators can be constrained by differential dijet cross section measurements 
in, \emph{e.g.},~\cite{ATLAS:2015nsi}. Such analyses do not make use of any 
$b$-tagging information and are therefore completely blind to jet flavour. 
Judging by the $\mathcal{O}(10^{-2})$ TeV$^{-2}$ sensitivity obtained by this analysis to flavour-universal, colour-singlet four fermion operators, we do not expect $\ttbb$ to provide competitive bounds. 
Apart from 4 fermion operators, the $\ttbb$ final state can also be affected by 
other operators that modify the interactions of the QCD sector. Namely the 
triple gluon and the chromomagnetic dipole operators, 
    \begin{align}
        O_G   = & \,f_{\sss ABC}G^{A\nu}_\mu G^{B\rho}_\nu G^{C\mu}_\rho,\\
        O_{tG}= & \left(\bar{Q}\,\sigma^{\mu\nu} \,T_A\,t_R\right)
        \tilde{\phi}\,G^A_{\mu\nu} +\text{h.c.},\\
        O_{bG}= & \left(\bar{Q}\,\sigma^{\mu\nu} \,T_A\,b_R\right)
        \phi\,G^A_{\mu\nu}  +\text{h.c.}.
    \end{align}
The first has been shown to be strongly constrained by multi-jet 
measurements~\cite{Krauss:2016ely} to lie within [-0.04,0.04] TeV$^{-2}$. The 
top quark chromomagnetic operator contributes directly to top pair production 
and is constrained individually to the range [-0.30, 0.64] 
TeV$^{-2}$~\cite{Buckley:2015lku}. The $b$-quark dipole operator will 
contribute to the dijet cross section and likely result in very strong limits. 
We argue that the $\ttbb$ process will not provide more stringent information 
on these operators and therefore do not include them in our study. In any case, 
the focus of this analysis is to point out the sensitivity to previously 
unconstrained directions in the SMEFT parameter space.

There are 12 independent 4-fermion operators involving only heavy quarks. Following 
the basis choice recommended by the LHC Top Working Group~\cite{AguilarSaavedra:2018nen}:
\begin{subequations}
\begin{align}
\nonumber
&\text{Operator} &\ttbb& &\tttt&\\[-1em]
\cline{1-5}
 & O^{1}_{QQ} = \frac{1}{2}\left( \bar{Q} \ \gamma_{\mu} \  Q \right)  \left( \bar{Q}  \ \gamma^{\mu} \  Q \right),
&\cmark& &\cmark& \label{eq:cQQ1}\\
& O^{8}_{QQ} = \frac{1}{2}\left( \bar{Q} \ \gamma_{\mu} \  T^{A} \  Q \right)  \left( \bar{Q}  \ \gamma^{\mu} \  T^{A} \  Q \right),
&\cmark& &\cmark&  \label{eq:cQQ8}\\
&O^{1}_{tb} = \left( \bar{t}  \ \gamma_{\mu}  \ t \right)  \left(\bar{b} \ \gamma_{\mu} \  b \right), 
&\cmark& &\xmark&\label{eq:ctb1} \\
&O^{8}_{tb} = \left( \bar{t}  \ \gamma_{\mu} T^{A} \  \ t \right)  \left(\bar{b} \ \gamma_{\mu} \ T^{A} \  b \right), 
&\cmark& &\xmark& \label{eq:ctb8}\\
&O^{1}_{tt} = \left( \bar{t}  \ \gamma_{\mu}  \ t \right)  \left(\bar{t} \ \gamma_{\mu} \  t \right), 
&\xmark& &\cmark&\label{eq:ctt1} \\
&O^{1}_{bb} = \left( \bar{b}  \ \gamma_{\mu}  \ b \right)  \left(\bar{b} \ \gamma_{\mu} \  b \right), 
&\xmark& &\xmark&\label{eq:cbb1} \\
& O^{1}_{Qt} = \left( \bar{Q} \ \gamma_{\mu} \  Q \right)  \left( \bar{t}  \ \gamma^{\mu} \  t \right), 
&\cmark& &\cmark&\label{eq:cQt1}\\
& O^{8}_{Qt} = \left( \bar{Q} \ \gamma_{\mu} \  T^{A} \  Q \right)  \left( \bar{t}  \ \gamma^{\mu} \  T^{A} \  t \right), 
&\cmark& &\cmark& \label{eq:cQt8}\\
& O^{1}_{Qb} = \left( \bar{Q} \ \gamma_{\mu} \  Q \right)  \left( \bar{b}  \ \gamma^{\mu} \ b \right), 
&\cmark& &\xmark&\label{eq:cQb1}\\
& O^{8}_{Qb} = \left( \bar{Q} \ \gamma_{\mu} \  T^{A} \  Q \right)  \left( \bar{b}  \ \gamma^{\mu} \  T^{A} \  b \right),  
&\cmark& &\xmark&\label{eq:cQb8}\\
& O^{1}_{QtQb} = \left( \bar{Q} \ t \right) \varepsilon 
                 \left( \bar{Q} \ b \right) , 
&\cmark& &\xmark& \label{eq:cQtQb1}\\
& O^{8}_{QtQb} = \left( \bar{Q} \  T^{A} \  t \right) \varepsilon
                 \left( \bar{Q} \  T^{A} \  b \right).
&\cmark& &\xmark&\label{eq:cQtQb8}
\end{align}
\label{eq:operators}
\end{subequations}
$Q$ represents the left-handed SU(2) doublet of third generation quarks (top and bottom), $t$ and $b$ represent the right-handed top and bottom quarks, $T^{A}$ denotes the $SU(3)$ generators and $\varepsilon$ is the totally antisymmetric Levi-Civita tensor in $SU(2)$-space. We additionally specify whether each operator contains $\ttbb$ and $\tttt$ interactions. 
It should be noted that a subset of the color singlet operators appearing in Eq.~\eqref{eq:cQQ1}--\eqref{eq:cQtQb8} have been indirectly constrained through RG induced contributions to electroweak precision
observables \cite{deBlas:2015aea}. Our study presents the first direct constraints on the full set of four heavy quark operators containing $\ttbb$ interactions.

\begin{figure}[h!]
\center
\includegraphics[width=.3\textwidth]{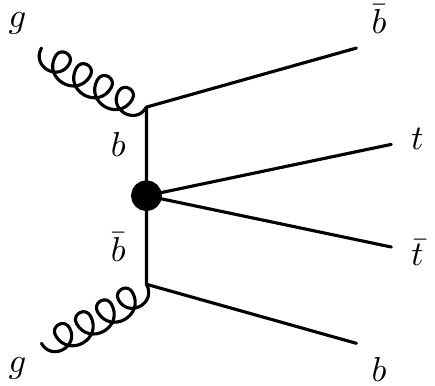}
\hspace{2cm}
\includegraphics[width=.3\textwidth]{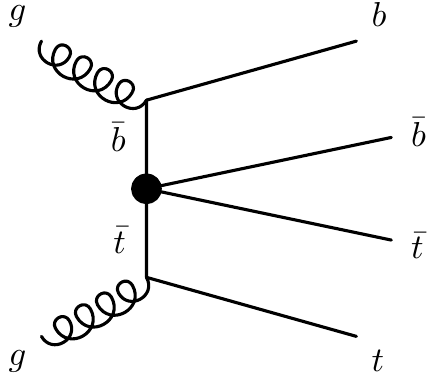}
\caption{\label{fig:diagrams}
Dominant EFT contributions to $\ttbb$ production.}
\end{figure}

The Wilson coefficient corresponding to each of these operators as they appear in the Lagrangian will be denoted by replacing the $O$ in the name of the operator by a $C$. We absorb the $1/\Lambda^{2}$ factor into the definition of the Wilson coefficients and assume $\Lambda = 1$ TeV throughout this work. The dependence of the $\ttbb$ cross section on the Wilson coefficients, in general, forms a 10-dimensional quadratic function in this Wilson coefficient space
\begin{align}
    \sigma_{\ttbb} = \sigma_{\ttbb}^{SM}\left(1+\sum_i p_1^i C_i +\sum_{i\leq j} p_2^{ij} C_iC_j\right).
\end{align} 
Both the SM and EFT contributions to this process are predominantly mediated by the $gg\to\ttbb$ subprocess. The dominant Feynman diagrams involving a single insertion of the EFT vertex are shown in Figure \ref{fig:diagrams}.

\subsection{Complementarity to four top production}
Out of the 12 operators in Equations \eqref{eq:operators}, those that contain a
$t\bar{t}t\bar{t}$ component can also be constrained by four top production
processes, for example in References \cite{ATLAS-CONF-2016-104,Zhang:2017mls}.
These operators are: $O^{1}_{QQ}$, $O^{8}_{QQ}$, $O^{1}_{Qt}$, $O^{8}_{Qt}$ and $O^{1}_{tt}$ (which does not contribute to $\ttbb$).
Of the first two operators $O^{1}_{QQ}$ and $O^{8}_{QQ}$, only one linear
combination can be probed by the four-top process.  This can be seen by writing
down their respective interaction terms:
\begin{flalign}
	&O_{QQ}^1=
	\frac{1}{2}\left[ \left(\bar t_L\gamma^\mu t_L\right)\left(\bar t_L\gamma_\mu t_L\right) 
	+\left(\bar b_L\gamma^\mu b_L\right)\left(\bar b_L\gamma_\mu b_L\right)\right]
	+\left(\bar t_L\gamma^\mu t_L\right)\left(\bar b_L\gamma_\mu b_L\right)
	\\
	&O_{QQ}^8=
	\frac{1}{6}\left[ \left(\bar t_L\gamma^\mu t_L\right)\left(\bar t_L\gamma_\mu t_L\right) 
	+\left(\bar b_L\gamma^\mu b_L\right)\left(\bar b_L\gamma_\mu b_L\right)\right]
	+\left(\bar t_L\gamma^\mu T^At_L\right)\left(\bar b_L\gamma_\mu T^Ab_L\right)
\end{flalign}
where the first term in $O_{QQ}^8$ has a color-singlet structure because it has
been Fierzed.  As a result, in four-top production only one combination of the
two operator coefficients
\begin{flalign}
	C_{QQ}^{(+)}=\frac{1}{2}C_{QQ}^1+\frac{1}{6}C_{QQ}^8
\end{flalign}
is probed. In contrast, in $\ttbb$ production both degrees of freedom are probed
independently, because the $\ttbb$ terms in $O_{QQ}^1$ and in $O_{QQ}^8$ have
different color structures.  This lifts the flat direction $C_{QQ}^8=-3C_{QQ}^1$
in the $C_{QQ}^1$-$C_{QQ}^8$ plane, left from the four-top measurements.
The underlying reason is that, while the color singlet and octet structures for
a $t_Lt_Lt_Lt_L$ interaction term are equivalent due to Fierz identity, it is
not the case for $t_Lt_Lb_Lb_L$ interaction. The projected individual LHC sensitivities with 300 fb$^{-1}$ from four-top production on the operators it shares with $\ttbb$, translated from Ref.~\cite{Zhang:2017mls} are summarised in Table~\ref{tab:4top_projections}. The final column represents the best sensitivities obtained from our $\ttbb$ study for comparison.

\begin{table}
    \centering
\begin{tabular}{|c|c|c|c|c|}
    \hline
    Operator & \makecell{4-top \\ ($M_{cut} = 2$ TeV)} & \makecell{4-top \\ ($M_{cut} = 3$ TeV)} & \makecell{4-top \\ ($M_{cut} = 4$ TeV)} & \makecell{this work \\ ($M_{cut} = 2$ TeV)}
    \tabularnewline\hline
    $C^{1}_{QQ}$&$[-3.9,3.5]$&$[-2.9,2.6]$&$[-2.8,2.5]$&$[-2.1,2.3]$
    \tabularnewline
    $C^{8}_{QQ}$&$[-11.8,10.5]$&$[-8.8,7.8]$&$[-8.4,7.4]$&$[-4.5,3.1]$
    \tabularnewline
    $C^{1}_{Qt}$&$[-3.2,3.3]$&$[-2.4,2.4]$&$[-2.2,2.3]$&$[-2.1,2.3]$
    \tabularnewline
    $C^{8}_{Qt}$&$[-7.4,5.8]$&$[-5.4,4.3]$&$[-5.1,4.1]$&$[-3.9,3.8]$
    \tabularnewline
    \hline
\end{tabular}
\caption{\label{tab:4top_projections}
Projected individual confidence intervals quoted in Ref.~\cite{Zhang:2017mls} from 4-top production. These are derived assuming an upper limit of the signal strength, $\mu < 1.87$ is obtainable at the LHC with 300 fb$^{-1}$, as estimated in Ref.~\cite{Alvarez:2016nrz}. The limits are reported as a function of an upper bound on the total invariant mass of the events, $M_{cut}$. The last column compares these intervals to best projections from $\ttbb$ production obtained in our work.
}
\end{table}

One of the interesting points of this study will be to compare the sensitivity of $\ttbb$ to the existing and future limits from four top. In this context, one major difference between the two processes is the comparative rarity of four top production. In 13 TeV $pp$ collisions, its cross section is of order 9 fb, compared to the $\sim$3 pb prediction for $\ttbb$. The limited statistics of four top measurements at the LHC will most likely mean that it will only ever be measured at inclusive level for the foreseeable future. $\ttbb$ production does not suffer from this and the methods developed in this paper are designed to exploit the sufficiently large statistics present in 300 fb$^{-1}$ of integrated luminosity to enhance the relative sensitivity to the EFT parameter space.
As mentioned, operators \eqref{eq:ctb1}, \eqref{eq:ctb8}, \eqref{eq:cQb1}--\eqref{eq:cQtQb8} have never been directly constrained before and we obtain sensitivity to them in the $\ttbb$ topology. In summary, the EFT interpretation of $\ttbb$ measurements at the LHC presents the following advantages:
\begin{itemize}
    \item A sufficiently large inclusive cross section that allows for the use of differential information after 300 fb$^{-1}$ of integrated luminosity.
    \item It directly constrains 6 four heavy quark operators for the first time.
    \item It breaks the degeneracy in a blind direction of the parameter space with respect to four top measurements.
\end{itemize}

\subsection{EFT validity, power-counting and UV connection\label{subsec:UV}}
The most basic requirement to satisfy when interpreting a particular
measurement in the SMEFT framework is to ensure that one is probing scales below
the mass scale of new physics, $\Lambda_{NP}$. Above this scale, one expects resonant physics to
appear which cannot be captured by the EFT description. 
To control the energy scales being probed, we introduce a new parameter $M_{cut}$ and 
impose that all energies associated to our process
be less than this value. 
As a result, our EFT description can approximate UV completions for which
\begin{align}
    \Lambda_{NP}>M_{cut}.\label{eq:validity}
\end{align}
However, from the low-energy perspective, $\Lambda_{NP}$ is unknown.
since the scale of new physics is degenerate with the value of the Wilson
coefficient. 
Hence, in order to make quantitative statements on the EFT validity, one has 
to make some assumptions on the power counting rule of 
SMEFT~\cite{Pomarol:2014dya,Contino:2016jqw} and therefore on the nature of the UV completion. 
From now on we assume that there is one single BSM coupling denoted $g_*$ 
associated with $\Lambda_{NP}$, as done in the SILH description of Higgs EFT~\cite{Giudice:2007fh}. 
The EFT operators can then be expanded in terms of the following building blocks
\begin{equation}
	\mathcal{L}_{\mathrm{EFT}}=\frac{\Lambda_{NP}^4}{g_*^2}
	\mathcal{L}\left( \frac{D_\mu}{\Lambda_{NP}},\frac{g_*H}{\Lambda_{NP}},
	\frac{g_*f_{L,R}}{\Lambda_{NP}^{3/2}},\frac{gF_{\mu\nu}}{\Lambda_{NP}^2}\right)\,
	\label{eq:pc}
\end{equation}
From this power-counting prescription, we see that
the four-fermion operator coefficients are of order $C=g_*^2/\Lambda_{NP}^2$.
The validity of the EFT description in Eq.~\eqref{eq:validity} can then be rewritten as
$
|C_{i}| M_{cut}^{2} \lesssim g_*^2
$.
This is minimally conservative when $g_*$ takes its largest value $g_* \sim 4 \pi$, i.e.
\begin{align}
|C_{i}| M_{cut}^{2} \lesssim (4\pi)^{2}. \label{eq:pert}
\end{align}
In this limit, such a condition
 is equivalent to the model independent requirement of quantum perturbativity 
 in the EFT~\cite{AguilarSaavedra:2018nen}.
The latter is examined through the contributions
involving more and more operator insertions with higher and higher numbers of
loops. The convergence of the series requires that $C_{i}M_{cut}^2/(4\pi)^2$ be
less than a constant, which should be roughly of order one.
In section~\ref{subsec:mcut},
we will make use of condition \eqref{eq:pert}
to identify an
appropriate value for the upper bound on the energy scale of the process $M_{cut}$.
The experimental sensitivities on $C_i$ that we will eventually find in our 
analysis (see Figure \ref{fig:summary_Limits}) are more stringent than \eqref{eq:pert} and hence
will be in a valid regime assuming the simple power counting of \eqref{eq:pc}. 
On the other hand, they will typically correspond to strong coupling values for $g_*$.

With the power counting introduced in Eq.~\eqref{eq:pc}, one can also investigate 
the relative size of interference and quadratic terms in dim-6 and dim-8 
SMEFT operators\footnote{Dimension 7 operators generate baryon or lepton number violation \cite{Degrande:2012wf,Kobach:2016ami} and we do not consider them here.}.
We will show that assuming a strongly coupled UV completion implies that the 
dim-8 interference terms are sub-leading with respect to the dim-6 quadratic 
ones, even though they are formally of the same order in the EFT expansion.
Concretely,
the dim-6 interference and quadratic
terms in the cross section for $gg\to t\bar tb\bar b$ production are
\begin{flalign}
	&\mbox{dim-6 interference:}\ \frac{g_s^6g_*^2E^2}{\Lambda_{NP}^2},
	\label{eq:dim61}
	\\
	&\mbox{dim-6 quadratic term:}\ \frac{g_s^4g_*^4E^4}{\Lambda_{NP}^4}, 
	\label{eq:dim62}
\end{flalign}
where $E$ is the largest energy scale characterizing the process and can be at 
most $E \sim M_{cut}$ in our analysis.
The two terms in \eqref{eq:dim61} and \eqref{eq:dim62}
could have similar size, 
and the quadratic terms could even dominate over the interference,
if $g_*$ is large enough such that
$(g_*/g_s)^2E^2/\Lambda_{NP}^2 \gtrsim 1$.  
This is actually the relevant regime 
for the typical constraints 
that we will find in this work on the Wilson coefficients (see again Figure \ref{fig:summary_Limits}),
meaning that both terms should
be included.

On the other hand, the dim-8 interference is subleading, even though it is
suppressed by the same power of $\Lambda_{NP}$ as the dim-6 squared terms.  A
dim-8 four-fermion operator would have the schematic form $ffffD^2$, and,
according to Eq.~(\ref{eq:pc}), a coefficient of order $g_*^2/\Lambda_{NP}^4$,
which is not enhanced by higher powers of $g_*$.  This gives a contribution of order
\begin{flalign}
	&\mbox{dim-8 interference:}\ \frac{g_s^6g_*^2E^4}{\Lambda_{NP}^4},
	\label{eq:dim81}
\end{flalign}
which is subleading compared to the dim-6 interference in Eq.~(\ref{eq:dim61}), as
far as $M_{cut}$ is below the scale $\Lambda_{NP}$, i.e.~the validity criterion is satisfied.
The dim-8 interference contribution is also definitely subleading with respect to the dim-6 quadratic contributions \eqref{eq:dim62} as soon as $g_* > g_s$.

Apart from four-fermion operators, a general dim-8 operator could involve more fields and thus have more powers of $g_*$ in its
coefficient.  For the process of interest, the relevant operators are those
that lead to contact $gttbb$ and $ggttbb$ interactions, and should have the
schematic forms $ffffG_{\mu\nu}$ and $ffffD_\mu D_\nu$.  Note that in Eq.~(\ref{eq:pc}),
the coupling that comes with $G_{\mu\nu}$ is $g_s$, instead of $g_*$.  This is
of course a model-dependent assumption, but seems natural, as the coupling
between a gauge boson and a BSM particle is likely to be its own gauge
coupling. Based on this assumption, the coefficients of the operators
$ffffG_{\mu\nu}$ and $ffffD_\mu D_\nu$ are of the order
$g_*^2g_s/\Lambda_{NP}^4$ and $g_*^2/\Lambda_{NP}^4$ respectively, and thus
their interference contributions to the $gg\to t\bar tb\bar b$ amplitude, from
either $gttbb$ or $ggttbb$ vertices, are the same as Eq.~(\ref{eq:dim81}), and
also subleading to Eq.~(\ref{eq:dim61}), again if the validity criterion is
satisfied.

In summary, we assume that the operators obey the power-counting depicted
in Eq.~(\ref{eq:pc}), apply the analysis cut $M_{cut}<\Lambda_{NP}$ to ensure
EFT validity, and truncate the EFT expansion at dim-6 at the amplitude level,
which amounts to including the dim-6 quadratic contribution while neglecting
dim-8 operators and beyond\footnote{For similar discussions on power counting arguments  in the context of helicity selection rules in the SMEFT, see Refs.~\cite{Biekoetter:2014jwa,Cheung:2015aba,Falkowski:2016cxu,Azatov:2016sqh}}. 
This corresponds to including the contributions
with sizes given in Eqs.~(\ref{eq:dim61}) and (\ref{eq:dim62}), and neglecting
any additional contributions suppressed by
$E^2/\Lambda_{NP}^2<M_{cut}^2/\Lambda_{NP}^2<1$. In Appendix~\ref{app:AM} we
give a concrete BSM example with a strongly coupled new particle, and derive its
relevant operators in $gg\to \ttbb$, to illustrate that the above power counting
assumption is satisfied and that our strategy would indeed capture the dominant
BSM contributions.

%% file: simulation.tex
\section{Analysis strategy}
\label{sec:strategy}
Before describing our workflow, we comment on the latest experimental 
method adopted by CMS~\cite{CMS:2014yxa,Sirunyan:2017snr} to measure the $\ttbb$ 
cross section. Rather than directly selecting $\ttbb$ events, a selection is 
performed to obtain an inclusive $t\bar{t}+2$ jet sample. The fractional 
$\ttbb$ yield is then extracted by fitting the multivariate $b$-jet 
discriminants of the two additional jets. The $b$-tagging calibration is the 
main 
source of systematic uncertainty in this procedure. In our analysis, we only 
consider the $\ttbb$ component, assuming that it can reliably be extracted by 
this method. This rests on the expectation that the $b$-jet discriminant 
information is not affected by the presence of EFT operators, nor by the 
additional kinematical selection requirement we impose in 
Section~\ref{sec:EFTPhaseSpace}. Indeed, these discriminants are designed to be 
as independent as possible of the jet $p_T$ and 
$\eta$~\cite{CMS-PAS-BTV-15-001,CMS-PAS-BTV-16-002}. Although we employ fast 
detector simulation, these do not include information on the $b$-jet 
discriminant, rather parametrising the $b$-tagging probability. Explicit 
verification of this assumption is therefore beyond the scope of this study.\\\\

\noindent \textbf{Simulation}\\
We begin by describing our signal sample generation for the $\ttbb$ process that 
will be used through our sensitivity study. We obtain our signal samples using 
the Universal FeynRules Output (UFO) model 
{\sc dim6top}~\cite{AguilarSaavedra:2018nen} that includes both the SM and the 
four heavy quark operators of Eqs.~\eqref{eq:cQQ1}--\eqref{eq:cQtQb8}. We also 
validate our generation with an independent implementation of the same 
operators using the FeynRules package~\cite{Alloul:2013bka}. The $\ttbb$ final 
state in which both top quarks decay leptonically, is simulated at LO in the 
four-flavour scheme\footnote{This choice was motivated by the recent studies on simulating $t\bar{t}$ + $b$-jet production 
at the LHC~ \cite{Jezo:2018yaf}. It is known that such multi-scale processes are currently difficult to simulate. The theory uncertainty of 
the best SM prediction at  NLO+PS is of order 20-30\%~ \cite{Jezo:2018yaf}. In this work we assume that this precision can be improved 
by the end Run III of the LHC.} from proton-proton collisions at 13 TeV center-of-mass 
energy using MadGraph5\_aMC@NLO 2.6.0~\cite{Alwall:2014hca} (MG5\_aMC@NLO). 
The so-called ``visible'' phase space as quoted in the CMS measurement of 
the $\ttbb$ cross section is mimicked as closely as possible by requiring the 
two charged leptons (electrons or muons) to have transverse momentum ($p_{T}$) 
$>$ 20 GeV and pseudorapidity ($\eta$) between $-2.4$ and $2.4$, and the four 
particle-level $b$-jets to satisfy  $p_{T} > $ 20 GeV and $|\eta| < $ 2.5. The 
angular separation\footnote{$\Delta R = \sqrt{(\Delta\phi)^{2} + (\Delta\eta)^{2}}$, where $\phi$ is the azimuthal angle 
difference between two objects and $\eta$ the pseudorapidity difference between two objects.} in $\Delta R$ between 
different jets or between jets and leptons is required to be larger than 0.5.  
Where necessary, parton shower/hadronisation is simulated with 
Pythia8~\cite{Sjostrand:2007gs} and object reconstruction is modelled with the 
Delphes~\cite{deFavereau:2013fsa} detector simulation software, using the 
default CMS card.\\

\noindent \textbf{Event reconstruction}\\
The parton level, ``visible'' phase space prediction will be compared to the 
CMS inclusive measurement and future prospects for 300 fb$^{-1}$ will be 
estimated. We will then progressively refine the selection procedure, assuming 
300 fb$^{-1}$ of LHC data, in order to increase the sensitivity to the 
operators. This however implies that one has to step away from the unfolded 
cross section to the fiducial detector volume, and instead impose further 
selection requirements on the reconstructed objects. We impose an event 
selection following as closely as possible the CMS 
analysis~\cite{Sirunyan:2017snr}. Each event must have two reconstructed, 
isolated leptons (electrons or muons) with $p_{T} > $ 20 GeV and $|\eta| < $ 
2.4, which are arbitrarily assigned the labels \lepone \ and \leptwo.  Missing 
transverse energy has to be larger than 30 GeV. At least four jets must be 
present with  $p_{T} > $ 30 GeV and $|\eta| < $ 2.5, of which at least two 
are $b$-tagged. 

In the CMS analysis the jets with the highest $b$-tagging discriminator are 
identified as the $b$-jets from the top quark decay. This information is 
however not available in Delphes which merely parametrises the $b$-tagging 
efficiency. Instead, out of the four highest-$p_{T}$ jets in the event, the 
one closest in $\Delta R$ to \lepone \ is assigned the label \bone \ and is 
considered to be the $b$-jet associated to the top quark decaying into 
\lepone \ and \bone \ and the same association is applied to identify 
\btwo \ associated to \leptwo. Finally the two remaining jets are ordered by 
decreasing $p_{T}$ and then assigned labels \adone \ and \adtwo. 

We obtain a reconstruction and selection efficiency from Delphes that is 
roughly a factor of two smaller than the one quoted by CMS. This is mostly due 
to the parametrised lepton reconstruction and isolation requirements as well as 
the jet reconstruction. It should be noted that the definition of the visible 
phase space by CMS includes the presence of at least four particle-level jets 
(clustered from generated particles rather than reconstructed objects), whereas 
our fiducial phase space prediction does not include parton-shower and jet 
clustering effects. Nevertheless, this should not affect the results of our 
analysis as long as the acceptance and efficiency of the event selection are 
the same for SM and EFT contributions. It has indeed been checked that these 
are the same up to Monte Carlo statistical uncertainties. 
We therefore identify the parton-level predictions with the visible phase space 
measurement of CMS and safely use the outlined event selection without biasing 
the results of this study. The $M_{\text{cut}}$ requirement discussed in Section~\ref{subsec:UV} is imposed on all combinations of invariant masses of final state particles (see list in Table~\ref{tab:inputs}) as well as the scalar sum of transverse momenta, $H_T$.\\

\noindent \textbf{Sensitivity analysis}\\
Having obtained reconstructed samples that should be similar to those obtained 
by current and future analyses in this final state, we proceed to estimate the 
sensitivity of various selection methods to the Wilson coefficients, one at a 
time, in Section~\ref{sec:sensitivity}. To access the energy growth of the EFT 
contributions, we first consider a cut on the invariant mass of the 4 $b$-jets 
in the final state ($M_{4b}$). Next, we construct a 
multi-class discriminant using a shallow neural network (NN), trained to 
identify classes corresponding to the SM point, left-handed top EFT operators 
and right-handed top EFT operators. The discriminant should draw from the full 
20-dimensional phase space of the 8-body final state and learn to distinguish 
samples with different top helicities in the final state through the angular 
correlations among the top decay products.
The sensitivities are first evaluated by requiring a lower threshold on the 
value of the discriminant. Additionally we also evaluate the sensitivity by 
performing a template fit to the full discriminant distribution. Finally, 
in Section \ref{sec:MultipleOperators}, we highlight the advantages of the 
multi-class output structure of the NN, which lead to improved limits when 
multiple operators with different Lorentz structures are allowed to vary 
simultaneously. As a case study, we will illustrate the improved limits that 
can be obtained in a two dimensional parameter space spanned by a pair of left 
and right handed operators.

The evaluation of the sensitivity generally proceeds via the same general method. We first construct the functional dependence of an observable $O$ on each Wilson coefficient in Eq. \eqref{eq:operators}, one at a time, according to 
\begin{align}
O_{fit} = O_{SM} \left( 1 + p_{1} \cdot C_{i} + p_{2} \cdot C_{i}^{2} \right), \label{eq:xsec}
\end{align}
where $O_{fit}$ is the total observed value of the observable, $O_{SM}$ is the SM prediction, $C_{i}$ is the value of the Wilson coefficient and $p_{i}$ ($i \in [1,2]$) are parameters to be determined. $p_{1}$ signifies the fractional importance of the interference of the EFT with the SM and $p_{2}$ represents the fractional EFT squared contribution to the observable at quadratic order in the Wilson coefficient of the EFT operator. The observable, $O$, may be a cross-section or a number of events given a certain integrated luminosity observed in a signal region or extracted from a template fit. Taking the experimentally measured value or assuming the SM prediction is observed in future projections and combining statistical and estimated systematic uncertainties in quadrature, we construct a $\Delta\chi^2$  

\begin{align}
\Delta\chi^{2} (C_{i} | p_{1}, p_{2}) &= \chi^{2} (C_{i} | p_{1}, p_{2})-\chi^2_{min}\\
&=\frac{\left( O_{fit}(C_{i} | p_{1}, p_{2}) -  O_{obs} \right)^{2}}{\delta{O}^{2}}-\chi^2_{min}, \label{eq:chi2}
\end{align}
where $O_{fit}$ and $O_{obs}$ are the predicted and observed observables, $\delta{O}$ is the uncertainty on the observable and $\chi^2_{min}$ is the minimum value of the $\chi^2$ function in the EFT parameter space. Typically the uncertainty is composed of statistical and systematic uncertainties at the LHC as well as some MC statistical uncertainties.
The 95\% CL sensitivity interval on the individual Wilson coefficients $C_{i}$ is then determined by the region in which the $\chi^{2}$ value is lower than 3.84, corresponding to a p-value of 0.05 for a $\chi^{2}$ distribution with 1 degree of freedom in the Gaussian limit.  Only in Section \ref{sec:MultipleOperators}, where two EFT operators are allowed with non-zero Wilson coefficients simultaneously, the number of degrees of freedom is augmented to 2, with a corresponding threshold of 5.991 for the same p-value.\\

\noindent \textbf{Future projections}\\
All of the projected limits on the Wilson coefficients presented in this analysis assume the observation of the SM prediction with 300 fb$^{-1}$ of integrated luminosity and 10\% systematic uncertainty. This is based on the fact that the 35\% (see Section~\ref{sec:xsec}) systematic uncertainty of the CMS $\ttbb$ measurement is dominated by the $b$-tagging scale factors, which contribute 27\% on their own. These uncertainties have improved by a factor of $\sim$4 between $b$-tagging studies performed with 2.6 fb$^{-1}$~\cite{CMS-PAS-BTV-15-001} and 36.1 fb$^{-1}$~\cite{CMS-PAS-BTV-16-002}. Furthermore, they are found to remain stable up to around 1 TeV in jet $p_T$. The next most important source of uncertainty comes from theoretical modelling and is quoted at 17\% in the current measurement. 
This consists mainly of MC generator and parton shower scale variations.
Given the importance of this final state in the context of Higgs physics it is reasonable to expect that these uncertainties will be significantly reduced by the end of Run III.

%% file: sensitivity.tex
\section{Sensitivity to individual operators}
\label{sec:sensitivity}
In this section, only one Wilson coefficient is considered at a time, 
assuming all others are kept at a value of zero. We then obtain individual 
limits that reflect the sensitivity of the $\ttbb$ final state to each of the 
operators. We start by using the CMS inclusive cross section measurement including projections for LHC Run III. We investigate the resulting sensitivity as a function of $M_{\text{cut}}$ and motivate the value of 2 TeV that we use throughout this work. We then progressively exploit more 
kinematical information in the reconstructed final state. After first considering a selection on what was found to be the 
most discriminating variable, the invariant mass of the four $b$-jets, 
further improvements using machine learning classifiers to select the EFT 
enriched phase space are illustrated.

\subsection{Cross section in the fiducial detector volume}
\label{sec:xsec}
\begin{figure}[ht!]
\center
\includegraphics[height=.3\textheight]{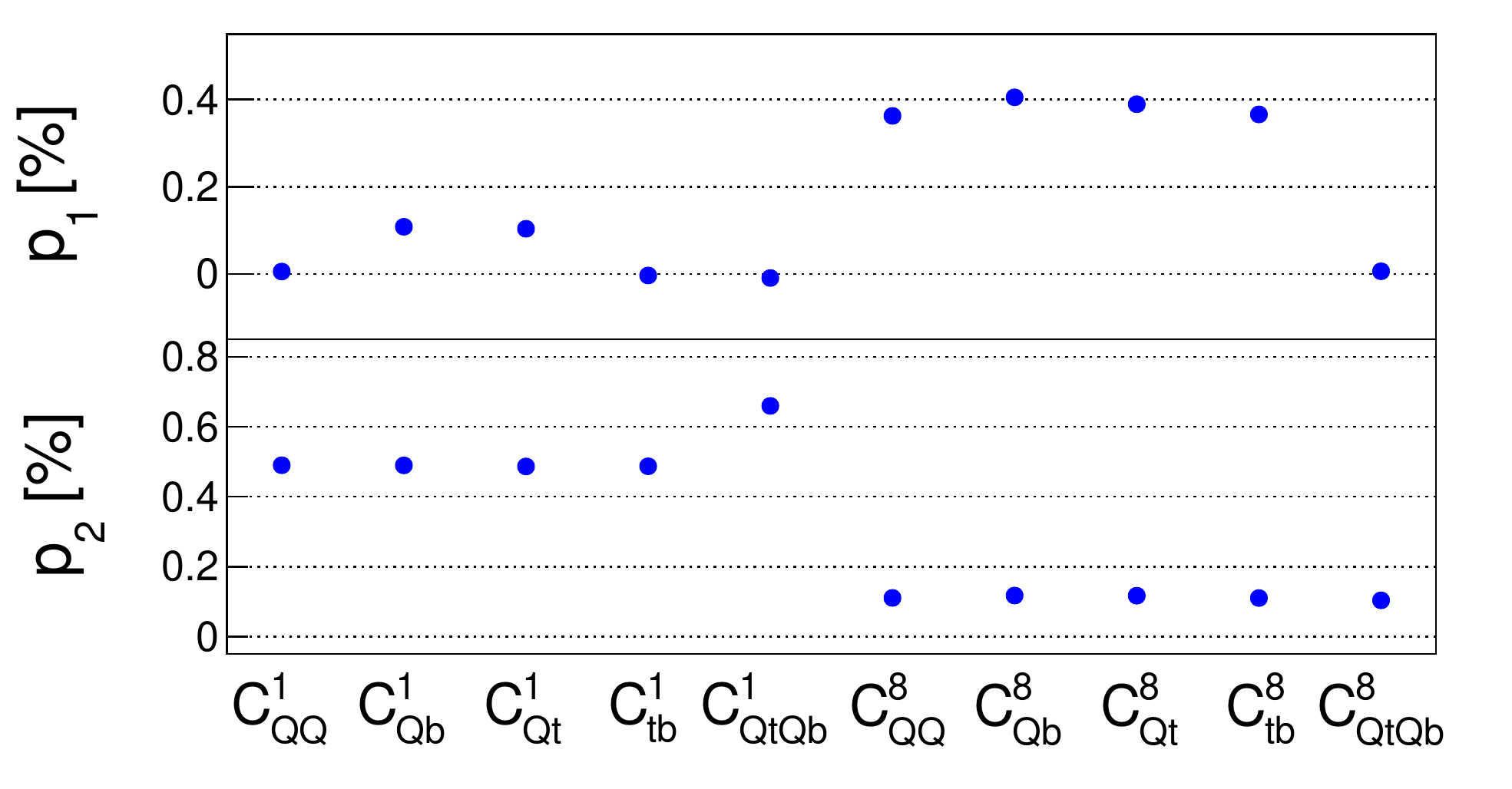}
\caption{\label{fig:xsec_coeff}
Coefficients of the fit to the cross section in the fiducial detector volume, for different EFT operators turned on one by one. The fit function has the form $\sigma_{fit} = \sigma_{SM} \left( 1 + p_{1} \cdot C_{i} + p_{2} \cdot C_{i}^{2} \right)$. In this notation $\sigma_{SM}$ represents the SM cross section, $p_{1}$ signifies the fractional importance of the interference of the EFT with the SM and $p_{2}$ represents the fractional pure EFT contribution to the cross section at quadratic order in the coupling strength of the EFT operator.
}
\end{figure}

The values of the coefficients of Eq.~\eqref{eq:xsec} for the fitted visible cross section for each operator can be found in Figure~\ref{fig:xsec_coeff}. 
One immediately observes a trend between color singlet operators and color octet operators for the values of $p_{1}$ and $p_{2}$. As expected, the singlet operators have comparatively small interference with the SM and their contribution to the cross section is dominated by the squared order in the Wilson coefficient. We observe a preferential interference of the singlet operators with opposite top and bottom chiralities ($O_{Qb}^{1}$ and $O_{Qt}^{1}$) while those that mediate same-chirality $\ttbb$ configurations ($O_{QQ}^{1}$ and $O_{tb}^{1}$) are suppressed. The color octet operators, however, clearly have a stronger interference with the SM because the SM processes leading to a $\ttbb$ final state are dominantly mediated by QCD. Their quadratic contribution to the cross section is smaller compared to the color singlet operators, which can be explained by a relative color factor of 2/9 in the EFT vertex, consistent with the observed values of $p_{2}$ between the two types of operators. This factor does not apply to the scalar current operators, due to the fact that they contain both $b\bar{t}t\bar{b}$ and $t\bar{t}b\bar{b}$ whose interference contribution will have a different color factor. The interference of these operators is $m_b$ suppressed. In any case, the interference terms only dominate the squared term for color octet operators when the Wilson coefficients are below roughly 3.3 [TeV$^{-2}$], which is below the sensitivity that can be achieved with this measurement, meaning that the squared order contributions dominate the limits in all cases.

\begin{figure}[ht!]
\center
\includegraphics[height=.3\textheight]{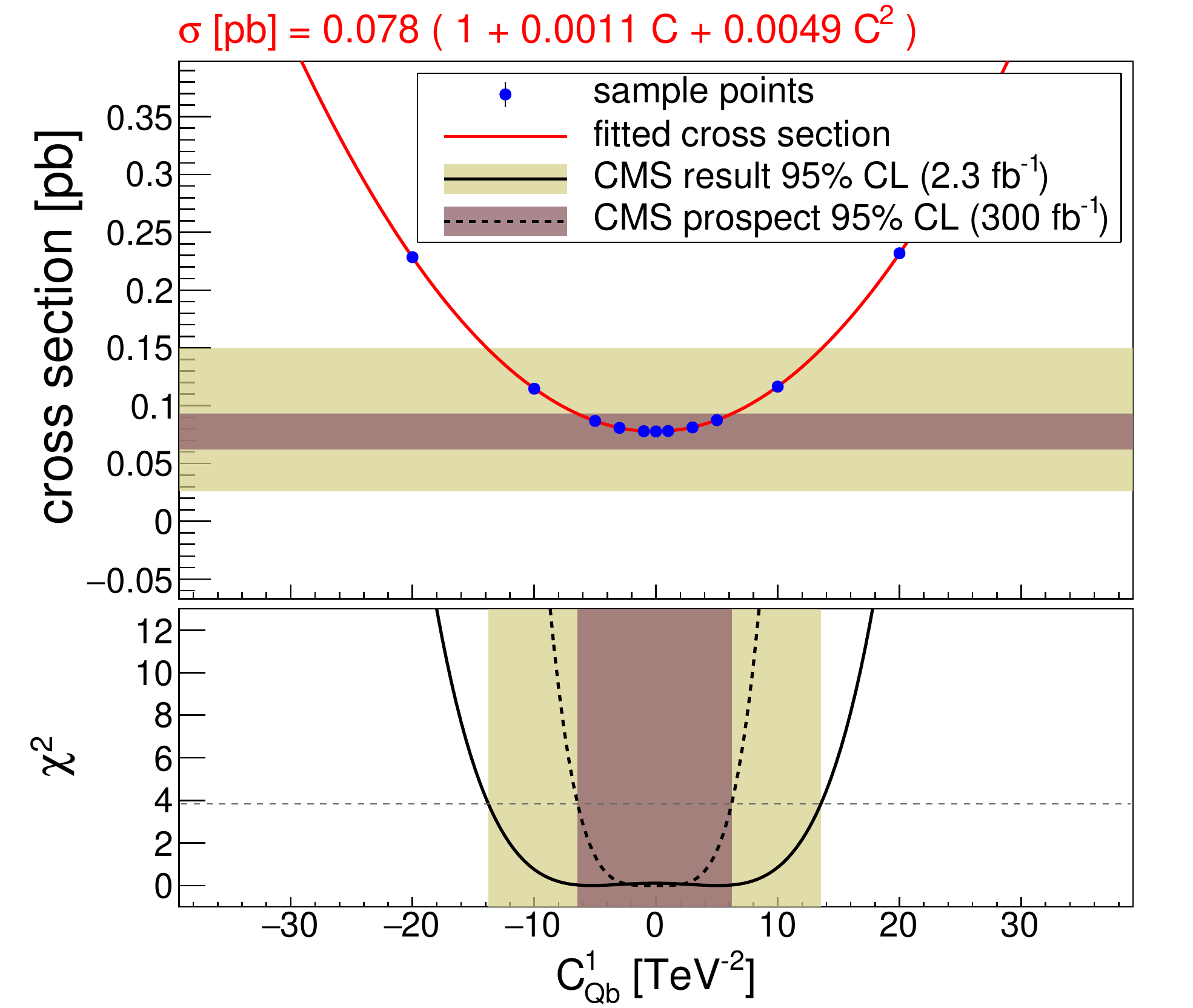}
\includegraphics[height=.3\textheight]{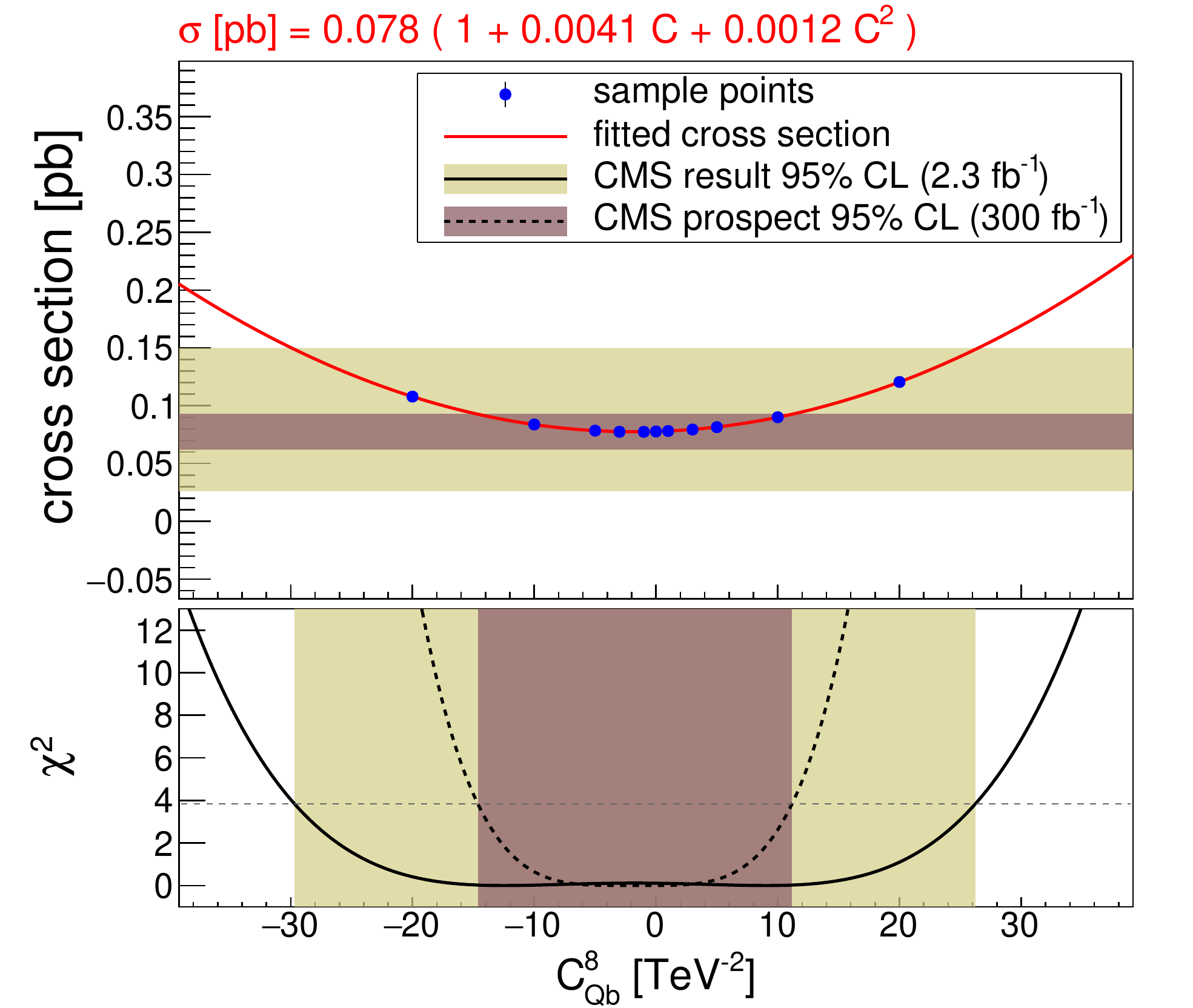}
\caption{\label{fig:xsec_limits}
Limits at 95\% CL on $C_{Qb}^{1}$ (left) and $C_{Qb}^{8}$ (right) using the CMS result with 2.3 fb$^{-1}$ (full line, light brown area) and prospects for 300 fb$^{-1}$ (dashed line, dark brown area).
}
\end{figure}

The measured $\ttbb$ cross section from proton-proton collisions at 13 TeV by CMS is found to be $\sigma_{\ttbb,CMS} = 0.088 \pm 0.012 \ (stat.) \pm 0.029 \ (syst.)$ pb. The LO computation of the SM $\ttbb$ cross section with MG5\_aMC@NLO in the visible phase space defined above yields a value of 78 fb. This is comparable to the NLO prediction with Powheg \cite{Nason:2004rx,Frixione:2007vw,Alioli:2010xd} of 70 $\pm$ 9 fb, as quoted in the CMS measurement and within the uncertainties of the CMS measurement. The total uncertainty is obtained by adding the statistical and systematical uncertainty in quadrature and is taken to be $\delta_{\ttbb,CMS} = 0.031$ pb (or 35\%). Indicative results for the sensitivity to  $C_{Qb}^{1}$ and $C_{Qb}^{8}$ are shown in Figure~\ref{fig:xsec_limits}, where in the top panel the red band shows the fitted cross section to the sample points with uncertainties (the fitted function is also quoted in red on top of the Figure). The light brown band represents the CMS measurement with uncertainties. In the bottom panel the full line is the resulting $\chi^{2}$ as a function of the Wilson coefficient and the light brown band shows the corresponding 95\% CL interval. The minima of the $\chi^{2}$ are not centered at 0, indicating the fact that the cross section obtained by the calculation of MG5\_aMC@NLO at leading order is slightly below the measured value by CMS, but still well within the uncertainty of the measurement.

Overall, both the linear and quadratic EFT contributions to the cross section stay below the percent level. 
We obtain limits of around [-14,14] (TeV$^{-2}$) and [-30,26] (TeV$^{-2}$) for the example color singlet and octet operators respectively. 
The projected sensitivities are shown with the dark brown bands and the dotted lines in Figure~\ref{fig:xsec_limits}, improving the limits to around [-6,6] (TeV$^{-2}$) and [-14,11] (TeV$^{-2}$) respectively. The limits on all Wilson coefficients are summarized in the left panel of Figure~\ref{fig:summary_Limits} (black and red lines for 2.3 fb$^{-1}$ and 300 fb$^{-1}$ respectively) at the end of this chapter.

\subsection{Quantifying the validity of the EFT
\label{subsec:mcut}}
\begin{figure}[ht!]
\center
\includegraphics[width=.48\textwidth]{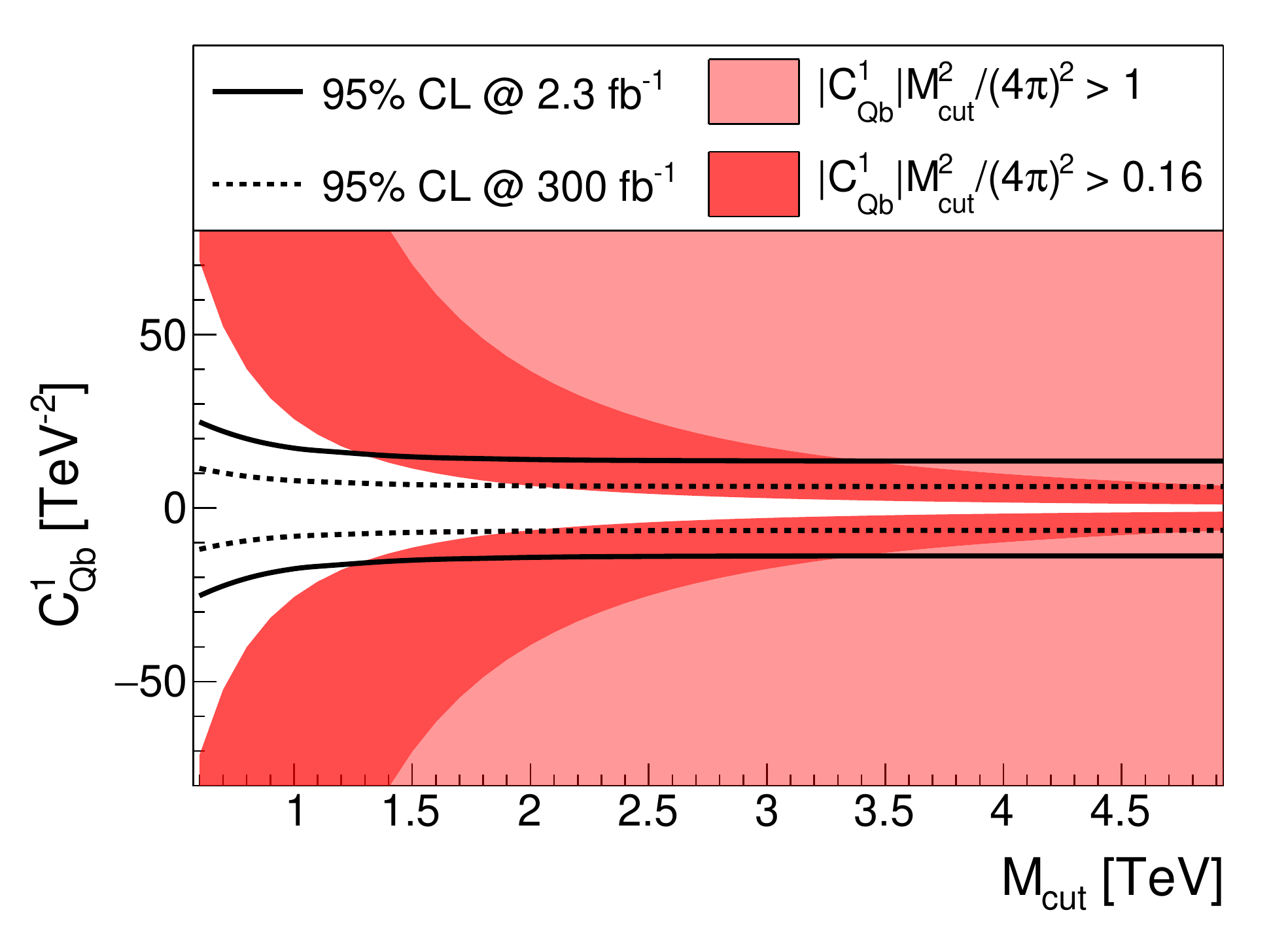}
\includegraphics[width=.48\textwidth]{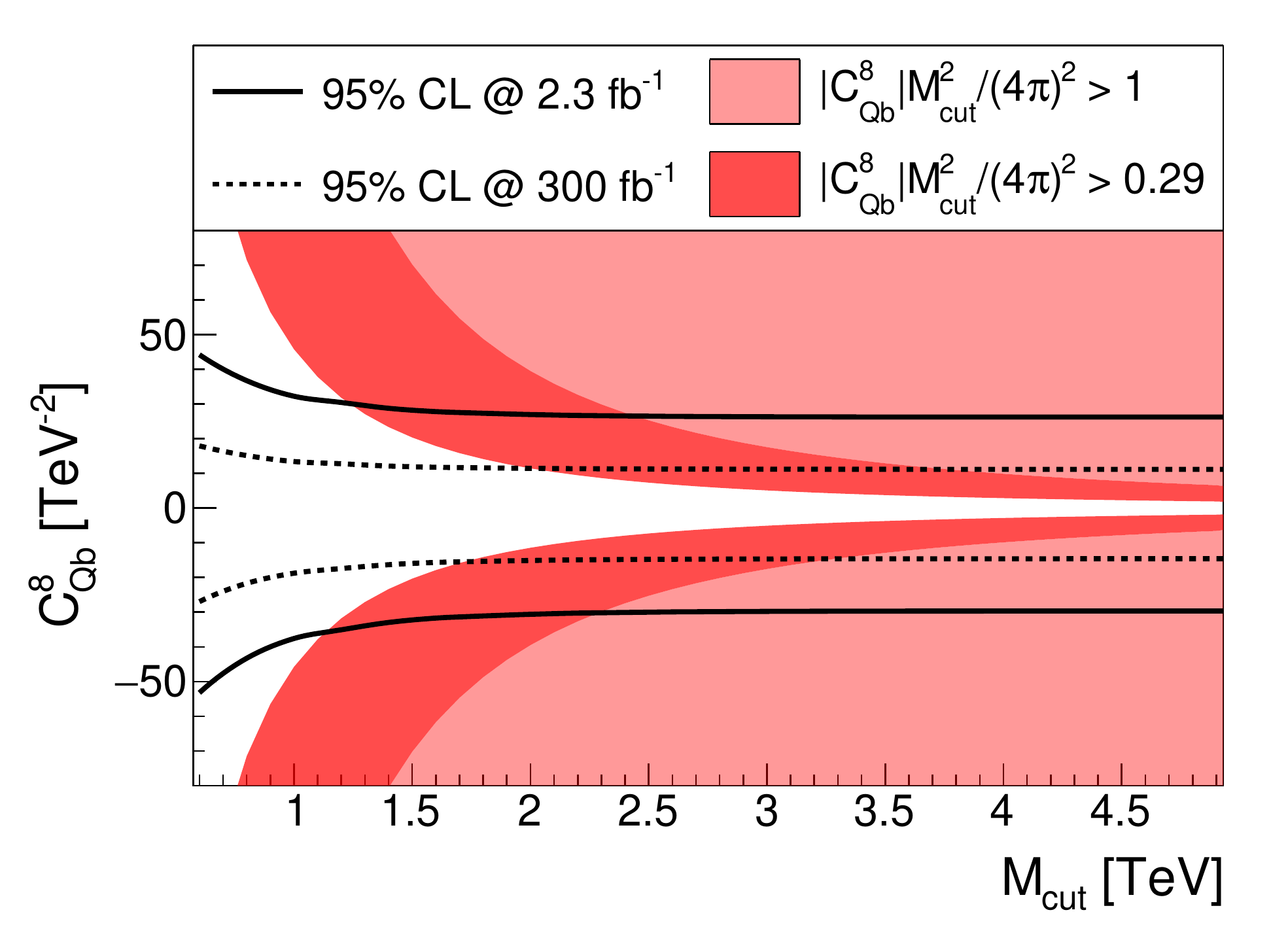}
\caption{\label{fig:limitsVsMcut}
Limits at 95\% CL on $C_{Qb}^{1}$ (left) and $C_{Qb}^{8}$ (right) as a function of the mass cut $M_{cut}$ for an integrated luminosity of 2.3 fb$^{-1}$ (full line) and projections to 300 fb$^{-1}$ (dashed line). The non-perturbative regime of the EFT in which $|C_{i}| M_{cut}^{2} > (4\pi)^{2}$ is indicated with the light pink shaded region. The darker red region represents a more stringent perturbativity requirement for which the upper limit on the Wilson coefficient (at 300 fb$^{-1}$) intersects the perturbativity threshold at $M_{cut}$ = 2 TeV.
}
\end{figure}

\begin{figure}[ht!]
\center
\includegraphics[height=.35\textheight]{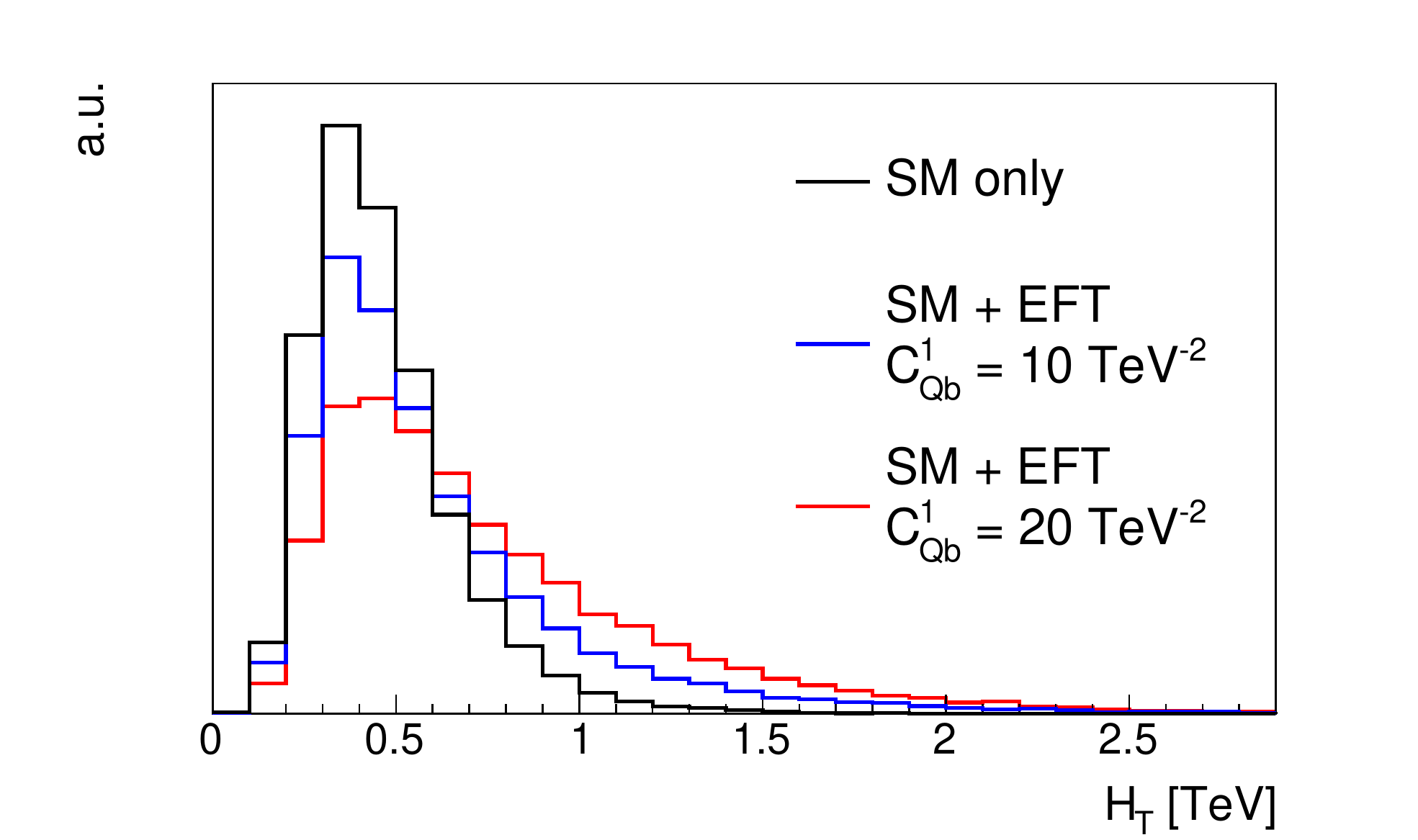}
\caption{\label{fig:HT}
Scalar sum of transverse momentum of all final state physics objects in the event ($H_{T}$).
}
\end{figure}

We now investigate the dependence of these limits on the value of $M_{\text{cut}}$. This will be used to assess the validity criteria as discussed in Section~\ref{subsec:UV}. 
As previously mentioned, we apply a cut on all energy (or mass) scales that appear in the events\footnote{This is a somewhat conservative approach, since the actual energy that is exchanged in an EFT vertex is typically lower than, {\it e.g.}, the summed transverse momenta of all final state particles. However, at dimension-8, $gg\ttbb$ contact terms are present, that would be sensitive to the total centre of mass energy of the scattering. }. Lowering the value of $M_{cut}$ enlarges the range of new physics mass scales compatible with EFT validity criterion at the price of a reduced sensitivity.
This is illustrated in Figure~\ref{fig:limitsVsMcut}, where the non-valid region defined by Eq.~\eqref{eq:pert} is indicated by the light pink shaded area as a function of the value of $M_{cut}$. The limits on the Wilson coefficient $C_{Qb}^{1}$ (left) and $C_{Qb}^{8}$ (right), both for the measurement at 2.3 fb$^{-1}$ (full black line) and for the prospects at 300 fb$^{-1}$ (dashed black line) are superimposed. 
The limits are almost insensitive to value of $M_{cut}$ down to 1.5 TeV and we therefore fix it to 2 TeV throughout the rest of this study. 
Given Eq.~\eqref{eq:validity}, the value of $M_{cut}$ serves as a hard lower bound on the scale of new physics that can be used to interpret limits on the Wilson coefficients. Imposing this cut trades sensitivity for a controlled interpretability of the results.
The dark pink shaded region illustrates a more stringent requirement than in Eq.~\eqref{eq:pert}, namely \( \frac{|C_{i}| M_{cut}^{2}}{(4\pi)^{2}} < \kappa^2\).
The specific value of $\kappa^2$ is chosen such that the edge of the new valid region intersects the projected upper limit for $M_{cut}=$2 TeV (at 300 fb$^{-1}$).
This provides a conservative estimate of the perturbative uncertainty of the EFT predictions at the edge of our sensitivity (see discussion around Eq.~\eqref{eq:pert}).
Finally in Figure~\ref{fig:HT}, the normalised distributions of the scalar sum of the transverse momentum of all visible objects, $H_{T}$, in the final state is shown comparing the SM contributions (black), with those of the $O_{Qb}^{1}$ operator with $C_{Qb}^{1}$ fixed at 10 TeV$^{-2}$ (blue) and 20 TeV$^{-2}$ (red) respectively. This is a representative variable for the typical energy scale of the $\ttbb$ events and indeed we see that only a small fraction of the events are present above $H_{T} = 2$ TeV.

\subsection{Tailoring the kinematical phase space}
\label{sec:EFTPhaseSpace}
%
In order to optimise the sensitivity of our process to the operators of interest, we go beyond inclusive level and consider observables with an enhanced dependence on the presence of EFT operators. A first step is to select a part of the phase space in which the EFT contributions are more abundant relative to the SM ones, as first proposed for this process in Ref. \cite{Degrande:2010kt}. 

\begin{figure}[ht!]
\center
\includegraphics[height=.35\textheight]{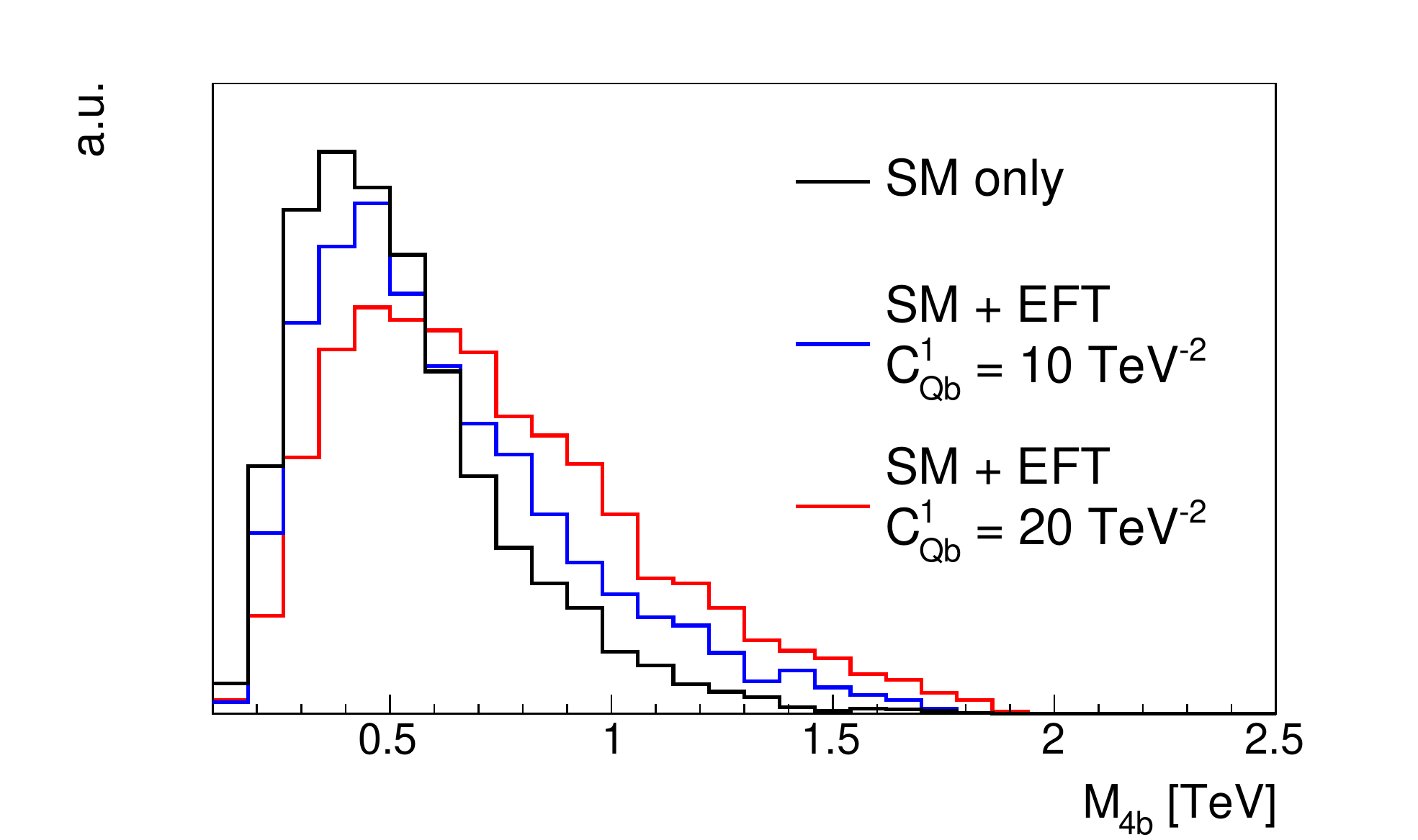}
\caption{\label{fig:M4b}
Invariant mass of the four leading jets in the event ($M_{4b}$) with and without the presence of EFT operators and before cutting on the maximal mass scale $M_{cut}$.
}
\end{figure}

\begin{figure}[ht!]
\center
\includegraphics[width=0.45\textwidth]{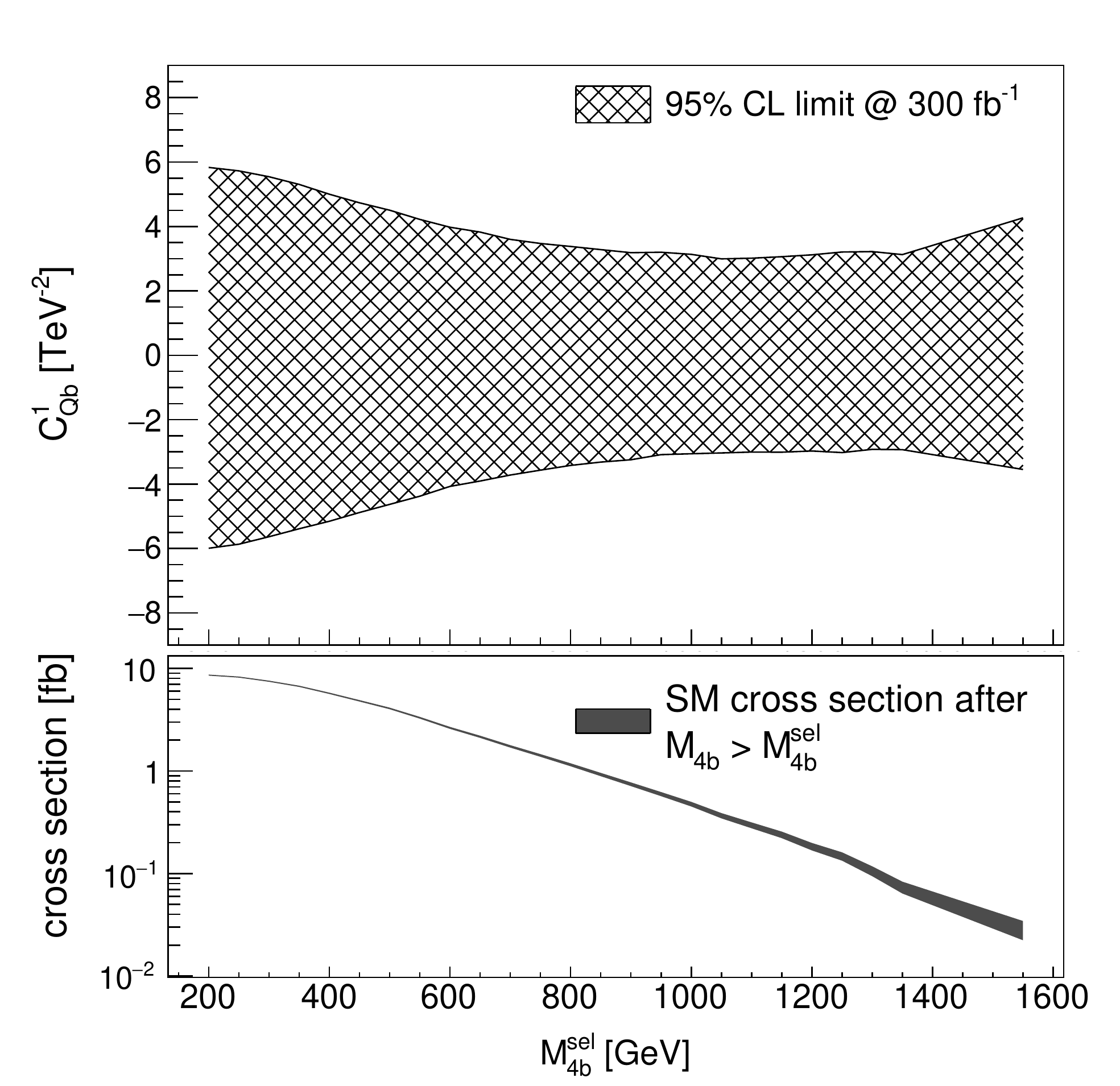}
\includegraphics[width=0.45\textwidth]{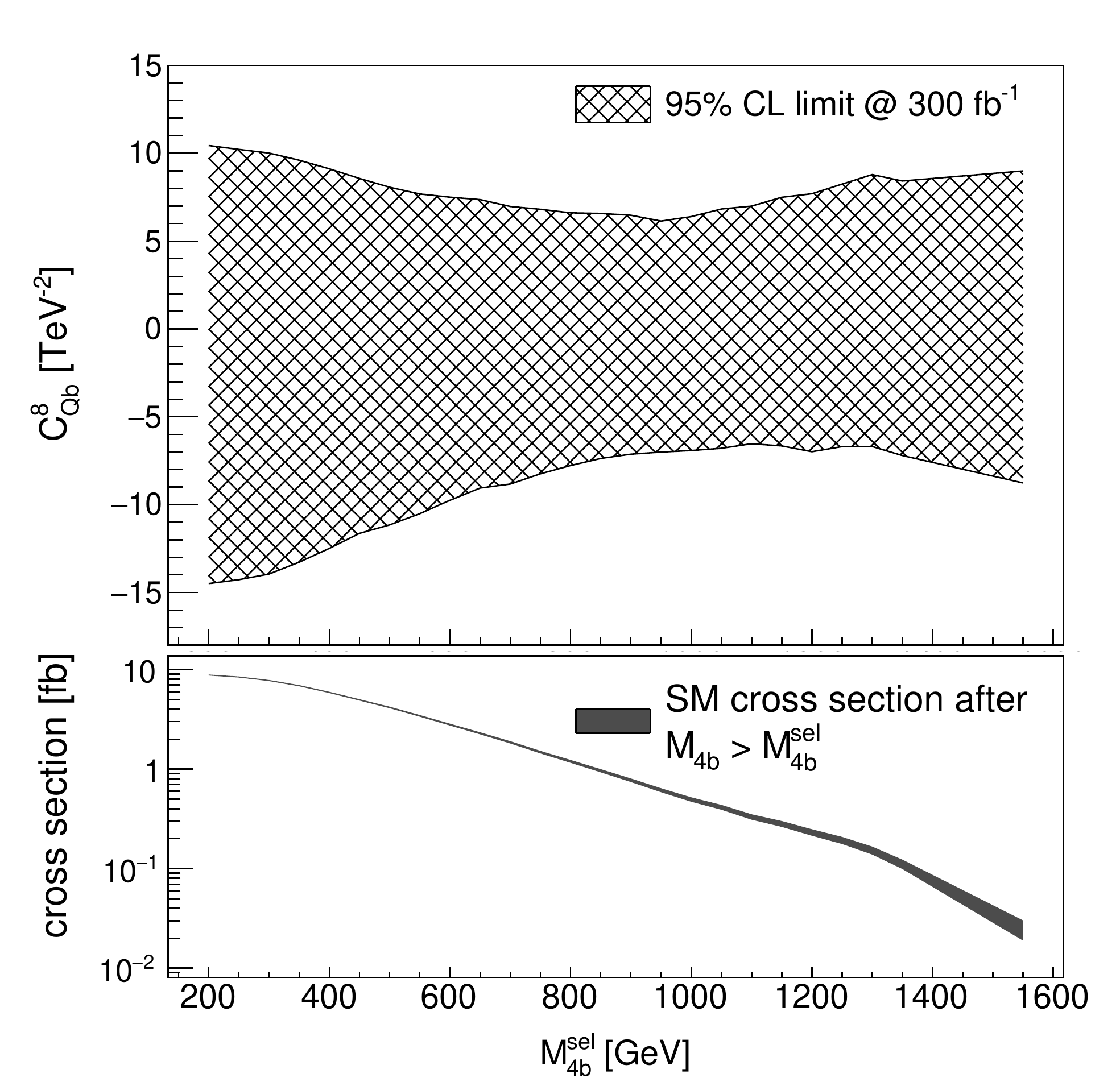}
\caption{\label{fig:limits_vs_M4b_singlet}
Individual limits at 95\% CL on $C_{Qb}^{1}$  and $C_{Qb}^{8}$ as a function of the threshold on the invariant mass of the 4 $b$-jets in the event ($M_{4b}^{sel}$). In the bottom panel the predicted SM cross section as a function of $M_{4b}^{sel}$ is shown, with the corresponding statistical uncertainty shown as a grey band.
}
\end{figure}

After the full event selection outlined in Section~\ref{sec:strategy} is applied, we define a set of reconstructed variables and identify those that show a clear difference in shape between the SM and EFT operators. Different such quantities were tested, including the transverse momenta, invariant masses and $\Delta R$ separation between final state objects. These variables are summarized in Table \ref{tab:inputs} of Appendix \ref{app:NN}. The separating strength of each variable is calculated by an ANOVA (Analysis of variance) F-statistic \cite{heiman2010basic}, which reflects the distance between the means of the SM and EFT distributions and is further defined in Appendix \ref{app:NN}. The invariant mass of the 4 $b$-jets in the final state ($M_{4b}$) was found to be the most discriminating variable and is illustrated in Figure \ref{fig:M4b}, comparing the shape of the SM (black) prediction to that of the $O_{Qb}^{1}$ operator with the Wilson coefficient fixed at 10 TeV$^{-2}$ (blue) and 20 TeV$^{-2}$ (red). This observable is able to capture the heightened energy dependence since at least two of the $b$-jets always originate from the EFT vertex, albeit sometimes via the decay of a top quark. Note that the tail of this distribution beyond 2 TeV is never included in our analysis due to the global $M_{cut}$ restriction on all energy scales introduced in Section~\ref{subsec:UV}. We define a signal cross section by applying the selection $M_{4b} > M^{sel}_{4b} = 1.1$ TeV, chosen to maximise the sensitivity to the Wilson coefficients, as shown in Figure \ref{fig:limits_vs_M4b_singlet} for $C_{Qb}^{1}$ and $C_{Qb}^{8}$.

\begin{figure}[ht!]
\center
\includegraphics[height=.3\textheight]{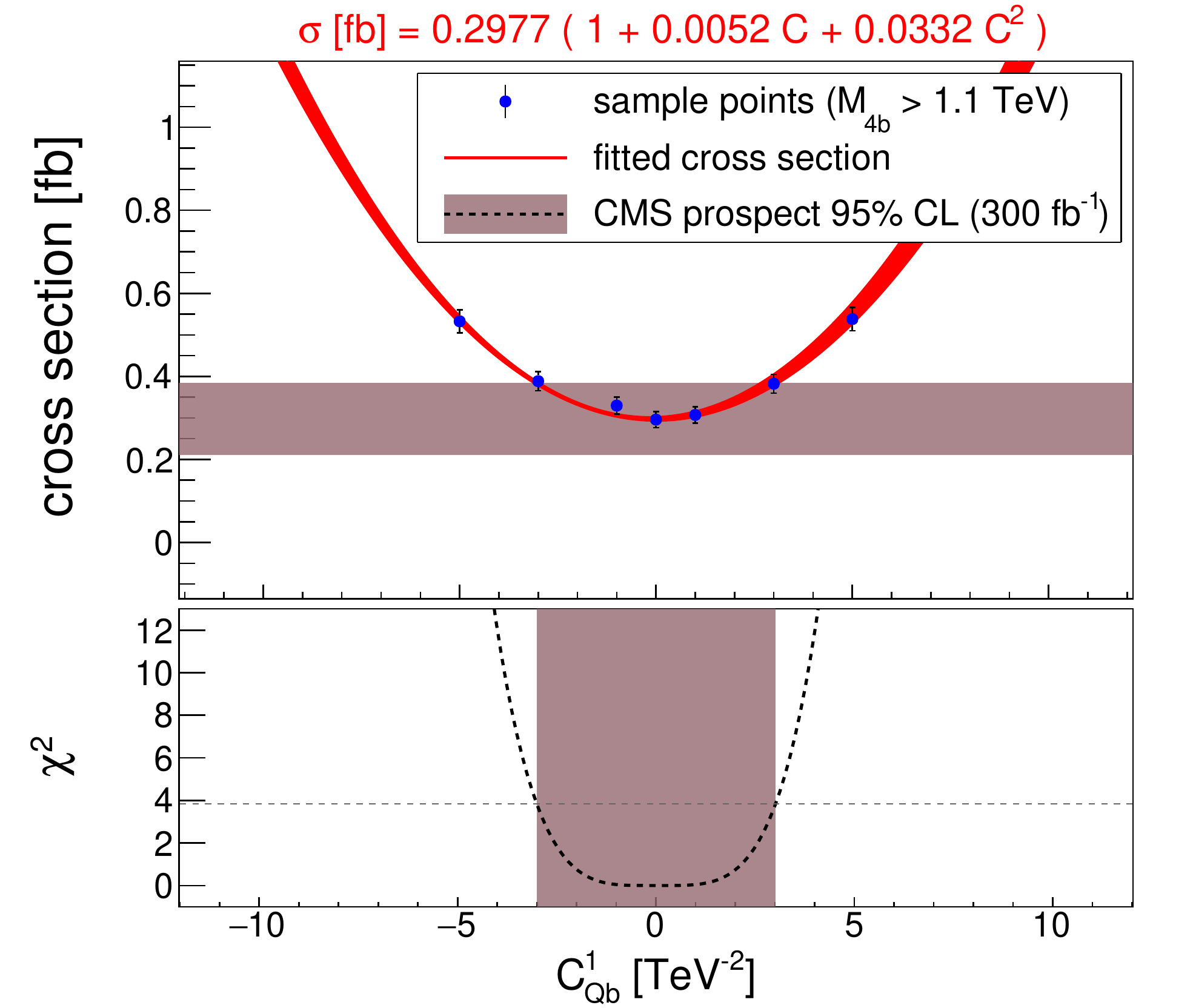}
\includegraphics[height=.3\textheight]{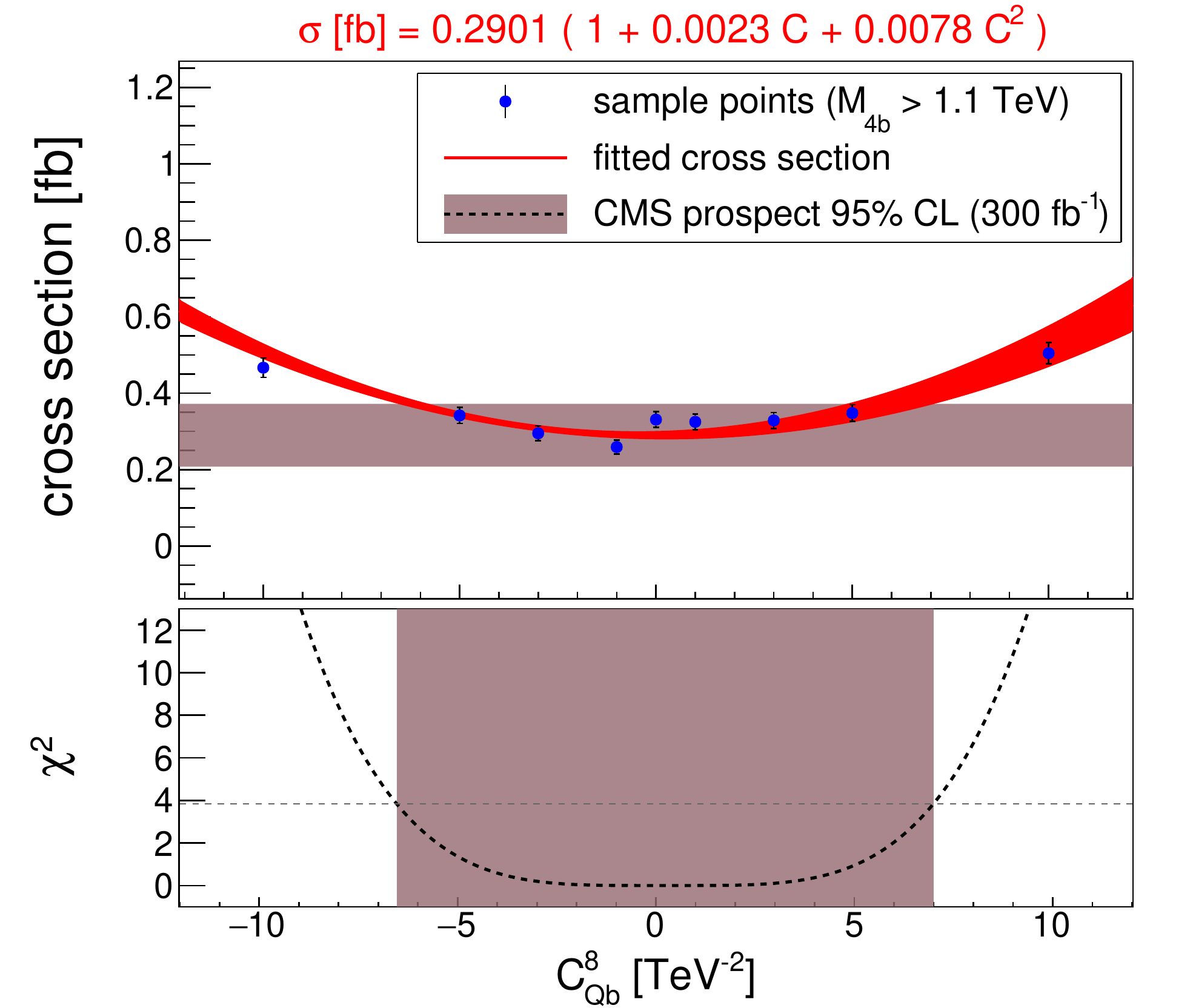}
\caption{\label{fig:xsec_limits_SensitiveVariable}
Limits at 95\% CL on $C_{Qb}^{1}$ (left) and $C_{Qb}^{8}$ (right) after applying a cut on $M_{4b} > $ 1.1 TeV, assuming an integrated luminosity of 300 fb$^{-1}$.
}
\end{figure}

After the reconstruction and the application of the above mentioned event selections (including $M_{4b} > 1.1$ TeV), we determine the functional dependence of the observed cross section on the value of the Wilson coefficients and the resulting 95\% CL interval on those coefficients. This is illustrated again for $C_{Qb}^{1}$ and $C_{Qb}^{8}$ in Figure \ref{fig:xsec_limits_SensitiveVariable}, from which it can be seen that the fit parameters $p_{1}$ and $p_{2}$ of Eq.~\eqref{eq:xsec} are in general larger than for the inclusive cross section, indicating a stronger dependence of the cross section in the presence of the operators in the selected phase space. The limits on the Wilson coefficients consequently improve, ranging between [-3,3] (TeV$^{-2}$) and between [-6,7] (TeV$^{-2}$) for the example color singlet and octet operators respectively. The results for all operators are summarized in Figure \ref{fig:summary_Limits} (blue) on the right. The sensitivity improves by around a factor of two compared to the unfolded cross section observable. 

We remark that the optimal choice of the value of $M_{cut}$ may be altered after the selection on $M_{4b} > 1.1$ TeV. To estimate the magnitude of this effect, the analysis was repeated with a value of $M_{cut} =$ 4 TeV and a mild improvement in the limits of at most 15$\%$ was observed, at the cost of a reduced regime of interpretability for the EFT. For a consistent comparison of the different methods that we present, we fix $M_{cut} =$ 2 TeV throughout the rest of this study.

\subsection{Neural network classifier
\label{sec:NN1D}}

Instead of selecting a favourable part of the phase space based on one variable, 
one can use 
machine learning algorithms
to optimally select a part of this higher-dimensional phase space. 
In this work we will demonstrate how a simple neural network (NN) can combine the information from a set of 
kinematical properties of the final states to separate SM events from those including an insertion of an 
EFT operator\footnote{We continue to use $M_{cut}=2$ TeV, as discussed in Section \ref{subsec:mcut}.}.
Afterwards, in Section \ref{sec:MultipleOperators}, we will demonstrate that by using multi-class outputs of the neural network, we are additionally able to distinguish among different classes of operators.
This will be shown to be especially beneficial in cases where more than one Wilson coefficient is switched on.

\subsubsection{Neural network design}
To illustrate the method, we defined a set of 18 kinematical variables\footnote{ Technically, there are 16 independent phase space variables after taking into account on-shell decay conditions imposed at generation-level. This will not impact the efficacy of the Neural Network classifier to learn to distinguish between signal and background. In fact, it can be considered a safeguard against accidentally including over-correlated inputs. Indeed, Fig.~\ref{fig:corrmat} in Appendix~\ref{app:NN} indicates sizeable correlations between the input features.}, consisting of transverse momenta of the final-state particles, invariant masses ($m_{inv}$) of combinations of two, four or even all six of these final-state particles and angular separations in $\Delta R$ between combinations of two particles. The list of variables can be found in Table \ref{tab:inputs} of Appendix \ref{app:NN},
and includes the invariant mass of the four 
$b$-jets used in Section \ref{sec:EFTPhaseSpace}. These variables are fed as input to a shallow neural network with one hidden layer, containing 50 neurons 
and 3 output classes. The outputs represent the probabilities (P) of an event belonging to one of the following three categories: a Standard Model event (SM), an event from an EFT operator with a left-handed top quark ($t_{L}$) current and an event from an EFT operator with a right-handed top quark ($t_{R}$) current. This means that the training is performed only on the squared order contributions from the EFT operators and therefore the resulting classifier does not learn about possible interference effects. The advantage of this is that the signal shapes of the quadratic pieces are independent of the values of the Wilson coefficients and three distinct sample classes can be used in training. The full parametric dependence (including interference) is, of course, included in the samples on which the discriminant is evaluated to obtain the limits.
A proper treatment of the interference during training would require a parametrized learning approach as the relative impact of interference and squared terms depends on the value of the Wilson coefficient. We leave for future work this interesting possibility which may improve sensitivity to certain regions of parameter space. The choice of splitting the EFT output class into two separate contributions is motivated by the fact that we expect to see differences between the kinematics of the decay products of left-handed and right-handed top quarks. For example, the $W$ bosons from right-handed top quark decays give a harder leptonic $p_{T}$ spectrum compared to those from left-handed top quarks. The complete set of distributions of the input variables, comparing the three categories are shown in Appendix~\ref{app:NN} and suggest that a considerable amount of information is present that could be used to distinguish them. This will allow us to demonstrate that the network can not only identify events including an insertion of an EFT operator, but can additionally identify the nature of the EFT operator itself.
The neural network was implemented with the Keras~\cite{chollet2015keras} software using the TensorFlow~\cite{tensorflow2015-whitepaper} backend. For more information on the neural network architecture and training, the reader is referred to Appendix~\ref{app:NN}.

\begin{figure}[ht!]
\center
\includegraphics[width=.535\textwidth]{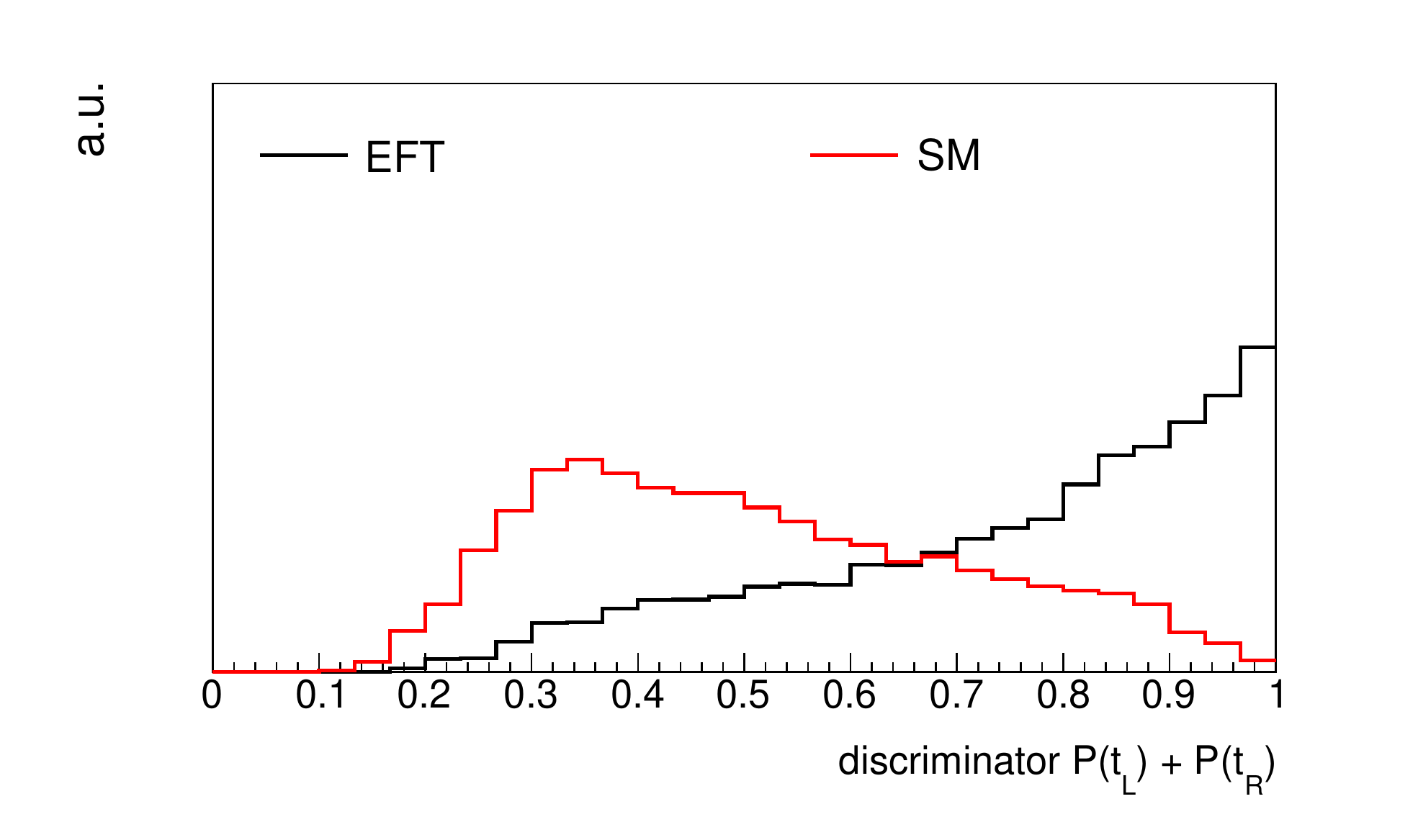}
\includegraphics[width=.365\textwidth]{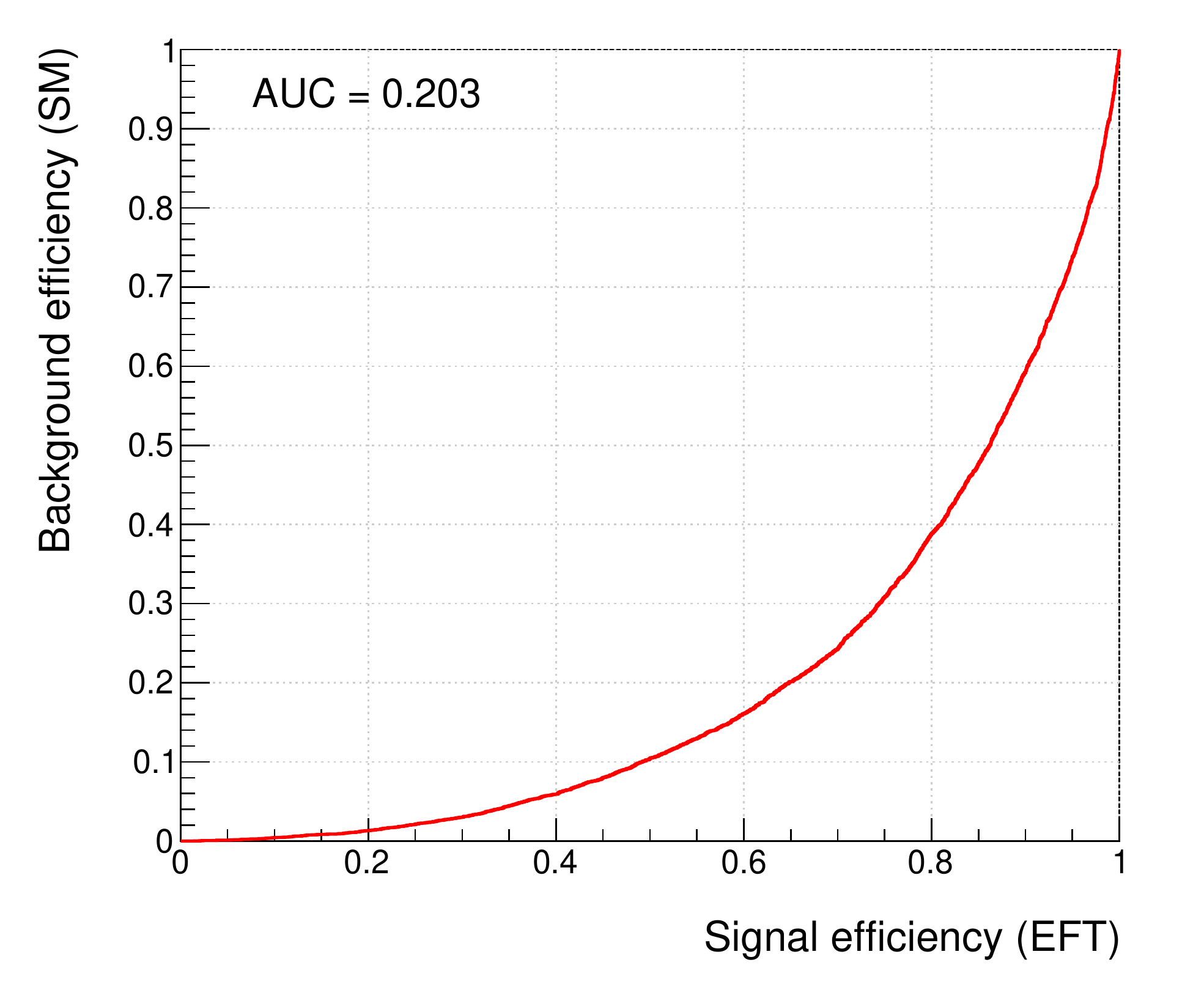}
\includegraphics[width=.535\textwidth]{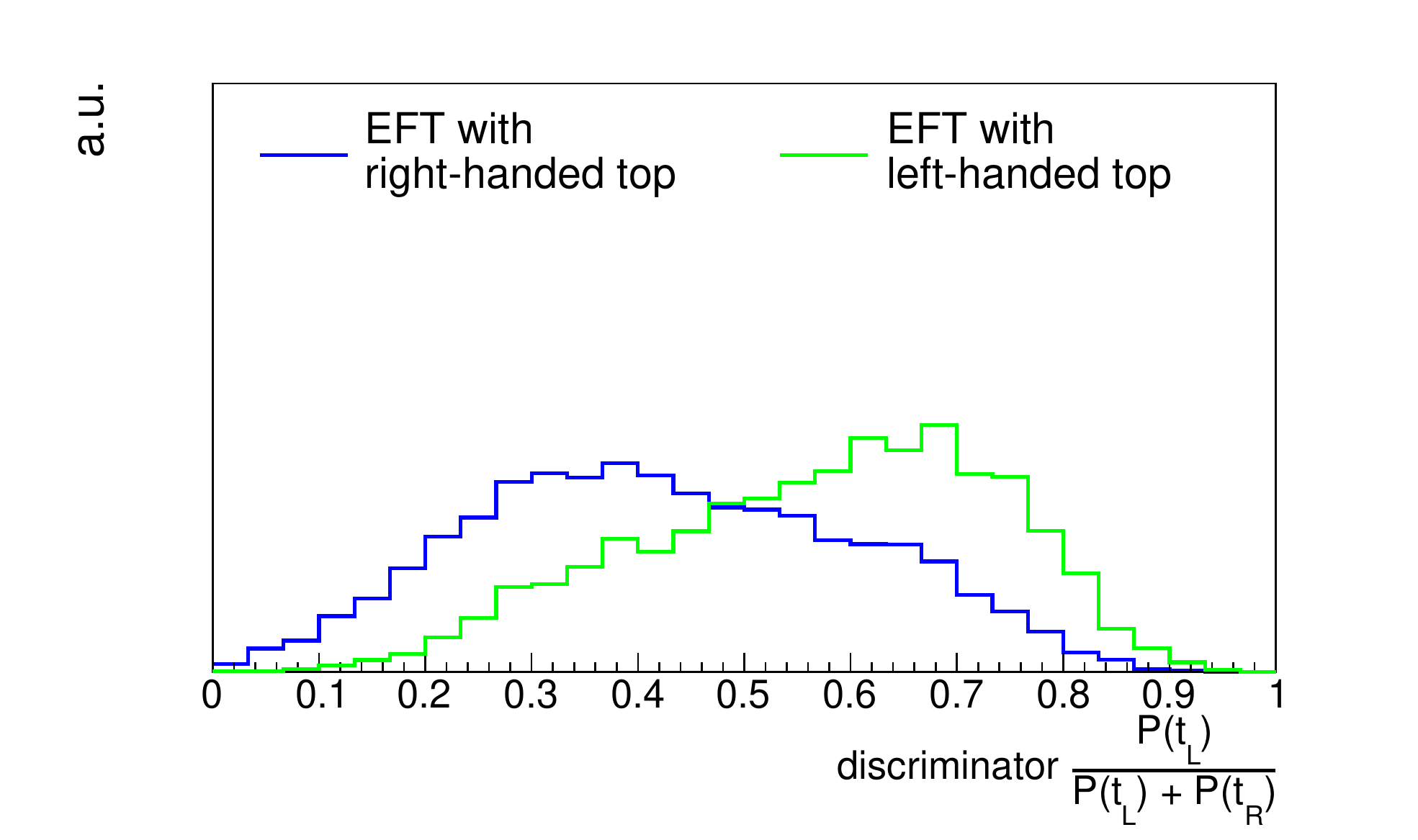}
\includegraphics[width=.365\textwidth]{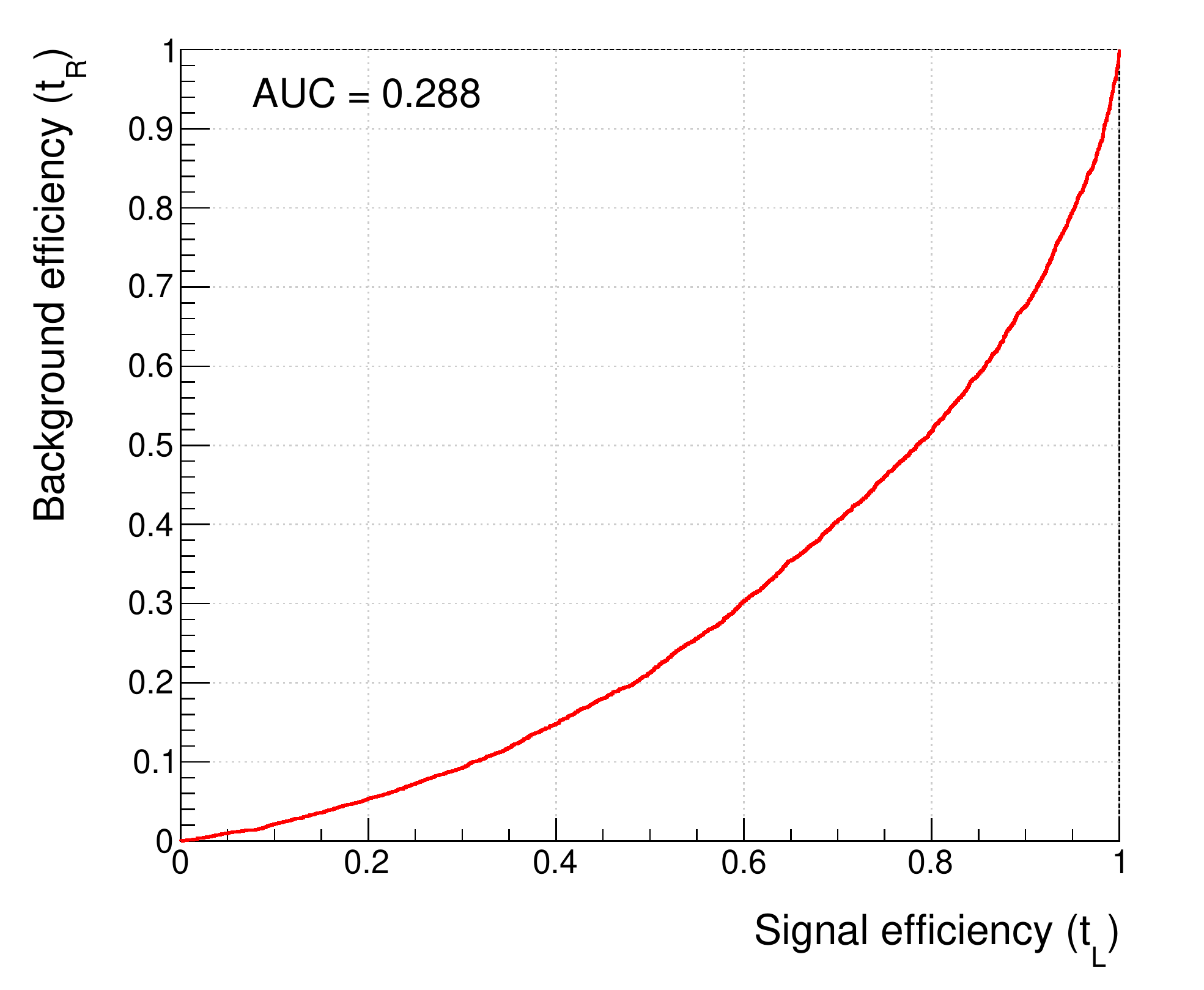}
\caption{\label{fig:Discr}
Discriminator distributions of combined outputs of the neural network that discriminate the SM processes from the EFT processes (top left) and the EFT operators with a $t_{L}$ current form the ones with a $t_{R}$ current (bottom left). The ROC curves corresponding to each of these distributions are shown on the right and the area under the ROC curve (AUC) is displayed.
}
\end{figure}

From the combination of the three outputs of the network, different observables can be constructed, each targeting a specific discrimination between two categories. The different options used in this work are summarized in Table \ref{tab:Discriminators}. In case only one operator is considered at a time, as will be discussed in Section \ref{sec:SingleOperator}, the combined NN output is constructed to optimally separate SM events from EFT operators in its category (upper two rows of Table \ref{tab:Discriminators}). However, when more than one operator is allowed to vary simultaneously, and  contributions from both the $t_{L}$ and $t_{R}$  categories are present as will be discussed in Section \ref{sec:MultipleOperators}, a combination of two observables is used (bottom rows of Table \ref{tab:Discriminators}). By adding the output probabilities of the left-handed and right-handed top quark EFT outputs ($P(t_{L}) + P(t_{R})$) one obtains a good discrimination between SM events and EFT events in general. This is illustrated in the top of Figure \ref{fig:Discr}, where on the left the normalized distributions of this combined output are shown for SM events in red and for events with a single insertion of an EFT operator in black. The corresponding receiver operating curve (ROC) is shown on the top right, showing on the x-axis the efficiency of selecting events with an insertion of an EFT operator and on the y-axis the selection efficiency for selecting a pure SM event.

Similarly, an observable can be constructed to distinguish between the second and third category, namely between events from operators containing $t_{L}$ or $t_{R}$ currents. This variable is defined as \( \frac{P(t_{L})}{P(t_{L}) + P(t_{R})} \) and is displayed in Figure \ref{fig:Discr} on the bottom left, together with the ROC curve on the bottom right. It can be clearly seen that the network has learned to differentiate between these two classes. We will use this distinction further along in Section \ref{sec:MultipleOperators} to illustrate a method which improves sensitivity when two Wilson coefficients are allowed to be non-zero at a time.

\renewcommand{\arraystretch}{3.}
\begin{table}[!ht]
 \centering
 \begin{tabular}{|c|c|c|}
 \hline
 & \multicolumn{1}{c|}{\makecell{\textbf{Desired} \\ \textbf{Discrimination}}} & \multicolumn{1}{c|}{\makecell{\textbf{Combined NN Output} \\ \textbf{used for limits}}} \\[7pt]
 \hline
 \hline
 \parbox[t]{10mm}{\multirow{1}{*}{\rotatebox[origin=r]{90}{\makecell{only $t_{L}$ \\ operator}}}} &  \large SM vs $t_{L}$  & \Large$\frac{P(t_{L})}{P(t_{L}) + P(SM)}$ \\[10pt]
 \hline
 \parbox[t]{10mm}{\multirow{1}{*}{\rotatebox[origin=r]{90}{\makecell{only $t_{R}$ \\ operator}}}}&  \large SM vs $t_{R}$  &  \Large$\frac{P(t_{R})}{P(t_{R}) + P(SM)}$\\[10pt]
 \hline
 \parbox[t]{10mm}{\multirow{2}{*}{\rotatebox[origin=r]{90}{\makecell{including both $t_{L}$ \\ and $t_{R}$ operators}}}} &  \large EFT vs SM  & $P(t_{L}) + P(t_{R})$\\[10pt]
 &  \large$t_{L}$ vs $t_{R}$  &  \Large$\frac{P(t_{L})}{P(t_{L}) + P(t_{R})}$\\[10pt]
 \hline
 \end{tabular}
 \caption{\label{tab:Discriminators}
 Definitions of the combined NN outputs used for deriving limits in different situations.
 }
 \end{table}
\renewcommand{\arraystretch}{1.}

\subsubsection{Network predictions for individual operators
\label{sec:SingleOperator}}

\begin{figure}[ht!]
\center
\includegraphics[height=.45\textwidth]{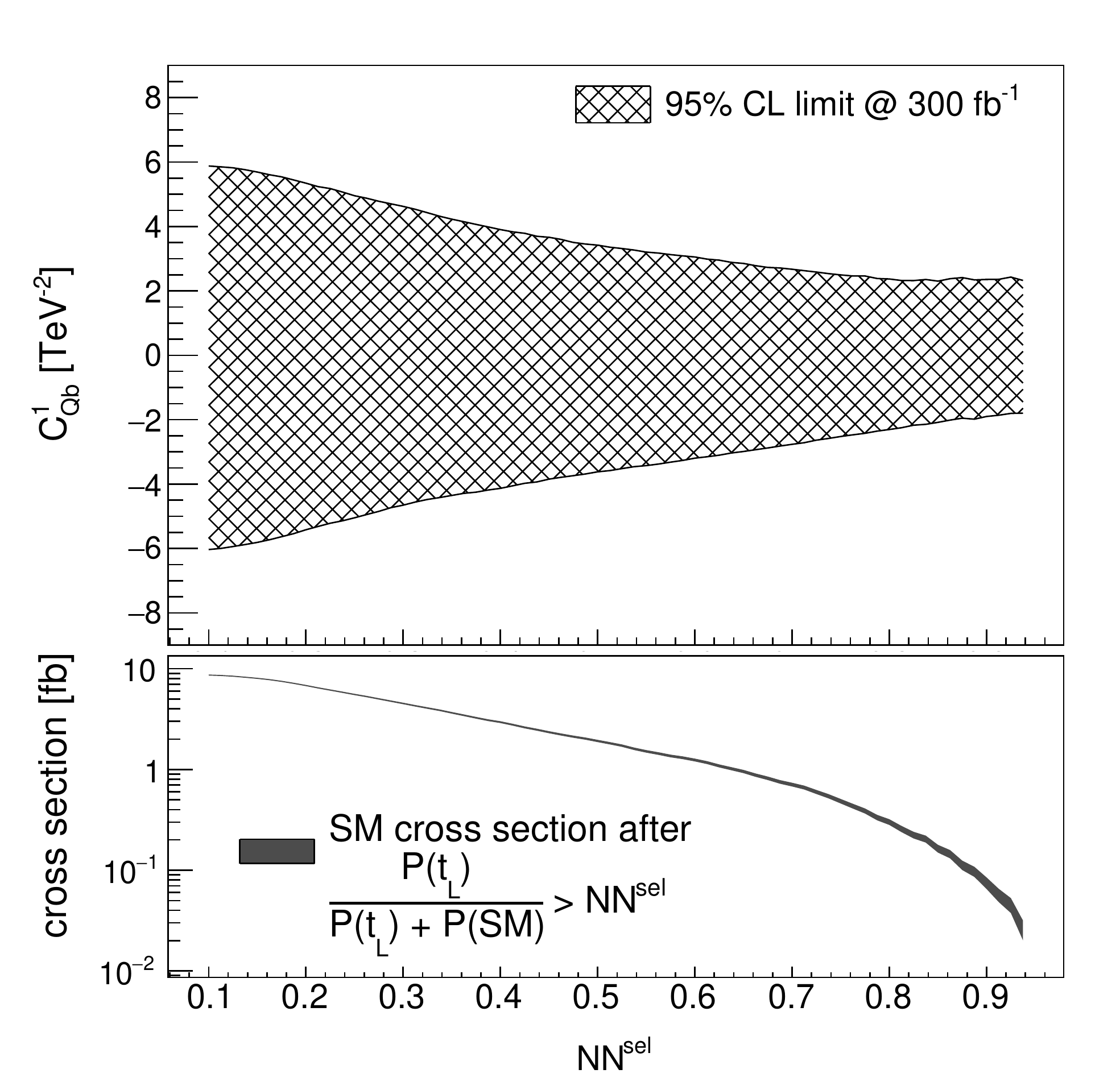}
\includegraphics[height=.45\textwidth]{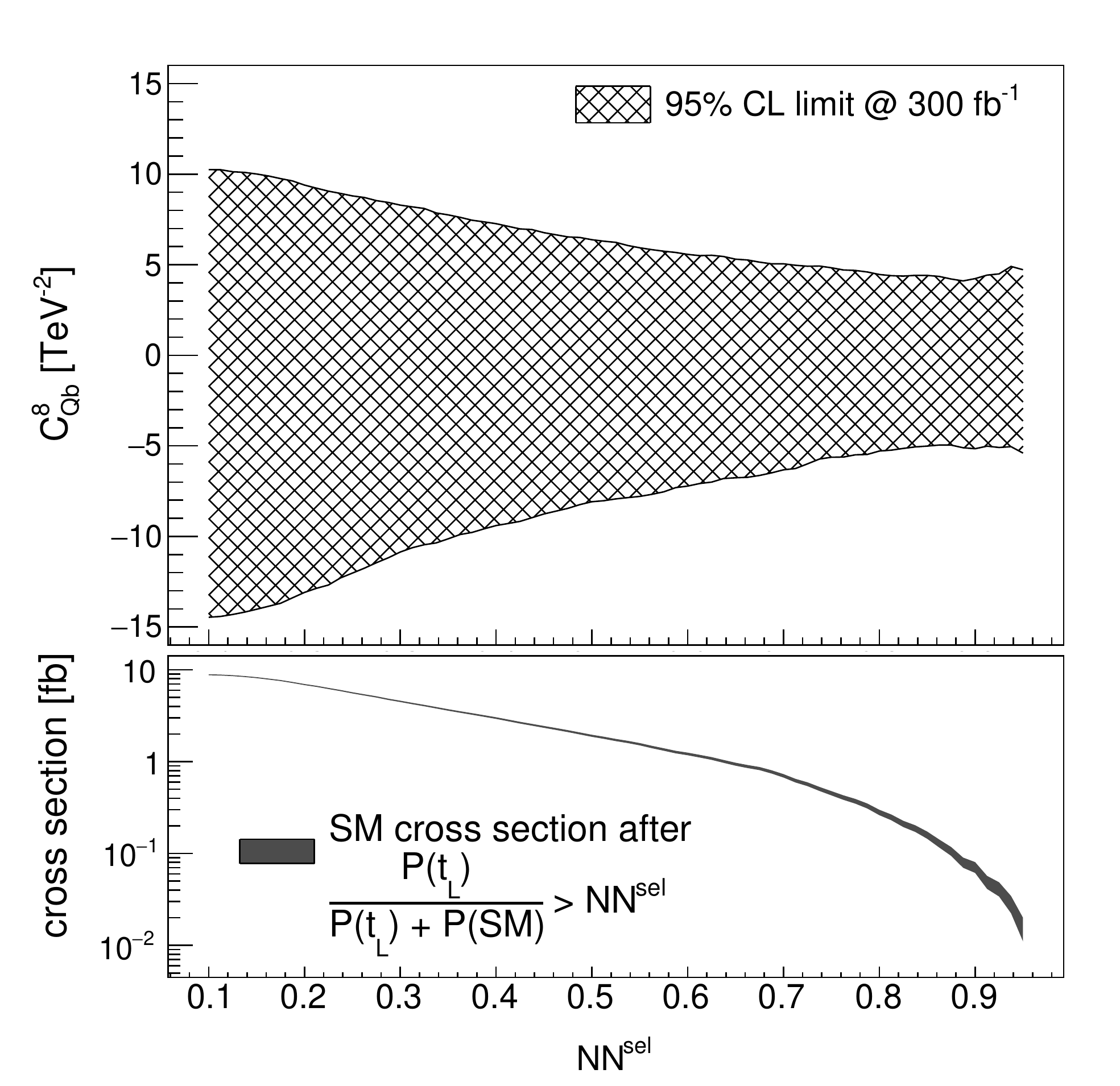}
\caption{\label{fig:limits_vs_NNoutput_singlet}
Limits at 95\% CL on $C_{Qb}^{1}$ (left) and  $C_{Qb}^{8}$ (right) as a function of the threshold on the network output. In the bottom panels, the effective (visible) SM cross section as a function of the NN output threshold is shown, with the corresponding statistical uncertainty shown as a grey band.
}
\end{figure}

\begin{figure}[ht!]
\center
\includegraphics[height=.3\textheight]{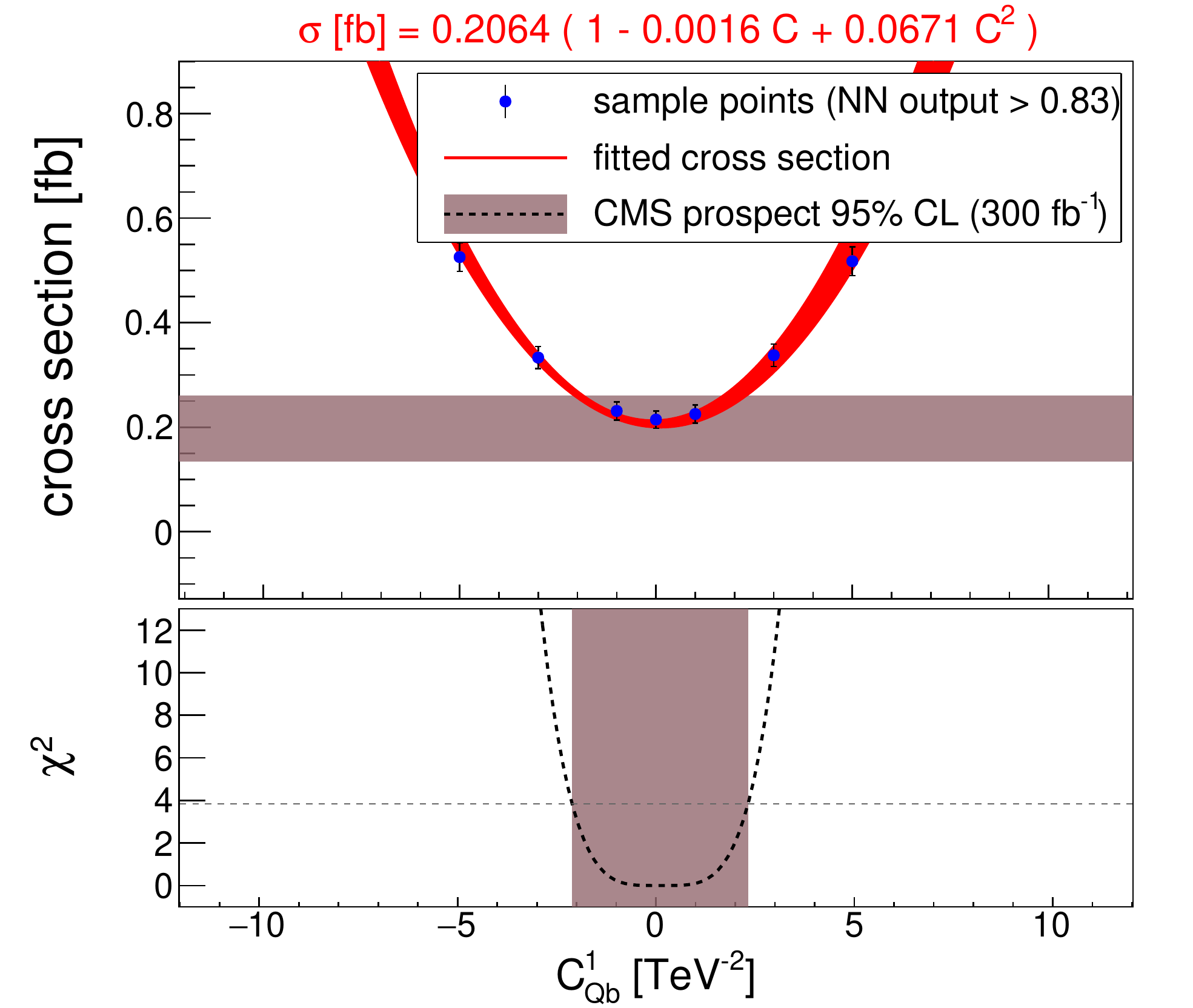}
\includegraphics[height=.3\textheight]{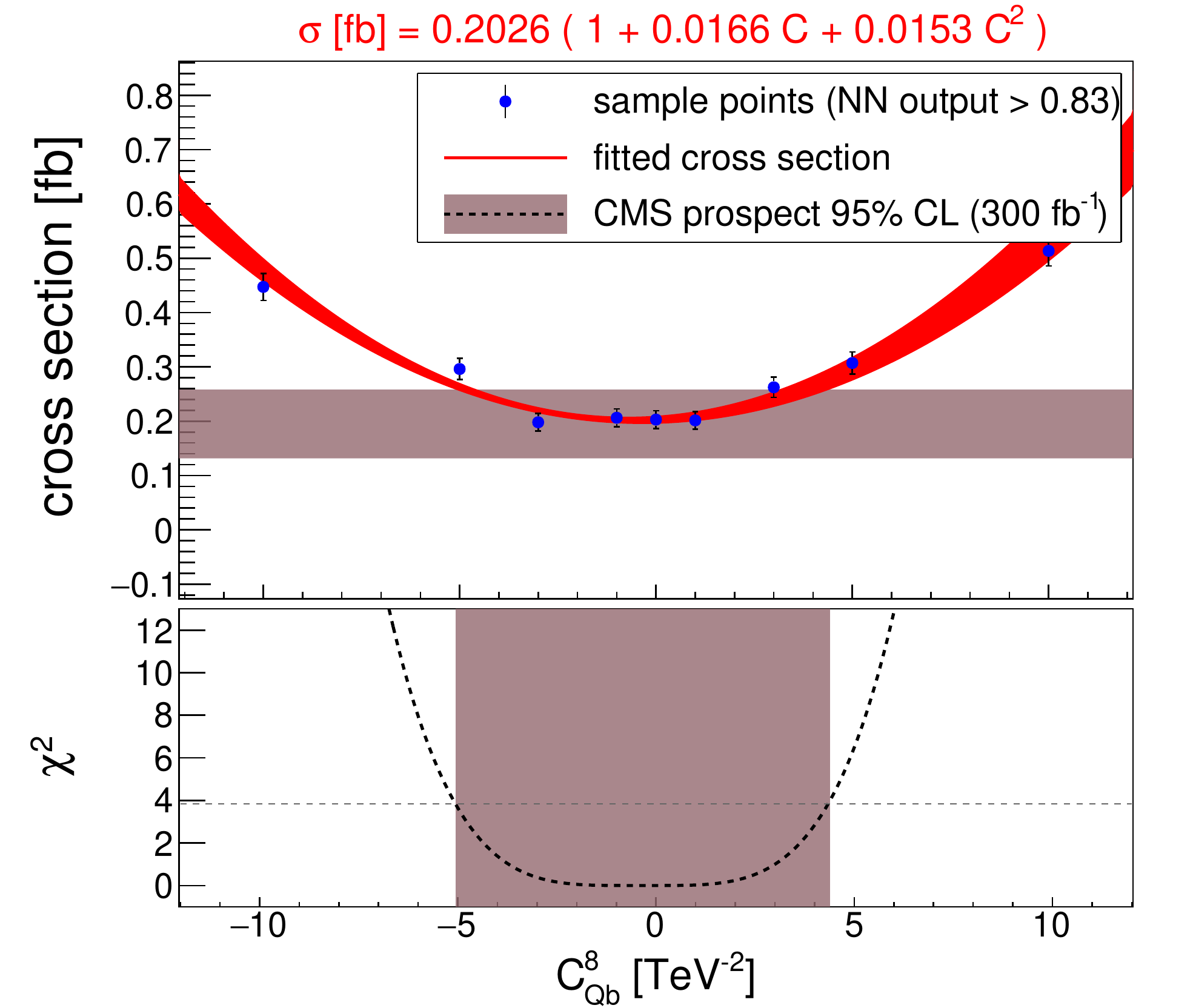}
\caption{\label{fig:xsec_limits_NNcut}
Limits at 95\% CL on $C_{Qb}^{1}$ (left) and $C_{Qb}^{8}$ (right) after requiring the network output to be above 0.83 and assuming an integrated luminosity of 300 fb$^{-1}$.
}
\end{figure}

\begin{figure}[ht!]
\center
\includegraphics[width=.48\textwidth]{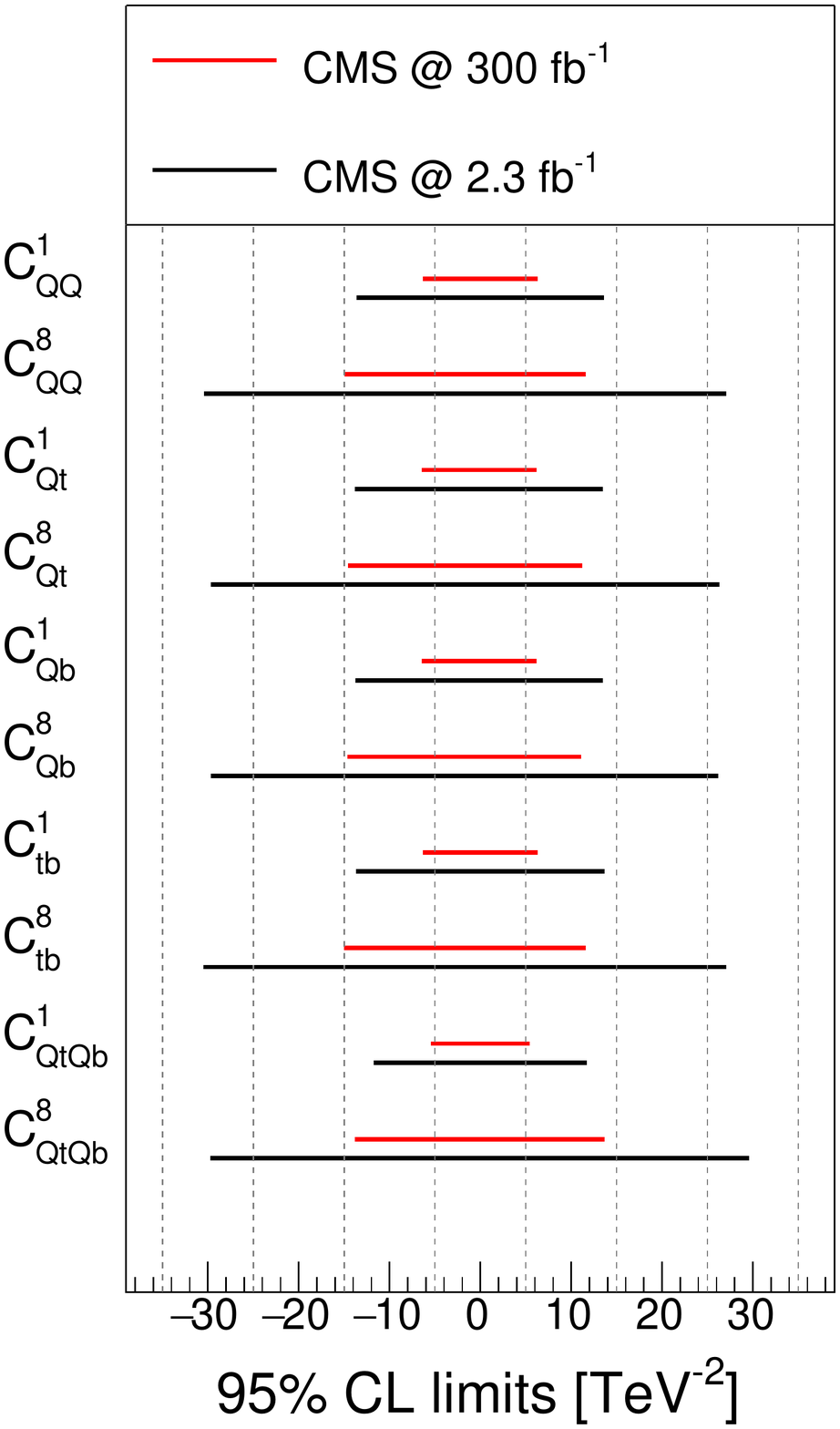}
\includegraphics[width=.48\textwidth]{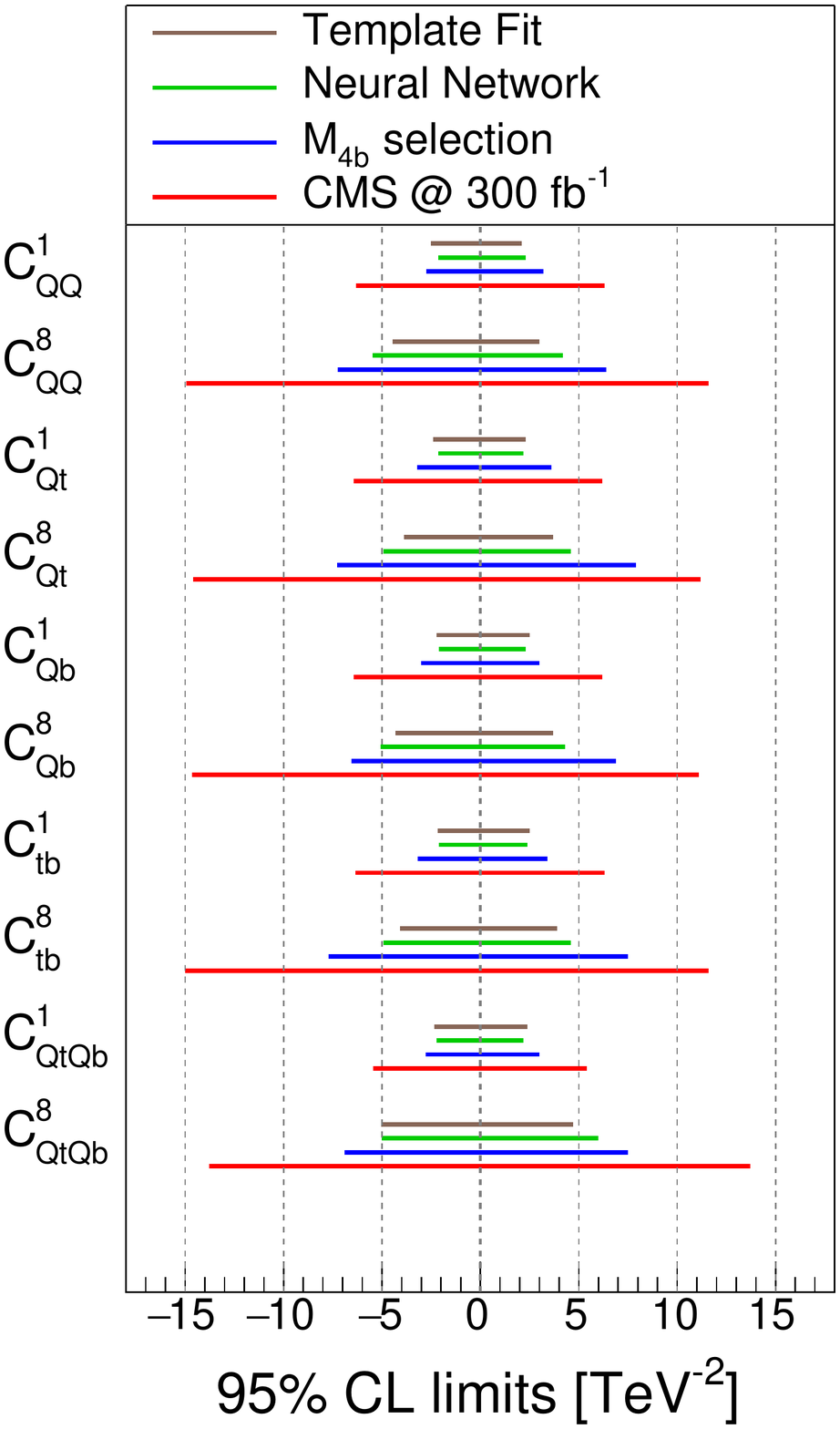}
\caption{\label{fig:summary_Limits}
(left) Summary of the individual limits at 95\% CL on all the Wilson coefficients, 
from the 13 TeV inclusive $\ttbb$ cross section measurement of CMS with 2.3 fb$^{-1}$ (black), as well as projections for 300 fb$^{-1}$ (red). 
(right) 
Corresponding limits with 300 fb$^{-1}$ obtained 
in this work: by making a selection on $M_{4b}$ (blue) and on the neural network output (green) and by applying template fitting techniques to the network outputs (brown).
For comparison we also include in red, the limits from the 300 fb$^{-1}$ projection shown on the left plot.
An upper cut on every energy scale of the process of $M_{cut}=2$ TeV has been applied throughout.
}
\end{figure}

\noindent \textbf{Selecting on the NN output}\\
Constraints on the Wilson coefficients are presented using the combined outputs of the network defined in the first two rows of Table \ref{tab:Discriminators} (depending on the chirality of the top quarks\footnote{For the scalar operators $O^{1}_{QtQb}$ and $O^{8}_{QtQb}$, where both left- and right-handed top quarks are involved, the choice was made to assign them to the $t_{R}$ category. This was motivated by the fact that the distributions of the kinematical variables show more similarity to this category.} in the operator). By selecting only events for which this value is larger than 0.83 (again chosen to optimise the constraints as shown in Figure \ref{fig:limits_vs_NNoutput_singlet}), limits are obtained on the individual Wilson coefficients. Examples are again shown in Figure~\ref{fig:xsec_limits_NNcut} for $C_{Qb}^{1}$ (left) and $C_{Qb}^{8}$ (right). This leads to a further improvement of the sensitivities on our example color singlet and octet operators to [-2.1,2.3] (TeV$^{-2}$) and [-5,4.5] (TeV$^{-2}$), respectively. Results for the other operators are again collected in Figure~\ref{fig:summary_Limits} (green lines).  \\

\noindent \textbf{Template fits to the NN output}\\
One can further use the separation power of the neural network by analysing the shape information of its outputs. 
The fully differential outputs can be 
used in a binned likelihood fit of the data to some predefined templates for the different categories of events. The advantage of such a fit lies in the fact that the relative normalization of the different categories can be deduced from the region in phase space where that category is dominantly abundant. This reduces the systematic uncertainty related to the normalization of the measured SM cross section and may, in practise, improve the limits that can be obtained on the Wilson coefficients. To illustrate the strength of such a fit, template histograms ($T^{1D}$) are defined for the three categories such that the NN outputs for a general point in our EFT parameter space can be parametrised as functions
of the event yields for the different event categories ($N_{SM}$ and $N_{L}$ or $N_{R}$)
\begin{align}
f_L\left(N_{SM},N_{L} \right) &= N_{SM} \cdot T^{1D}_{SM} + N_{L} \cdot T^{1D}_{L}, \label{eq:templatefit1DLeft} \\
f_R\left(N_{SM},N_{R} \right) &= N_{SM} \cdot T^{1D}_{SM} + N_{R} \cdot T^{1D}_{R} . \label{eq:templatefit1DRight}
\end{align}
These yields normalize each template ($T^{1D}_{SM},T^{1D}_{L}$ and $T^{1D}_{R}$) and are extracted by fitting to data. The RooFit package \cite{Verkerke:2003ir} incorporated in the ROOT data analysis framework \cite{Brun:1997pa} was used to perform the fit of pseudo-data to Eqs.~\eqref{eq:templatefit1DLeft} and \eqref{eq:templatefit1DRight}, generated for different values of the Wilson coefficients assuming 300 fb$^{-1}$ of integrated luminosity. The fitted yields are used as described in Section \ref{sec:strategy}  to obtain limits on the individual Wilson coefficients, which are summarised in Figure~\ref{fig:summary_Limits} (brown lines). This shows that a similar sensitivity can be achieved with this method. 

Overall, the relative gain in sensitivity with respect to $M_{4b}$ is less pronounced than going from the total cross section to $M_{4b}$. This suggests that the majority of the information in distinguishing between the SM and the EFT is contained in this variable. Nonetheless it is clear that the use of the NN outputs consistently improves the sensitivity. The next section will further highlight the benefits of a dedicated machine learning classification.

%% file: MachineLearning.tex
\section{Learning to pinpoint the effective operators
\label{sec:MultipleOperators}}
\begin{figure}[ht!]
\center
\includegraphics[width=.80\textwidth]{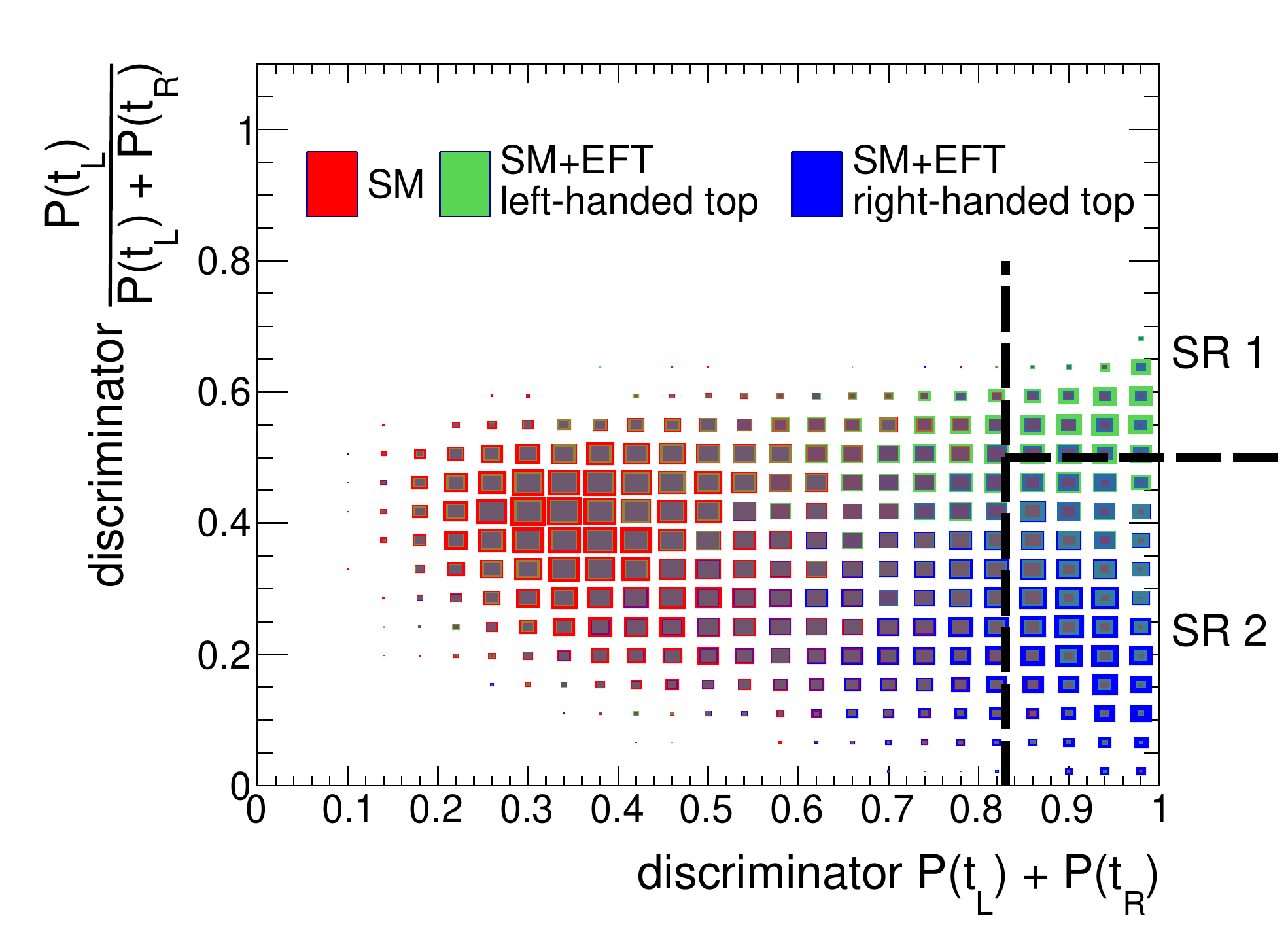}
\caption{\label{fig:2DplotDiscriminators}
 Normalized distributions of the combined NN outputs for events corresponding to SM (red), SM+EFT ($t_L$ operators) (green) and SM+EFT ($t_R$ operators) (blue). The Wilson coefficients are set to 20 (TeV$^{-2}$). The size of each box is proportional to the abundance of events of the corresponding sample.
The discriminators on the x and y axis are as defined in Table \ref{tab:Discriminators}.  
The dashed lines define SR1 and SR2. See text for more details.
}
\end{figure}

We finally illustrate the strength of the multi-class output structure of the network, which becomes apparent when both EFT operators with a $t_{L}$ current and with a $t_{R}$ current are given non-zero Wilson coefficients at the same time. We illustrate this with an example using events generated with both $C_{Qb}^{1}$ and $C_{tb}^{1}$ non-zero. To visualize the separation potential of the neural network between the three classes, Figure \ref{fig:2DplotDiscriminators} shows how the different classes are distributed in the plane of the combined neural network outputs outlined in the last two rows of Table \ref{tab:Discriminators}. The x-axis represents the summed probability $P(t_{L}) + P(t_{R})$ that is able to separate the SM events (red) from any kind of event that includes the insertion of an EFT operator. On the y-axis, the normalized probability \( \frac{P(t_{L})}{P(t_{L}) + P(t_{R})} \) is displayed, designed to distinguish between the $t_{L}$ (green) and the $t_{R}$ (blue) categories. These distributions show a clear concentration of SM events to the left, whereas the $t_{L}$ and $t_R$ contributions dominantly populate the upper and lower right hand corners, respectively.
 We therefore define two signal regions (SR1) and (SR2) as delimited in Figure~\ref{fig:2DplotDiscriminators}.

\begin{figure}[ht!]
\center
\includegraphics[width=.48\textwidth]{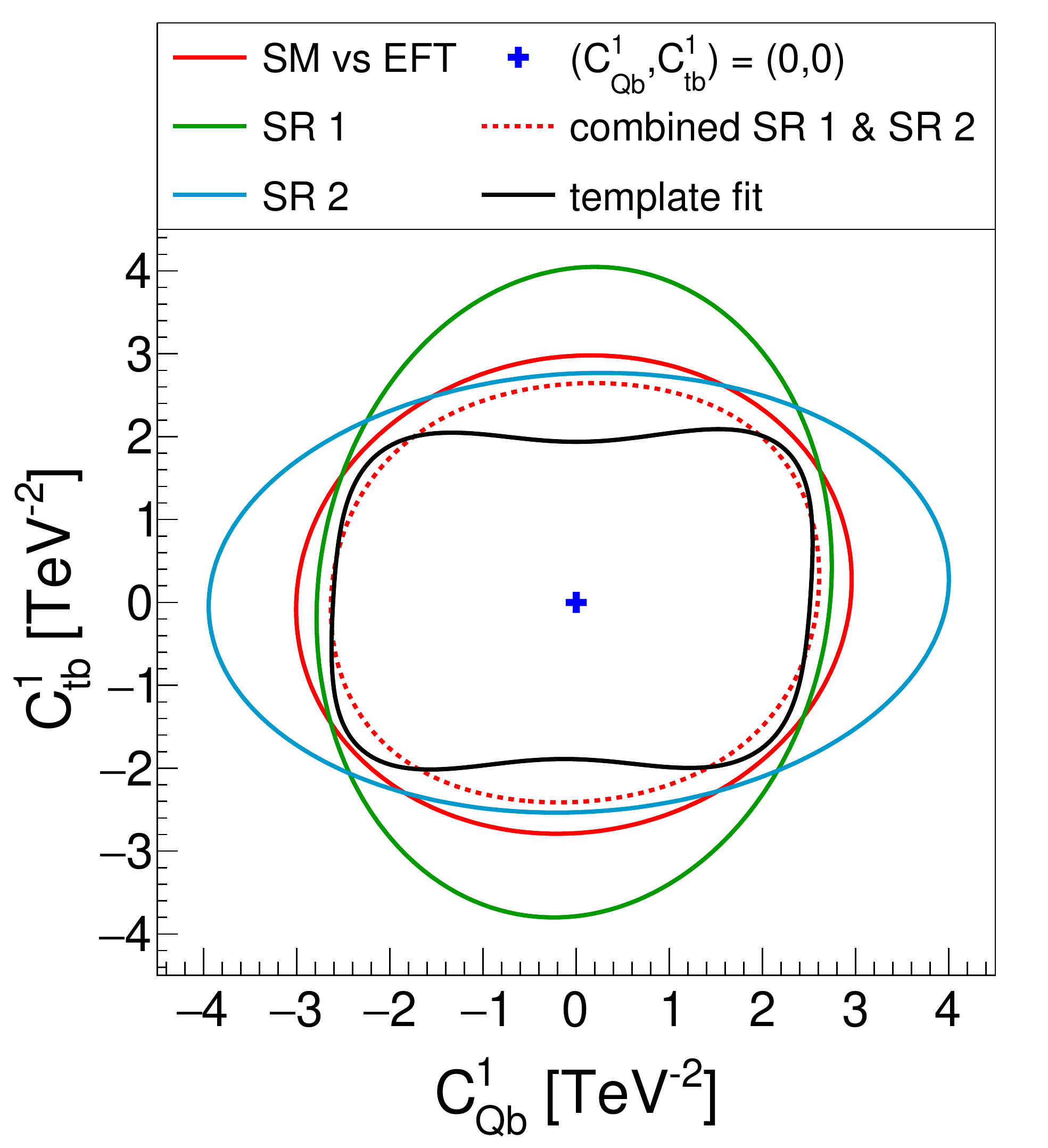}
\includegraphics[width=.48\textwidth]{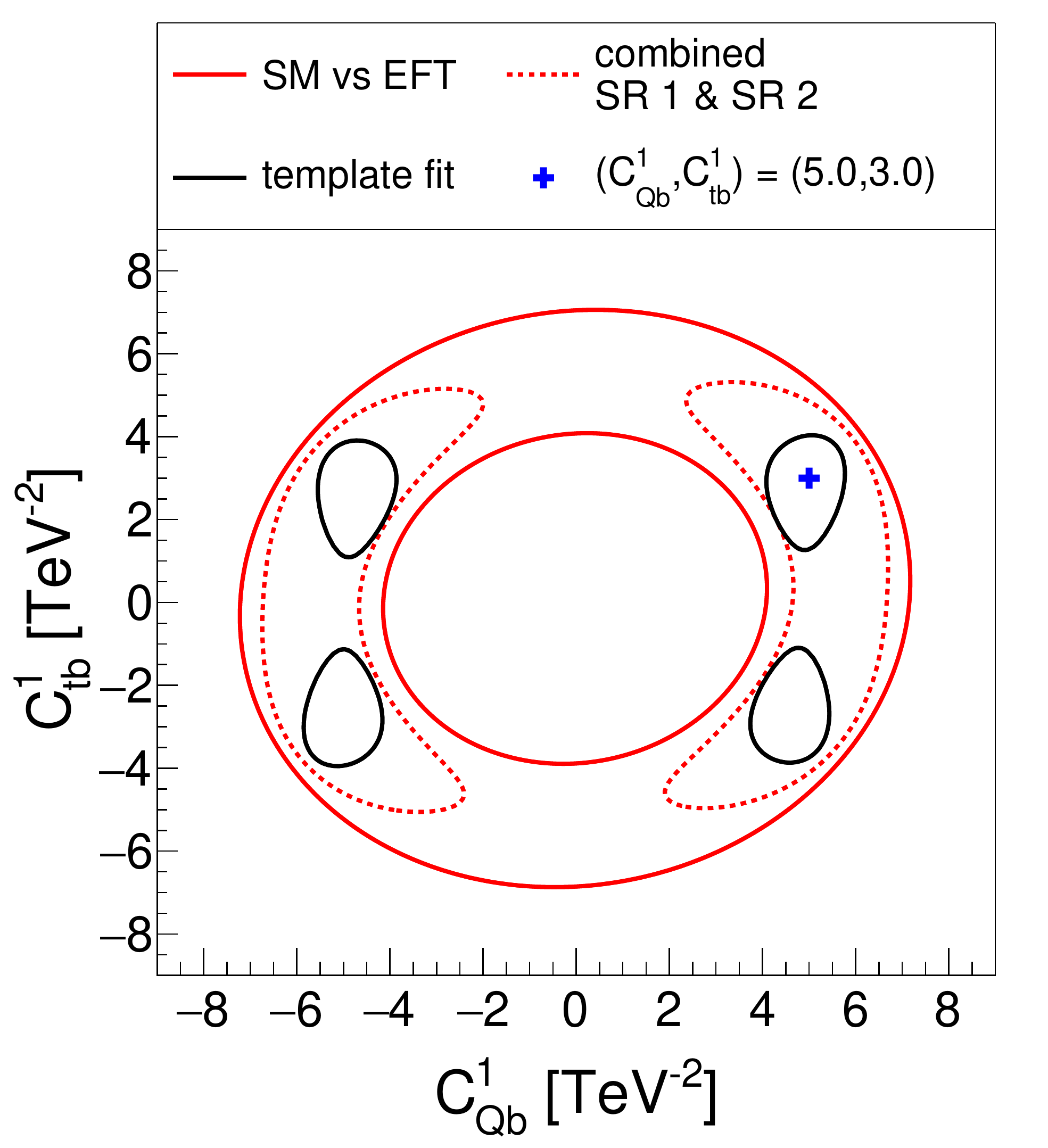}
\caption{\label{fig:Marginalized}
 (left) Two-dimensional limits at 95\% CL assuming a measurement consistent with the SM-only hypothesis (blue cross) and allowing two couplings, $C_{Qb}^{1}$ and $C_{tb}^{1}$ to vary simultaneously: (red) one dimensional cut on $P(t_{L}) + P(t_{R})$ output; (green) SR1; (blue) SR2; (red dashed) combination of SR1 and SR2; (black) two dimensional template fit. 
 (right) Same as on the left plot, but for the EFT signal injection hypothesis.
 See text and Figure \ref{fig:2DplotDiscriminators} for more details.
}
\end{figure}

Adopting a similar strategy to the individual operator case, we can make a single selection on $P(t_{L}) + P(t_{R})$ asking this value to be larger than 0.83. The observed cross section is now fitted according to the generalised function
\begin{align}
\sigma_{fit} = \sigma_{SM} \left( 1 + p_{A} \cdot C_{A} + p_{B} \cdot C_{B} + p_{AA} \cdot C_{A}^{2} + p_{BB} \cdot C_{B}^{2} + p_{AB} \cdot C_{A}C_{B} \right), \label{eq:xsec_2operators}
\end{align}
for two simultaneously non-zero Wilson coefficients. Under the assumption of observing the SM, this yields a two-dimensional contour of the 95\% CL limit 
on the Wilson coefficients $C_{Qb}^{1}$ and $C_{tb}^{1}$ as shown by the full red line in Figure \ref{fig:Marginalized} on the left. When the limits are obtained additionally selecting SR1(SR2) separately, one becomes more sensitive to $C_{Qb}^{1}$ ($C_{tb}^{1}$), as indicated by the green (blue) contours. By combining these two signal regions (red dashed contour), an increased sensitivity is observed compared to the one obtained by the one-dimensional selection on $P(t_{L}) + P(t_{R})$.

More interesting observations can be made in the case of a potential discovery of new physics. Under the hypothesis of observing an EFT signal, this strategy can help in the determination of which type of  operators are involved. To illustrate this effect, we 
inject a benchmark signal with 
$C_{Qb}^{1} = 5$ TeV$^{-2}$ and $C_{tb}^{1} = 3$ TeV$^{-2}$
into our pesudo-data.
The 2D limit obtained at 95\% CL by the one-dimensional selection on $P(t_{L}) + P(t_{R})$  is shown in red in Figure \ref{fig:Marginalized} on the right. The shape of the contour shows a symmetry around the central point (0,0), indicating that this selection is insensitive to the sign of the Wilson coefficient as well as to relative contribution of each operator. 
However, a combination of confidence intervals obtained in SR1 and SR2 (dashed red) is able to reduce the best fit region.
 It excludes at 95\% CL a value of 0 TeV$^{-2}$ for $C_{Qb}^{1}$, which was not possible with the one-dimensional selection. 

\begin{figure}[ht!]
\center
\includegraphics[width=.95\textwidth]{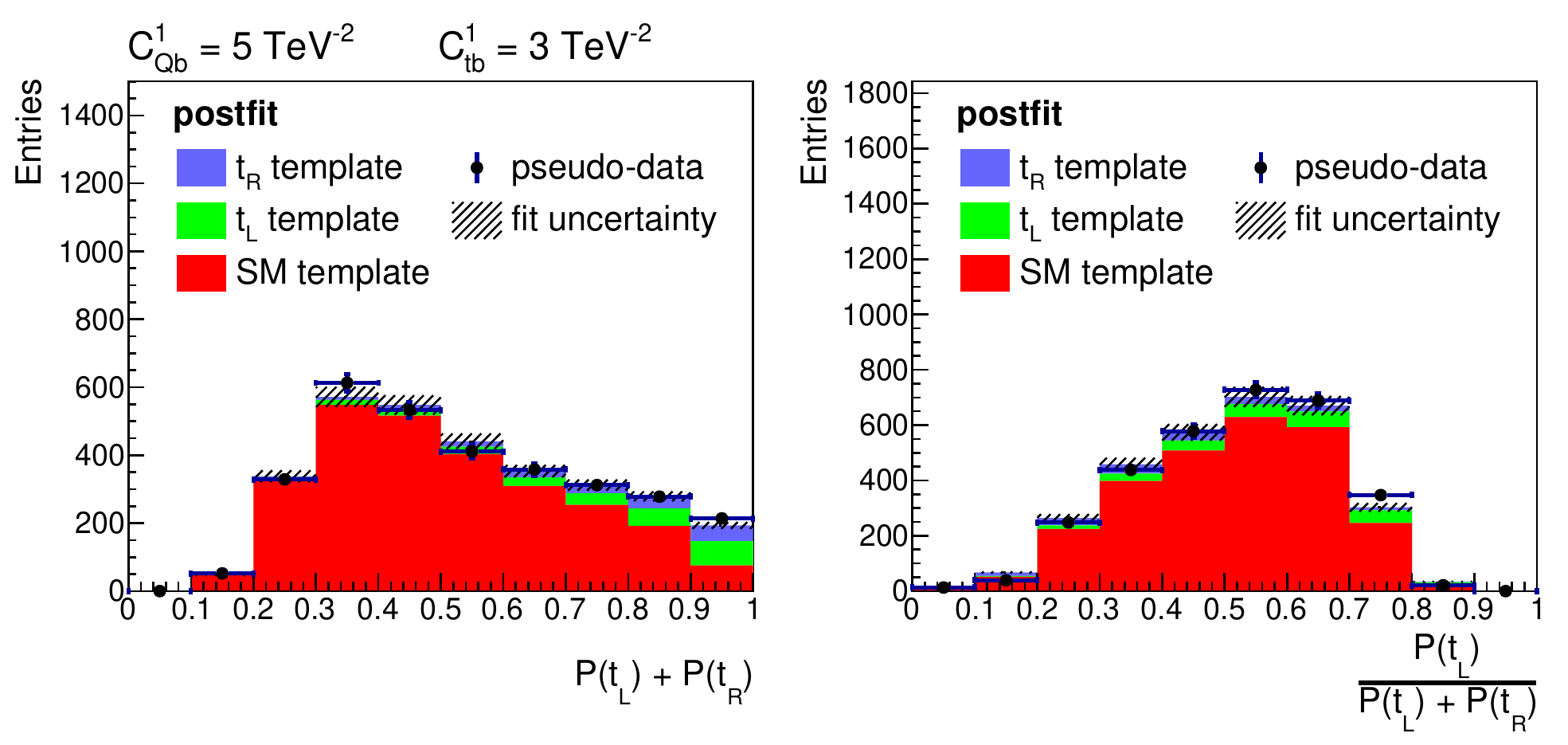}
\caption{\label{fig:TemplateFit}
Projected distributions of the fitted templates and one of the generated pseudo-datasets onto the \( P(t_{L}) + P(t_{R}) \) axis (left) and onto the  \( \frac{P(t_{L})}{P(t_{L}) + P(t_{R})} \) axis (right). The pseudo-experiments are generated from a sample with the Wilson coefficients $C_{Qb}^{1} = 5$ TeV$^{-2}$ and $C_{tb}^{1} = 3$ TeV$^{-2}$, and assuming an integrated luminosity of 300 fb$^{-1}$.
}
\end{figure}

The use of template fitting methods becomes even more interesting in this two dimensional example.
 A two-dimensional binned maximum likelihood fit to predefined templates ($T^{2D}$) is performed by fitting the function
\begin{align}
f_{2D}\left(N_{SM},N_{L},N_{R} \right) = N_{SM} \cdot T^{2D}_{SM} + N_{L} \cdot T^{2D}_{L}   + N_{R} \cdot T^{2D}_{R},  \label{eq:templatefit2D}
\end{align} 
to pseudo-data corresponding to the SM observation and also to the observation of a potential excess as above. A $\chi^{2}$ value is calculated from the sum of each of the EFT event categories separate ($t_{L}$ and $t_{R}$) for each sample point in the parameter space of Wilson coefficients. The 95\% CL contours of this distribution are shown in black in Figure \ref{fig:Marginalized} on the left (SM-only hypothesis) and the right (possible observation of a signal due to EFT operators). In the former case the more rectangular shape of the contour leads to the strongest observed limits in some parts of the parameter space. In the latter case it is clear that the template fitting procedure is able to pinpoint with more precision the values of the Wilson coefficients. By using template fits, a value of 0 (TeV$^{-2}$) for $C_{tb}^{1}$ is now also excluded at 95\% CL, which was not the case for combined limits in SR1 and SR2. 
Figure \ref{fig:TemplateFit} shows the projected distributions of the fitted templates
for \( P(t_{L}) + P(t_{R}) \) on the left and for \( \frac{P(t_{L})}{P(t_{L}) + P(t_{R})} \) on the right.

%% file: conclusion.tex
\section{Summary and conclusions}
\label{sec:summary}
In this work, we present new methods designed to  
exploit the full kinematical information to
interpret Standard Model searches in the SMEFT framework.
The high multiplicity and complexity of the final-state, in combination with the possible contributions from multiple effective operators, make machine learning classifiers a promising candidate to maximise our sensitivity. We identify the production of a top-quark pair in association with two $b$-jets as an interesting process, given its 8-body final state and its dependence on 10 four-heavy-quark operators of dimension six. Its production cross section is large enough to provide the required statistics for a differential analysis of the kinematical properties with 300 fb$^{-1}$ of integrated luminosity. We show that it provides sensitivity to previously unconstrained directions in the SMEFT parameter space and would therefore be an indispensable component in a future global fit for the top-quark interactions. We also present a discussion of various issues concerning the validity and perturbativity of the EFT and its eventual UV completion. Therein, we motivate making an upper cut of 2 TeV on all energy scales involved in the process, to provide a measure of control while hardly sacrificing any sensitivity to the operators. Using power-counting arguments, we show that, in the case of a strongly coupled UV completion with coupling $g_\ast > g_s$, the dominant SMEFT contribution arises quadratically in the dimension-6 operators while all others are parametrically suppressed. This is supported by an explicit example of an axigluon scenario.

We have presented a detailed analysis of the LHC sensitivity in this process to the Wilson coefficients of four-quark operators involving only third generation quarks. Starting with new limits from the current inclusive measurement, we progressively employ kinematical information and machine learning classifiers, obtaining a significant improvement in projected sensitivity at the end of Run III. 
%
%
%
To this end, we employ a multi-class architecture in which a shallow neural network is trained to classify events into three categories: SM, left-handed top quark operator and right-handed top quark operator. This strategy allows for the construction of optimal discriminants both for distinguishing individual operator types from the SM and/or among themselves.

The strength of our approach becomes apparent when multiple operators with different chiral structures are considered simultaneously. We illustrate this by considering the presence of two operators with opposite chiral structure at the same time. Using template fits to the two-dimensional neural network discriminant distribution is shown to provide the best limits in the two-dimensional Wilson coefficient space. Furthermore, in the scenario of a hypothetical excess in $\ttbb$, this method is able to pinpoint with the most accuracy the values of the responsible Wilson coefficients.

Our method can be extended to more advanced network architectures in combination with more optimal input variables and larger training datasets to further exploit the power of these machine learning algorithms to constrain the SMEFT \cite{Brehmer:2018eca,Brehmer:2018kdj}. Here we have presented one example where the kinematics of top decay are employed to discriminate between two broad classes of EFT operators. 
A comprehensive exploration of how far such a strategy could be pushed towards a discriminator capable of distinguishing individual SMEFT operators would be extremely interesting.

To conclude, we have presented a detailed investigation of the application of machine learning classifiers in extracting SMEFT signals in the $\ttbb$ final state. This process has shown itself to be a important component 
to constrain top EFT interactions. Furthermore, our study serves as a proof of principle that motivates the use of multi-class discriminants in the context of globally constraining the SMEFT at the LHC.


%% file: AxigluonAppendix.tex
\section{Axigluon model}
\label{app:AM}

We will use an axigluon model to illustrate that our power counting assumption
is satisfied and to justify the truncation of the SMEFT expansion at
dim-6, even with a strongly coupled theory.
\subsection*{Model setup}
Consider an axigluon model where the strong sector is extended to
$SU(3)_L\times SU(3)_R$, 
\begin{equation}
	\mathcal{L}=-\frac{1}{4}G_{1\mu\nu}G_1^{\mu\nu}
	-\frac{1}{4}G_{2\mu\nu}G_2^{\mu\nu}+\frac{f^2}{4}\mathrm{Tr}D_\mu\Sigma
	D^\mu\Sigma^\dagger,
\end{equation}
which is spontaneously broken to the diagonal
subgroup $SU(3)_c=SU(3)_{L+R}$ of QCD by the nonlinear sigma field $\Sigma$,
which transforms in the bifundamental representation:
\begin{equation}
	\Sigma\to U_L\Sigma U_R^\dagger.
\end{equation}
The physical fields are obtained by rotating the gauge fields $G_1$ and $G_2$
to the mass eigenstate basis:
\newcommand{\cth}{{c_\theta}}
\newcommand{\sth}{{s_\theta}}
\begin{equation}
	\left(\begin{array}{c}
		G_{1\mu}^A \\ G_{2\mu}^A
	\end{array} \right)
	=
	\left(\begin{array}{cc}
		\cth & -\sth \\
		\sth & \cth
	\end{array} \right)
	\left(\begin{array}{c}
		G_{\mu}^A \\ C_{\mu}^A
	\end{array} \right),
\end{equation}
where $C_{\mu}^A$ is the axigluon field with mass $M$.  The mixing angle
is given by
\begin{equation}
	\sth=\frac{g_1}{\sqrt{g_1^2+g_2^2}},
\end{equation}
where $g_1$ and $g_2$ are the coupling strength of the $SU(3)_L$ and $SU(3)_R$
gauge fields, respectively.  The QCD strong coupling is given by
\begin{equation}
	g_s=\frac{g_1g_2}{\sqrt{g_1^2+g_2^2}}.
\end{equation}
Below, we will demonstrate that the power counting assumption of
Eq.~(\ref{eq:pc}) is satisfied in this model.

\bigskip
\noindent\textbf{Gauge coupling of fermions}\\
\noindent The gluon and axigluon couplings to the fermions are given by the covariant
derivative:
\begin{flalign}
	D_\mu q=&\partial_\mu q-ig_sT^AG^A_\mu q-ig_*T^AC^A_\mu (1+C_A\gamma^5q)
\end{flalign}
where the couplings are
\begin{equation}
	g_s=\frac{g_1g_2}{\sqrt{g_1^2+g_2^2}},\ 
	g_*=\frac{\cth^2-\sth^2}{2\sth\cth}g_s,\ 
	C_A= \frac{1}{\cth^2-\sth^2}.
\end{equation}
The axigluon $C_\mu$ couples to the fermions with coupling strength $g_*$.
In the following we consider the limit $\sth\ll1$, where we have $g_*\gg
g_s$, so the theory is strongly coupled.  In this limit, the fermions couple
strongly to the heavy axigluons, leading to the 
	$\frac{g_*f}{\Lambda^{3/2}}$
in the power counting assumption of Eq.~(\ref{eq:pc}).  Note that under this
limit the axial coupling $C_A\approx 1$.

\bigskip
\noindent\textbf{Gauge coupling of axigluon}\\
\noindent The couplings between gluons and axigluons come from the kinetic terms of
$G_1$ and $G_2$.  In terms of mass eigenstates, we find the following
gauge interaction terms
		\begin{flalign}
			&CCG:\	-\frac{1}{2}g_sf^{ABC}\left( 
			\partial_\mu G_\nu^A - \partial_\nu G_\mu^A
			\right)C_\mu^B C_\nu^C	
			-g_sf^{ABC}\left( 
			\partial_\mu C_\nu^A-\partial_\nu C_\mu^A
			\right)G_\mu^B C_\nu^C,
\\&CCGG:\ 
			-\frac{1}{2}g_s^2 f^{ABC}f^{ADE}\left( 
			G_\mu^B G_\nu^C C^{D\mu} C^{E\nu}
		       +G_\mu^B C_\nu^C G^{D\mu} C^{E\nu}
		       +G_\mu^B C_\nu^C C^{D\mu} G^{E\nu}
			\right).
		\end{flalign}
This implies that the gluon couples to axigluon with strength $g_s$,
not $g_*$.
This is exactly what we have argued for the power counting rule for $G_{\mu\nu}$,
where the coupling strength for the $\frac{gG_{\mu\nu}}{\Lambda^2}$ term
in Eq.~(\ref{eq:pc}) is $g_s$ instead of $g_*$.

\subsection*{Matching }
We now derive the coefficients for the relevant operators, to explicitly
show that the assumption in Eq.~(\ref{eq:pc}) indeed applies to the matched
operator coefficients.

\bigskip
\noindent\textbf{Four-fermion operator}\\
\begin{figure}[!ht]
	\begin{center}
		\includegraphics[width=.6\linewidth]{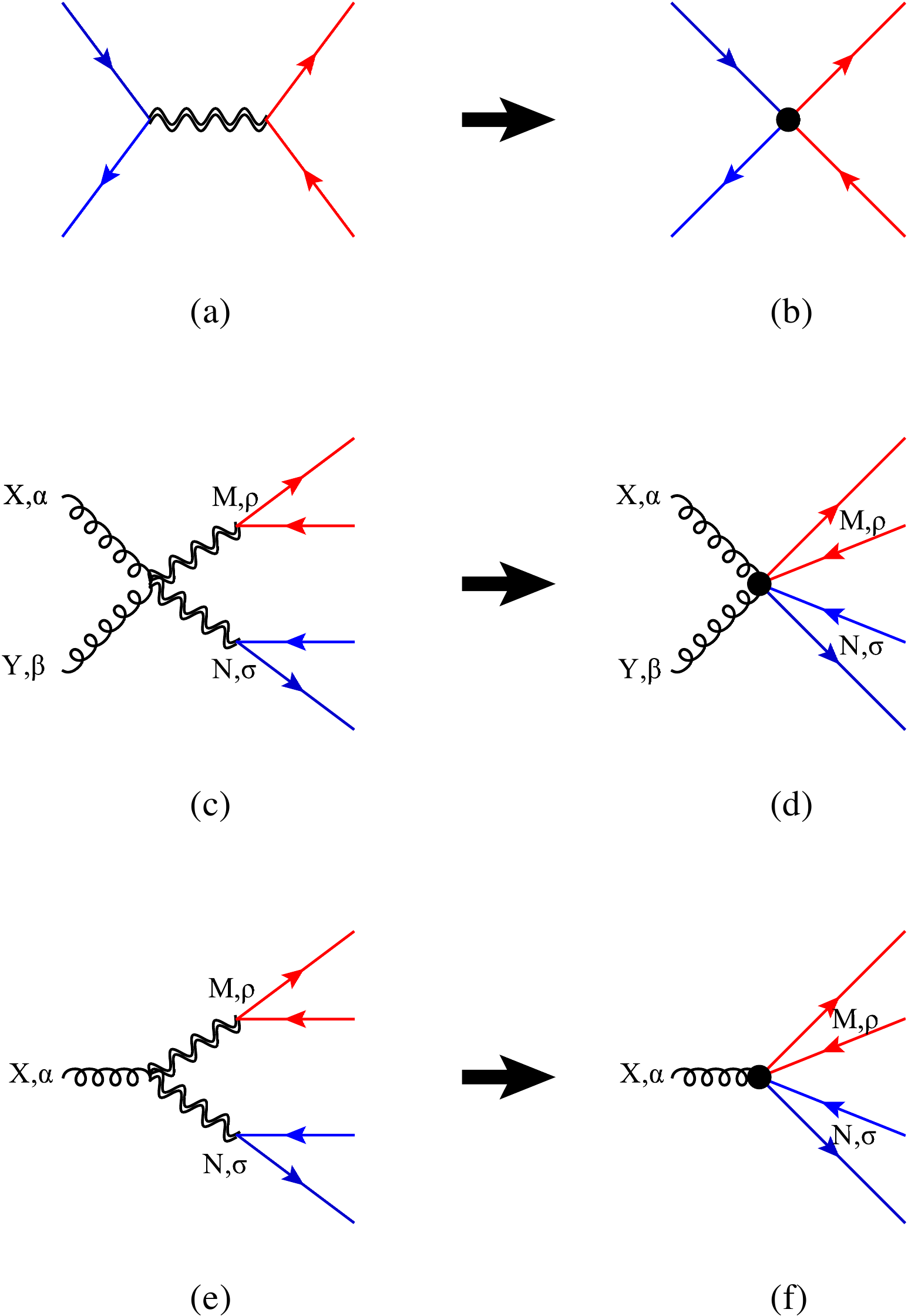}
	\end{center}
	\caption{Matching in the axigluon model. Red and blue fermion lines
	represent top-quark and bottom-quark currents.  Double wavy line
        represents the axigluon.  A blob represents the insertion of an
        effective operator. (a), (c), (e) are diagrams in the full theory,
        which are replaced by (b), (d), (f) in the EFT. $X,Y,M,N$ are color
indices.  $\alpha,\beta,\rho,\sigma$ are Lorentz indices.}
	\label{fig:matching}
\end{figure}
\noindent At leading order, the BSM contribution to the $\ttbb$ amplitude is given by 
Figure~\ref{fig:matching} (a).  The corresponding contribution is reproduced by
effective operators as in Figure~\ref{fig:matching} (b).  The full amplitude
can be expanded:
\begin{flalign}
	M_{ttbb}=&(ig_*)^2J_{t\mu}^A\frac{-i}{s-M^2}J_{b}^{A\mu}
	=\frac{-ig_*^2}{M^2}J_{t\mu}^AJ_{b}^{A\mu}\left( 1+\frac{s}{M^2}
	+\cdots\right),
	\label{eq:ttbbexpansion}
\end{flalign}
where
	$J_f^{A\mu}\equiv\bar u(f)\gamma^\mu(1+C_A\gamma^5)T^A v(f)$
is the top or bottom quark current.

The first term can be reproduced by the following dim-6 operator
(neglecting SU(2) as it is irrelevant for our purpose):
\begin{flalign}
	O^{(6)}_{4f}=\left[ \bar t\gamma^\mu(1+C_A\gamma^5)T^A t \right]
	\left[ \bar b\gamma_\mu(1+C_A\gamma^5)T^A b \right],
\end{flalign}
with coefficient
\begin{equation}
	\frac{C^{(6)}_{4f}}{\Lambda^2}=-\frac{g_*^2}{M^2},
\end{equation}
while the second term can be reproduced by the following dim-8 operator
\begin{flalign}
	O^{(8)}_{DD}=D^\mu\left[ \bar t\gamma^\nu(1+C_A\gamma^5)T^A t \right]
	D_\mu\left[ \bar b\gamma_\nu(1+C_A\gamma^5)T^A b \right],
\end{flalign}
with coefficient
\begin{equation}
	\frac{C^{(8)}_{DD}}{\Lambda^4}=-\frac{g_*^2}{M^4}.
\end{equation}

The above coefficients are exactly consistent with what we have expected from
Eq.~(\ref{eq:pc}), taking $\Lambda_{NP}=M$.
It implies that dim-8 four-fermion operators will not be enhanced by more powers
of $g_*$, relative to dim-6 operators, and thus the truncation of dim-8 operators is
well-motivated given that $E^2/M^2<1$ is ensured by $M_{cut}$.  This is also
obvious from Eq.~(\ref{eq:ttbbexpansion}), where the validity of the expansion
is guaranteed, if $s<M^2$. Note that this is independent of the relative size
of the dim-6 quadratic and interference terms, which relies on the size
of $g_*^2$.

\bigskip
\noindent\textbf{$ffffDD$ and $ffffG_{\mu\nu}$ operators}\\
\noindent We also have to check whether the dim-8 contribution from a contact $ggffff$
interaction could be enhanced by more powers of $g_*$.  The amplitude in the
full theory is given by Figure~\ref{fig:matching} (c). 
To reproduce the amplitude we find that two additional operators are
needed:
\begin{flalign}
	&O^{(8)}_{G}=f^{ABC}G_{\mu\nu}^{A}\left[ \bar
	t\gamma^\mu(1+C_A\gamma^5)T^B t \right] \left[ \bar
	b\gamma^\nu(1+C_A\gamma^5)T^C b \right],
	\\
	&O^{(8)}_{DD'}=D_\mu\left[ \bar t\gamma^\mu(1+C_A\gamma^5)T^A t \right]
	D_\nu\left[ \bar b\gamma^\nu(1+C_A\gamma^5)T^A b \right].
\end{flalign}
Together with $O_{DD}^{(8)}$, by equating the diagrams in
Figure~\ref{fig:matching} (c) and (d), we find
\begin{flalign}
	\frac{C^{(8)}_{G}}{\Lambda^4}=\frac{-2g_sg_*^2}{M^4},
\qquad
	\frac{C^{(8)}_{DD'}}{\Lambda^4}=\frac{g_*^2}{M^4}.
\end{flalign}
These coefficients are again consistent with the assumption of
Eq.~(\ref{eq:pc}), and so as we have argued, they all lead to subleading
contributions as they are not enhanced by more powers of $g_*$.
We have also checked that these three dim-8 operators reproduce the correct
$gttbb$ amplitude, as in Figures~\ref{fig:matching} (e) and (f).
Since the axigluons can only contribute through the three
one-light-particle-irreducible (1LPI) diagrams, i.e.~Figure~\ref{fig:matching}
(a), (c), and (e), up to $\mathcal{O}(\Lambda^{-4})$,
we can now conclude that truncating the SMEFT at dim-6 is justified in this
model, regardless of the size of $g^*$ and the relative size of dim-6 quadratic
term and dim-6 interference.

A final remark is that the operator $O_{DD'}^{(8)}$ is a redundant one.  Its
contribution to the $gg\to t\bar t b\bar b$ from Figure~\ref{fig:matching} (d)
and from Figure~\ref{fig:matching} (f) will cancel each other.  We include this
operator simply to have a diagram-by-diagram matching, i.e.~all three
one-light-particle-irreducible diagrams $ttbb+0g,1g$ and $2g$ are matched,
which is intuitively more transparent.

%% file: NeuralNetworkAppendix.tex
\section{Neural network setup}
\label{app:NN}

\noindent\textbf{Training and validation datasets}\\
\noindent The network was trained on 18 input variables, which are summarized in Table~\ref{tab:inputs}.  The F-values in this table denote the results of the ANOVA (Analysis of variances) F-statistics as defined in \cite{heiman2010basic}. This value scales with the absolute difference between the mean values of the SM distribution and the EFT distribution for a given variable. At the same time it is inversely proportional to the average variance of of the individual SM and EFT distributions.
Qualitatively, the F-value thus describes both the overlap between two distributions and the distance between their mean values, thereby providing information on which observables have strong separating power between SM and EFT contributions.

\begin{table}[ht!]
 \center
  \begin{tabular}{| l | c || l | c || l | c | }
    \hline
    $\Delta R$ & F-value & $m_{inv}$ &F-value & $p_{T}$ &F-value \\
    \hline
    \hline
    $\Delta R$(\lepone,\leptwo) & 274 & $m_{inv}$(\lepone,\leptwo) & 312 & $p_{T}$(\lepone) & 580 \\
    \hline
    $\Delta R$(\bone,\btwo) & 12 & $m_{inv}$(\bone,\btwo) &8455 & $p_{T}$(\leptwo) & 10\\
    \hline
    $\Delta R$(\bone,\lepone) & 1493 & $m_{inv}$(\bone,\lepone) &1505 & $p_{T}$(\bone) & 8500 \\
    \hline
    $\Delta R$(\btwo,\leptwo) & 714 & $m_{inv}$(\btwo,\leptwo) & 1673 & $p_{T}$(\btwo) & 8434 \\
    \hline
     $\Delta R$(\adone,\adtwo) & 309 & $m_{inv}$(\adone,\adtwo) &  6589 & $p_{T}$(\adone) & 9664 \\
    \hline
      & & $m_{inv}$(\bone,\btwo,\adone,\adtwo) & 14805 & $p_{T}$(\adtwo) & 5081\\
    \hline
     & & $m_{inv}$(\lepone,\leptwo,\bone,\btwo,\adone,\adtwo) & 12895 & & \\
    \hline
  \end{tabular}
   \caption{\label{tab:inputs}
Kinematical variables used in the neural network. The F-values denote the results of the ANOVA (Analysis of variances) F-statistics.
}
 \end{table}

 Events are simulated in three classes: events including only SM contributions, events that have a single insertion of an EFT operator with a left-handed top quark and events that have a single insertion of an EFT operator with a right-handed top quark. Each of these categories contains around 28,000 events for training and around 7,000 events for testing. It is important to note that when the network is used to calculate limits on the Wilson coefficients, it is applied to events which do not strictly belong to one of these three classes. Instead the events used for the determination of the limits are generated with both the SM contributions and the EFT contributions (including possible interference) included.\\

\begin{figure}[ht!]
\center
\includegraphics[height=.3\textheight]{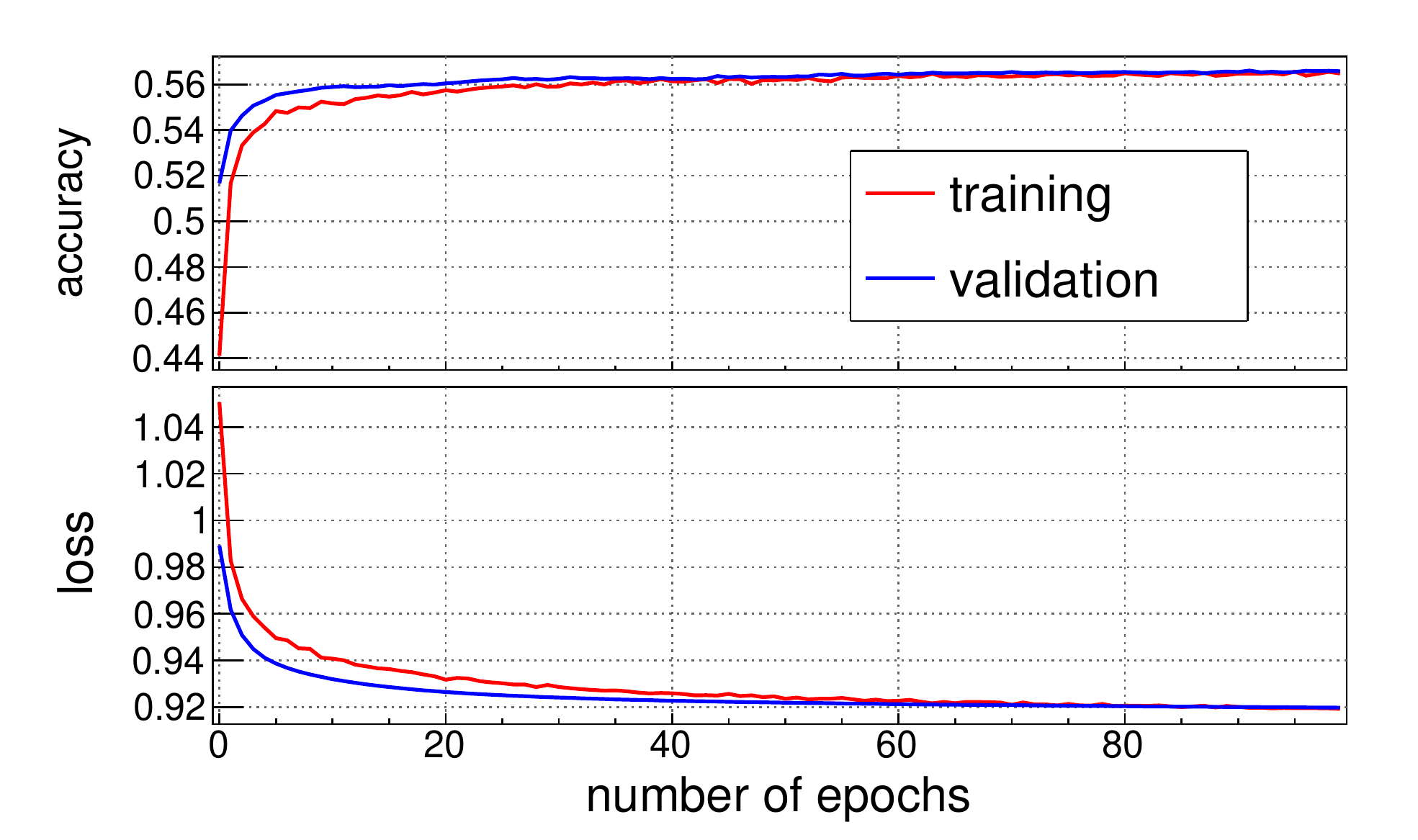}
\caption{\label{fig:training_curve}
Training curves of the neural network training, displaying the evolution of the accuracy (top) and value of the loss function (bottom) for increasing number of epochs. These curves are shown both for the training (red) and for an independent validation data set (blue). These curves converge towards each other and reach a plateau after about 100 epochs.
}
\end{figure}

\noindent\textbf{Network architecture}\\
\noindent The neural network was trained using Keras with the Tensorflow backend. The 18 input nodes are linked to a fully-connected dense layer with 50 neurons with a rectified linear unit activation. A dropout layer is added which randomly freezes 10\% of the neurons in this inner layer in every mini-batch to avoid overfitting. This layer is connected to the 3 ouputs with a softmax activation such that the outputs sum up to one. A categorical crossentropy loss function is used and the minimization of this loss function is performed with a stochastic gradient descent set to an initial learning rate of 0.005 and a decay of 10$^{-6}$. Nestrov momentum is used and is fixed to a value of 0.8. The training is performed in mini-batches of 128 events and is stopped after 100 epochs. The training curve is shown in Figure \ref{fig:training_curve}, showing a convergence to a plateau both for training (blue) and testing (green) datasets. The top panel shows the accuracy whereas the bottom panel shows the value of the loss function.

\begin{figure}[h!]
\center
\includegraphics[height=.6\textheight]{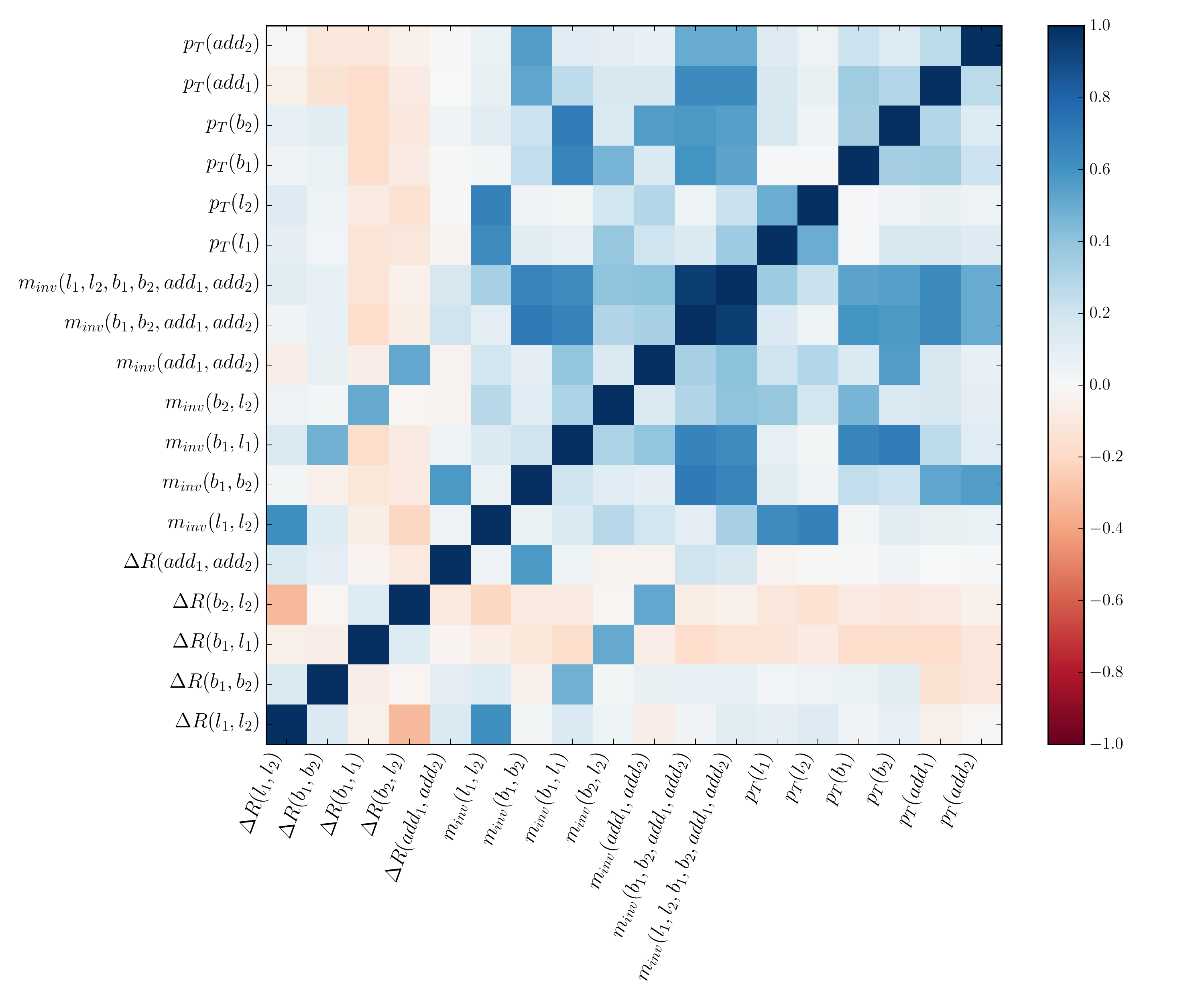}
\caption{\label{fig:corrmat}
Correlation matrix of the NN input variables.
}
\end{figure}

\newpage
\noindent\textbf{Variable Distributions}\\
\noindent Below the distributions of all the input variables of the neural network are shown for SM only events (red), for events with a single insertion of an EFT operator with a left-handed top quark (green) and for those with a right-handed top quark (blue). The correlation matrix is shown in Figure \ref{fig:corrmat}. \\[2ex]

\noindent
\begin{minipage}{\textwidth}
      \centering
      \begin{minipage}{0.3\textwidth}
       \includegraphics[width=\linewidth]{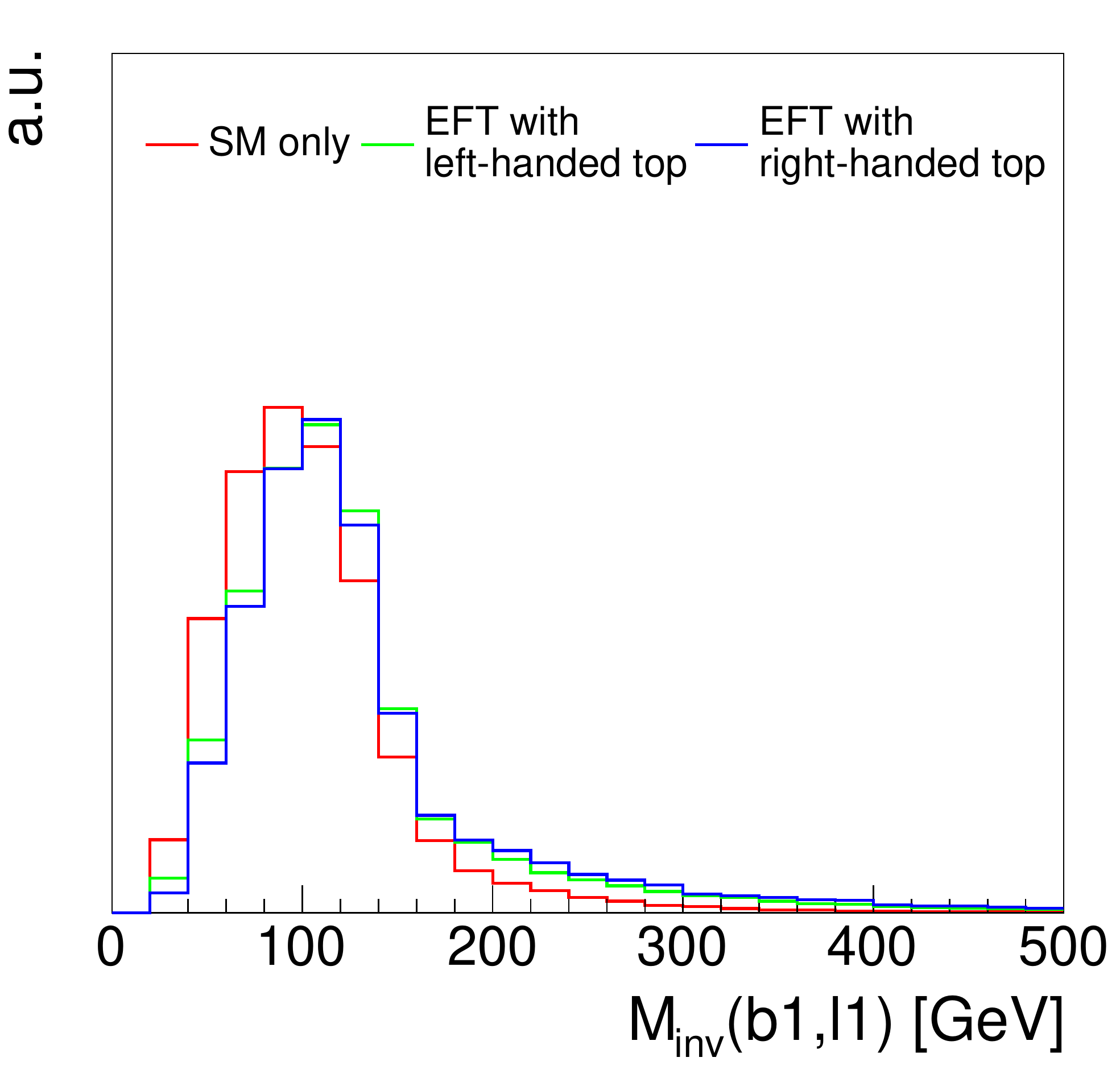}
      \end{minipage}
      \begin{minipage}{0.3\textwidth}
       \includegraphics[width=\textwidth]{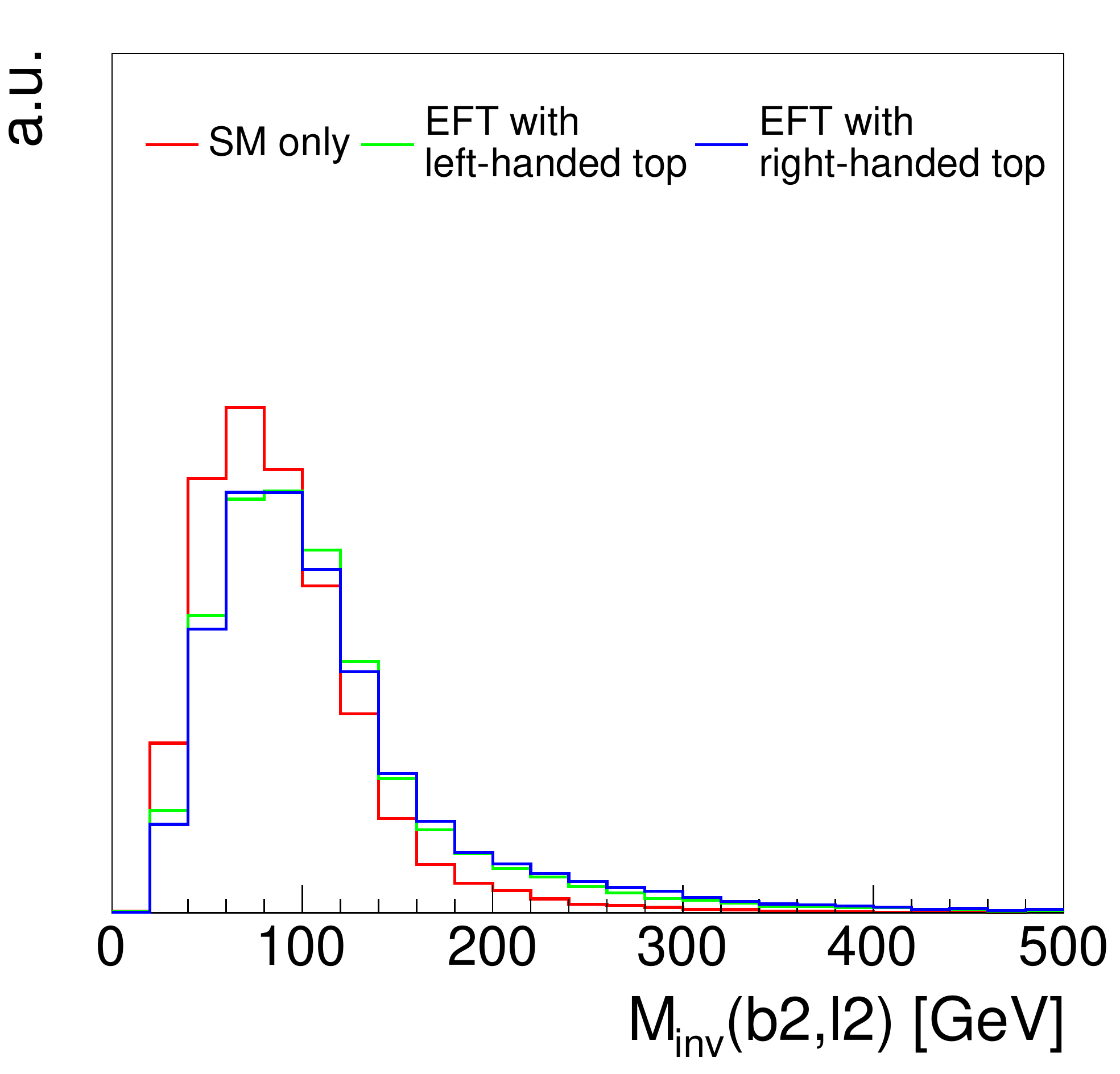}
      \end{minipage}
      \begin{minipage}{0.3\textwidth}
       \includegraphics[width=\textwidth]{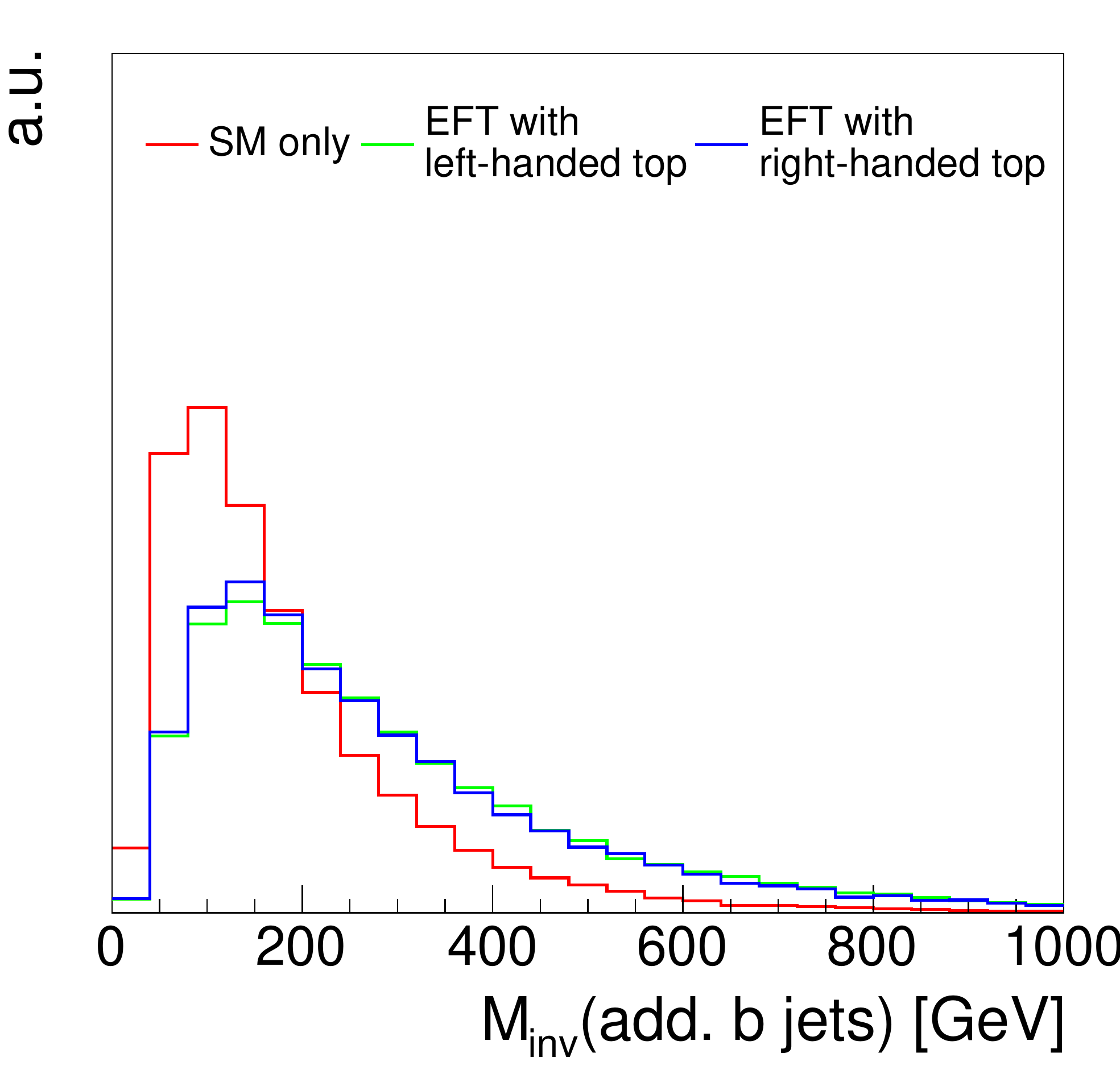}
      \end{minipage}
\end{minipage}\\
\begin{minipage}{\textwidth}
      \centering
      \begin{minipage}{0.3\textwidth}
       \includegraphics[width=\linewidth]{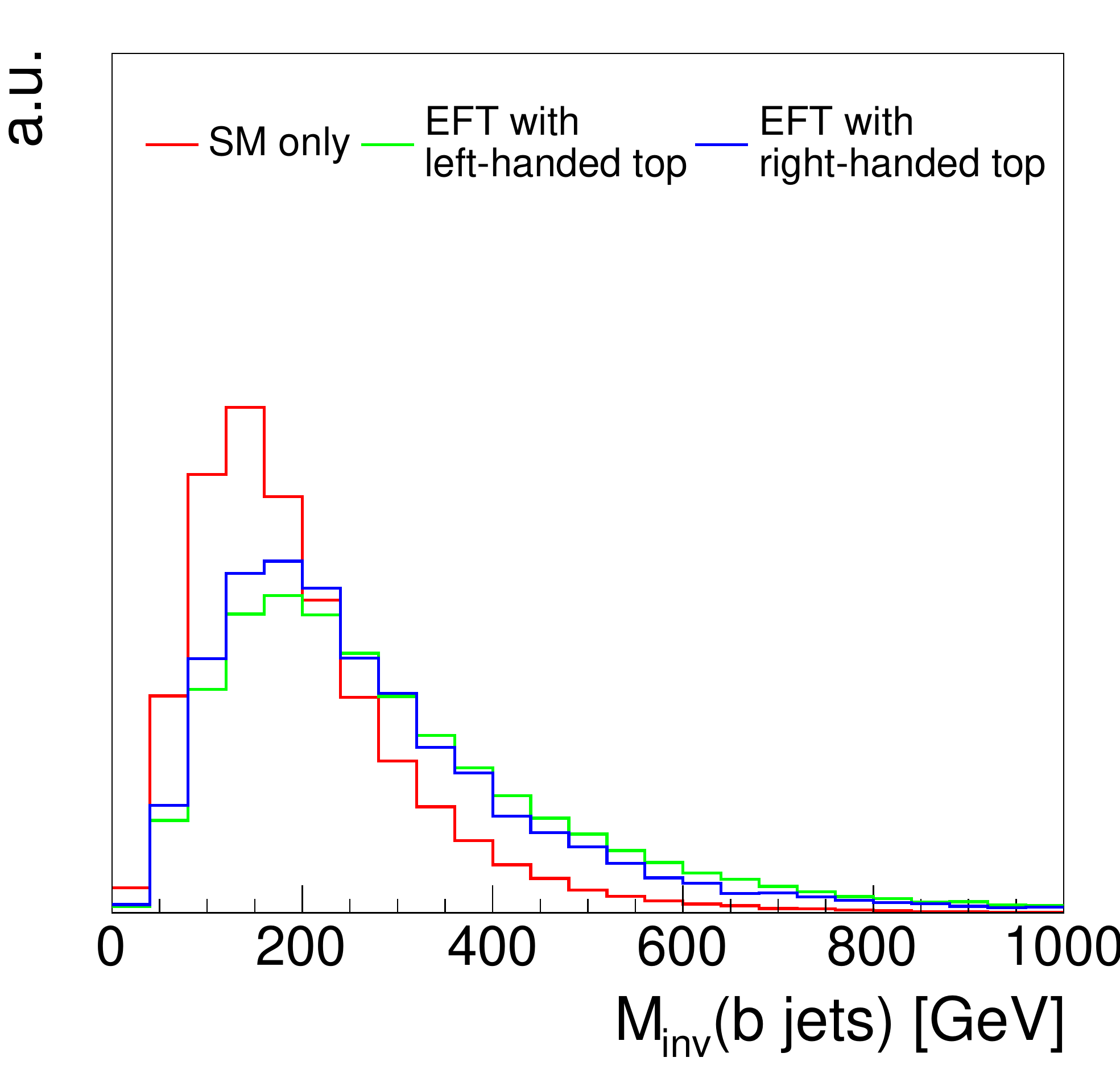}
      \end{minipage}
      \begin{minipage}{0.3\textwidth}
       \includegraphics[width=\textwidth]{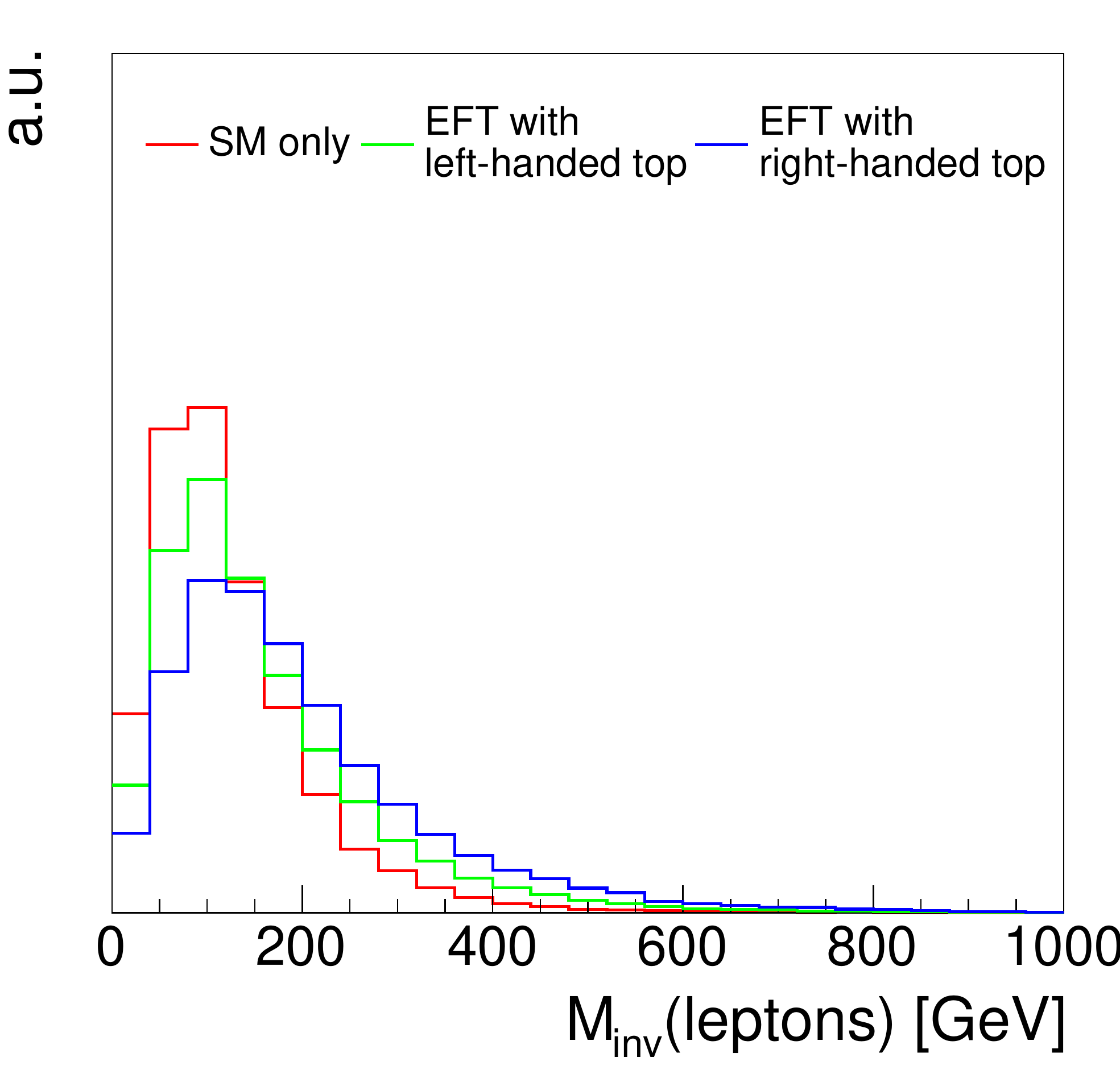}
      \end{minipage}
      \begin{minipage}{0.3\textwidth}
       \includegraphics[width=\textwidth]{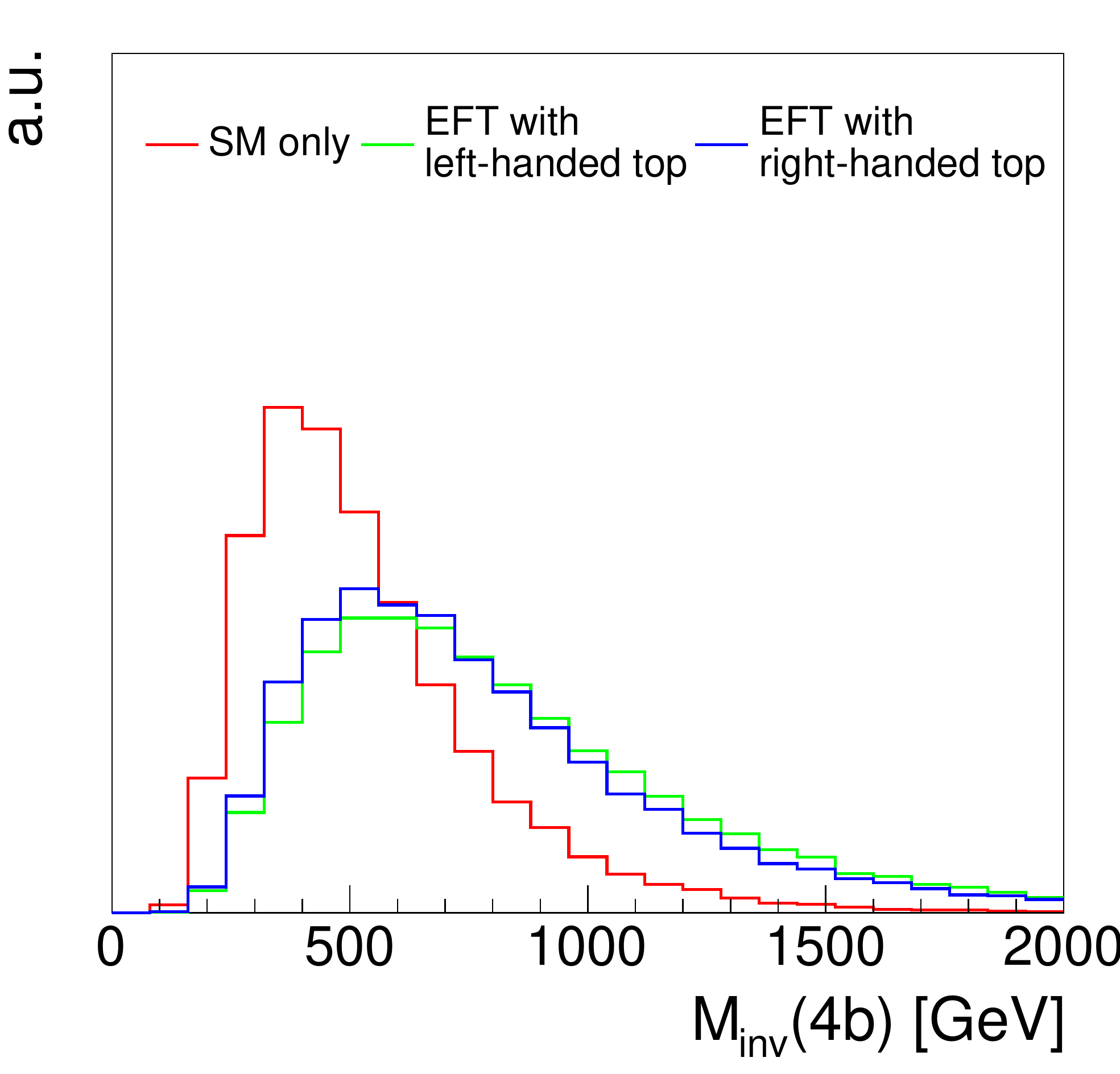}
      \end{minipage}
\end{minipage}\\
\begin{minipage}{\textwidth}
      \centering
      \begin{minipage}{0.3\textwidth}
       \includegraphics[width=\linewidth]{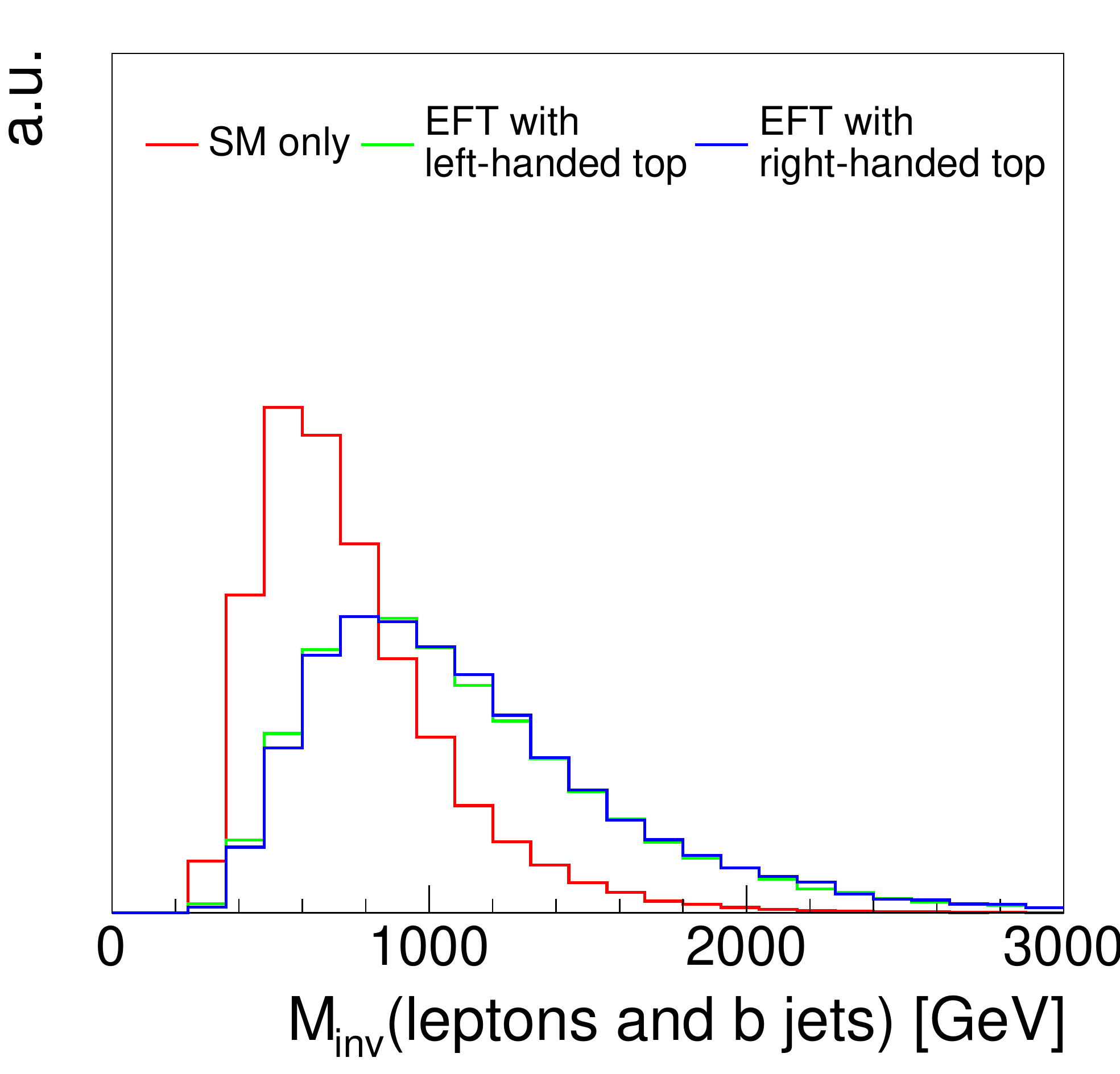}
      \end{minipage}
      \begin{minipage}{0.3\textwidth}
       \includegraphics[width=\textwidth]{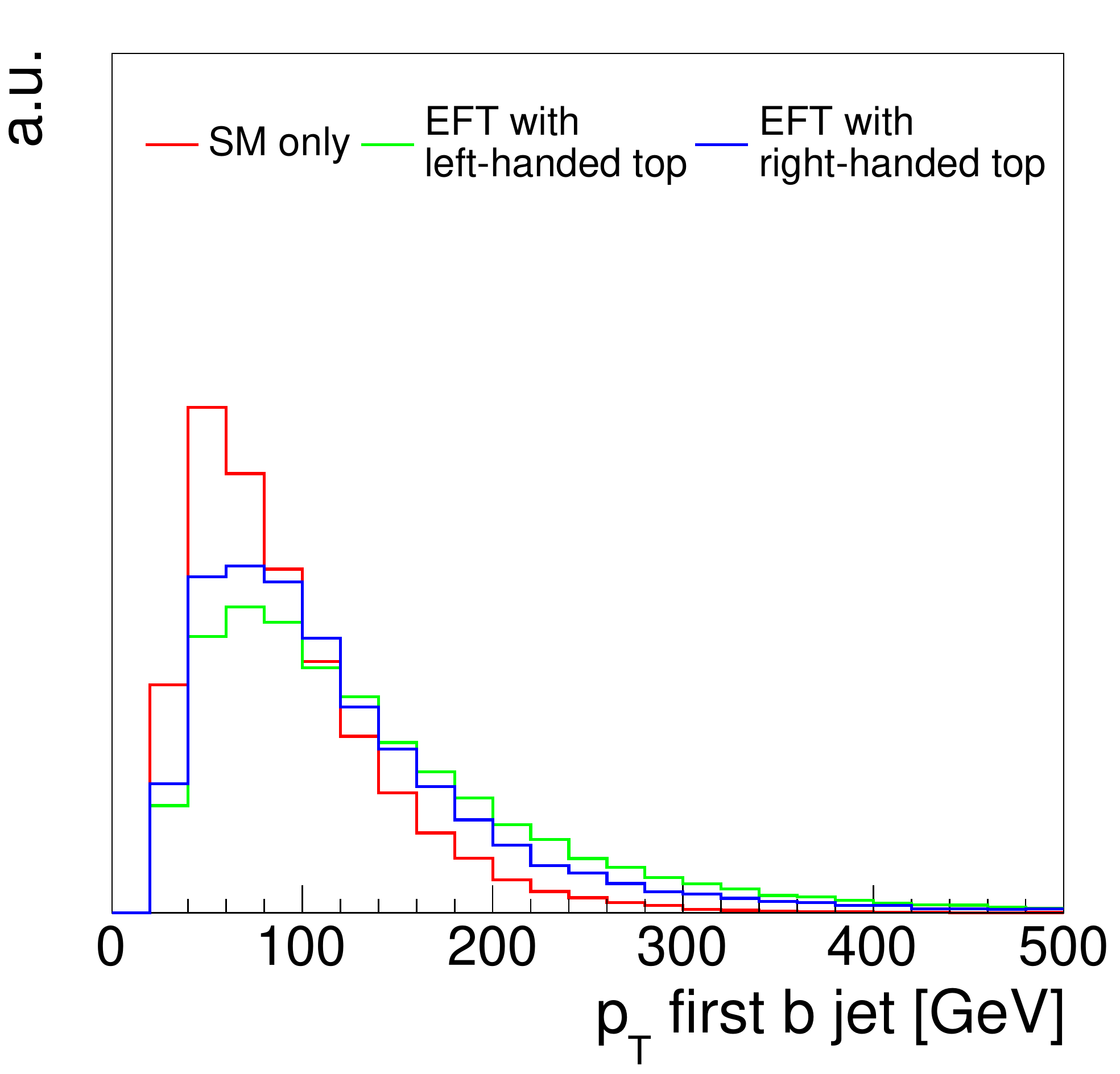}
      \end{minipage}
      \begin{minipage}{0.3\textwidth}
       \includegraphics[width=\textwidth]{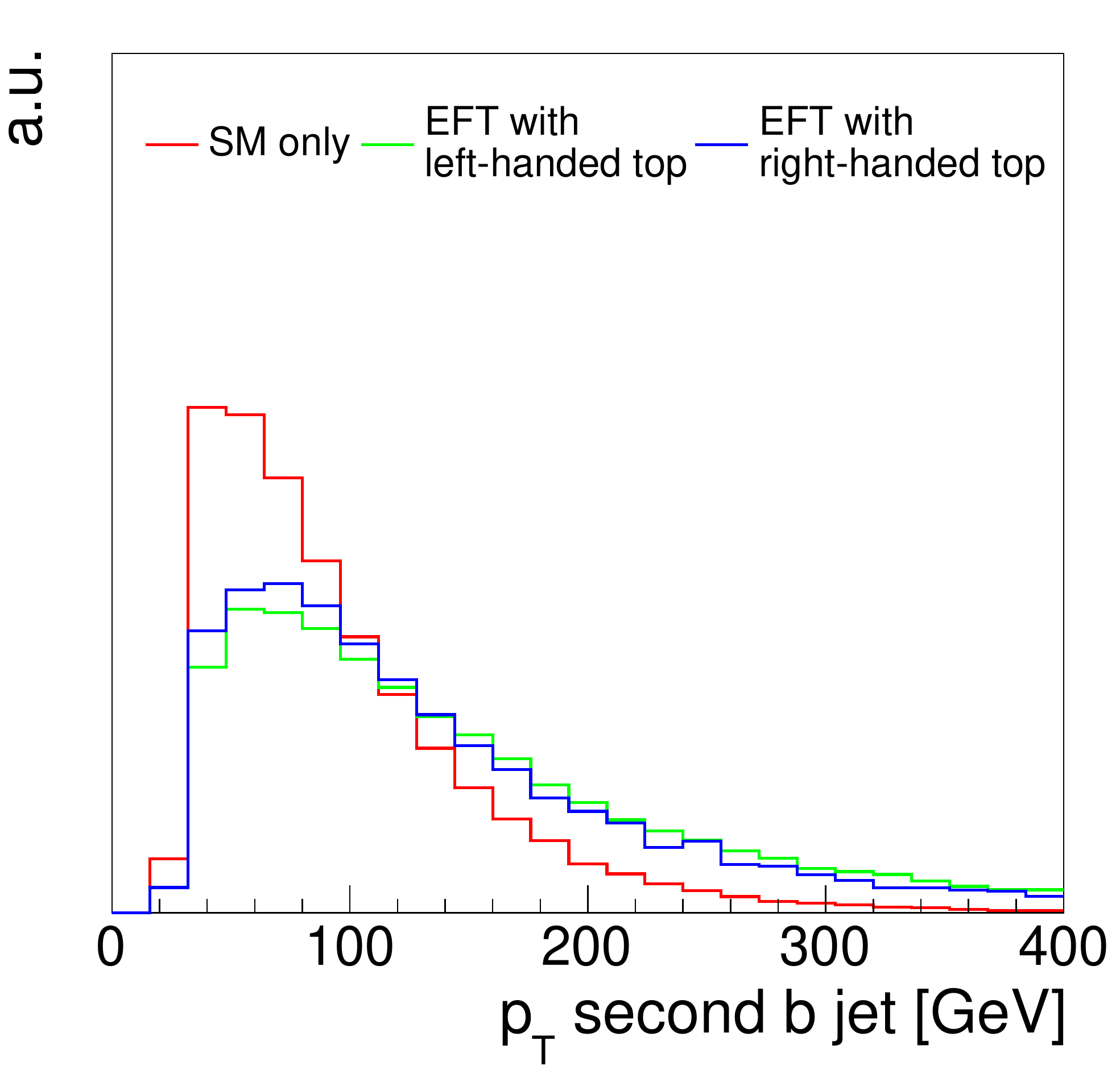}
      \end{minipage}
\end{minipage}\\
\begin{minipage}{\textwidth}
      \centering
      \begin{minipage}{0.3\textwidth}
       \includegraphics[width=\linewidth]{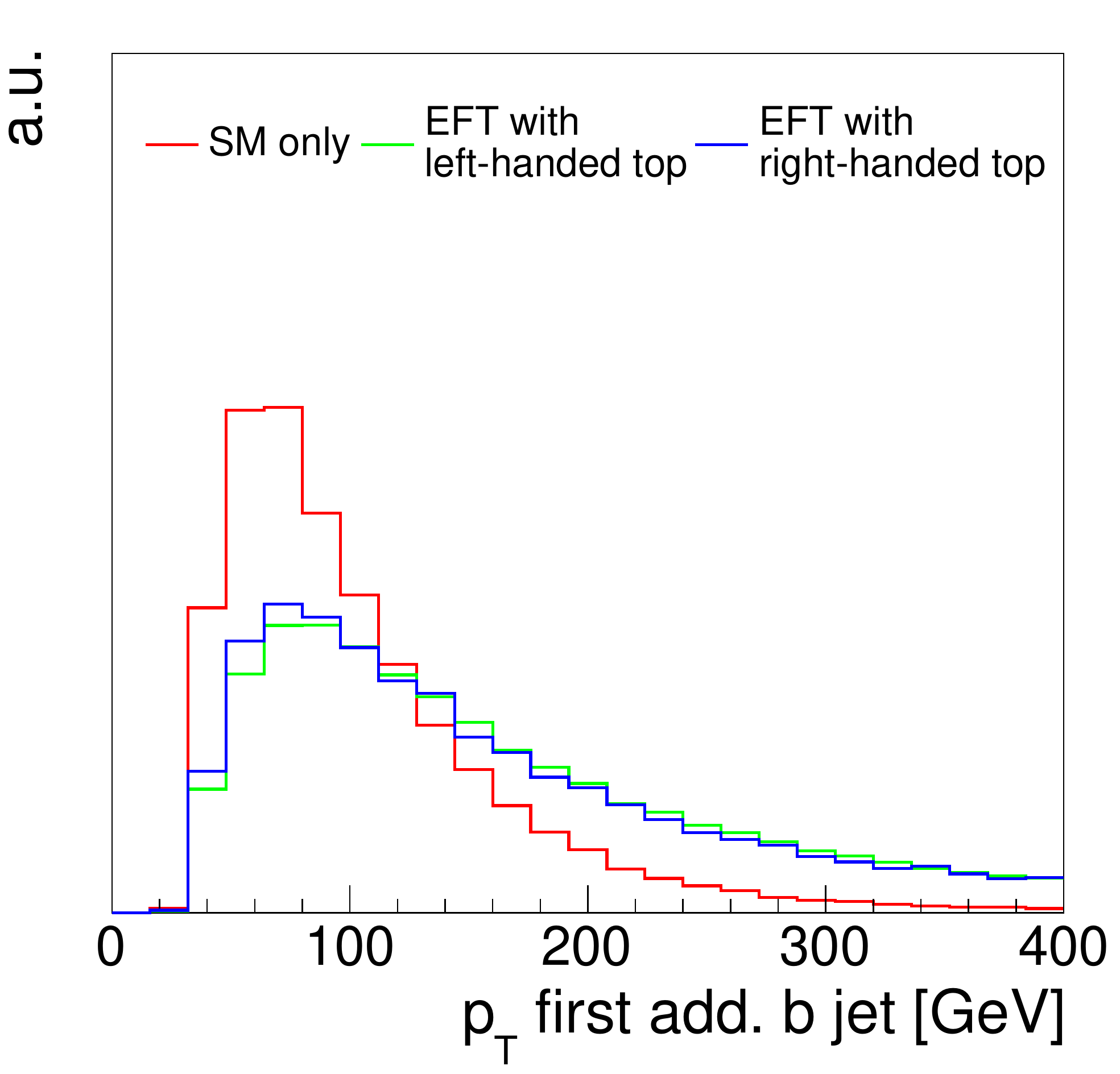}
      \end{minipage}
      \begin{minipage}{0.3\textwidth}
       \includegraphics[width=\textwidth]{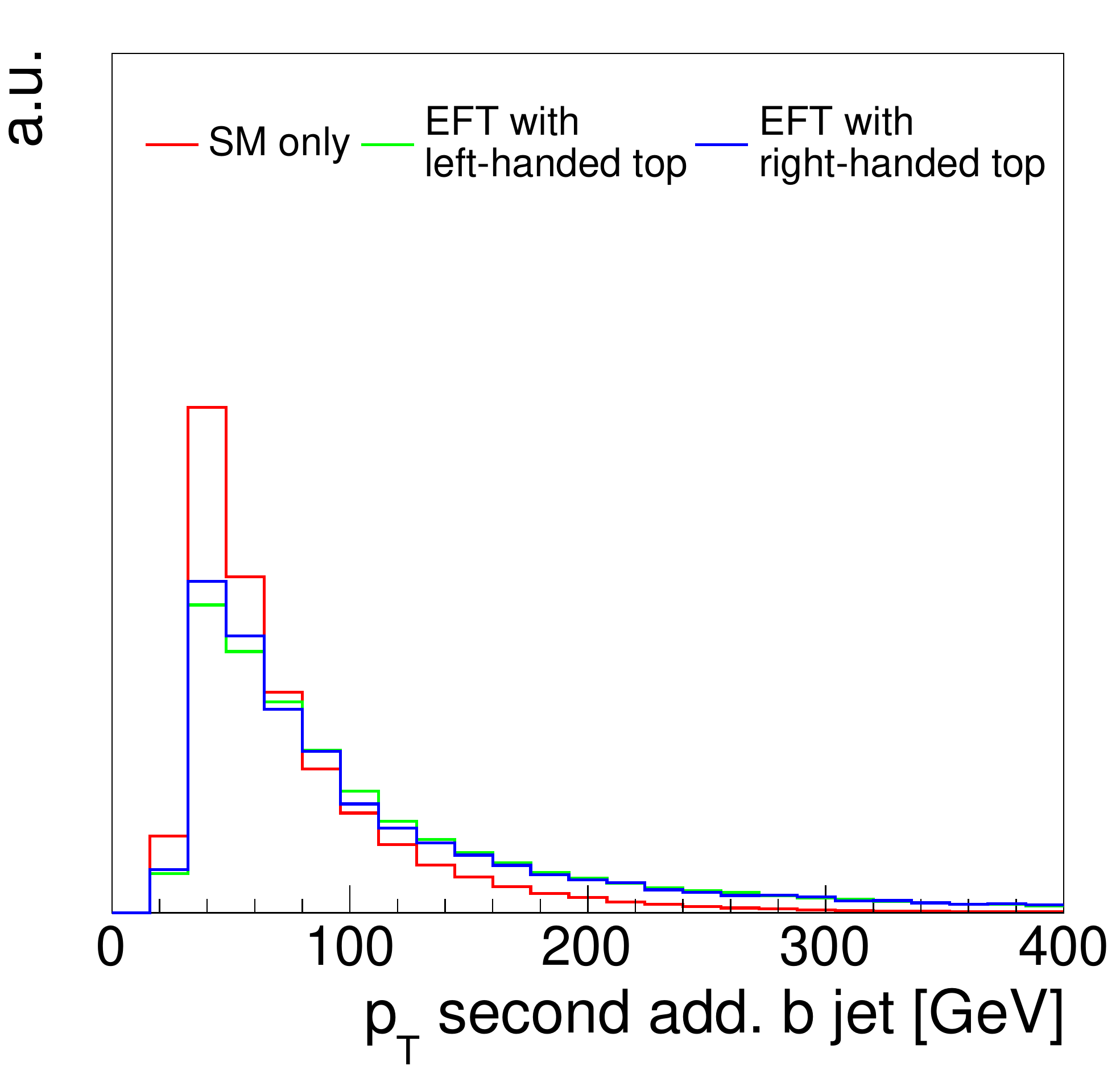}
      \end{minipage}
      \begin{minipage}{0.3\textwidth}
       \includegraphics[width=\textwidth]{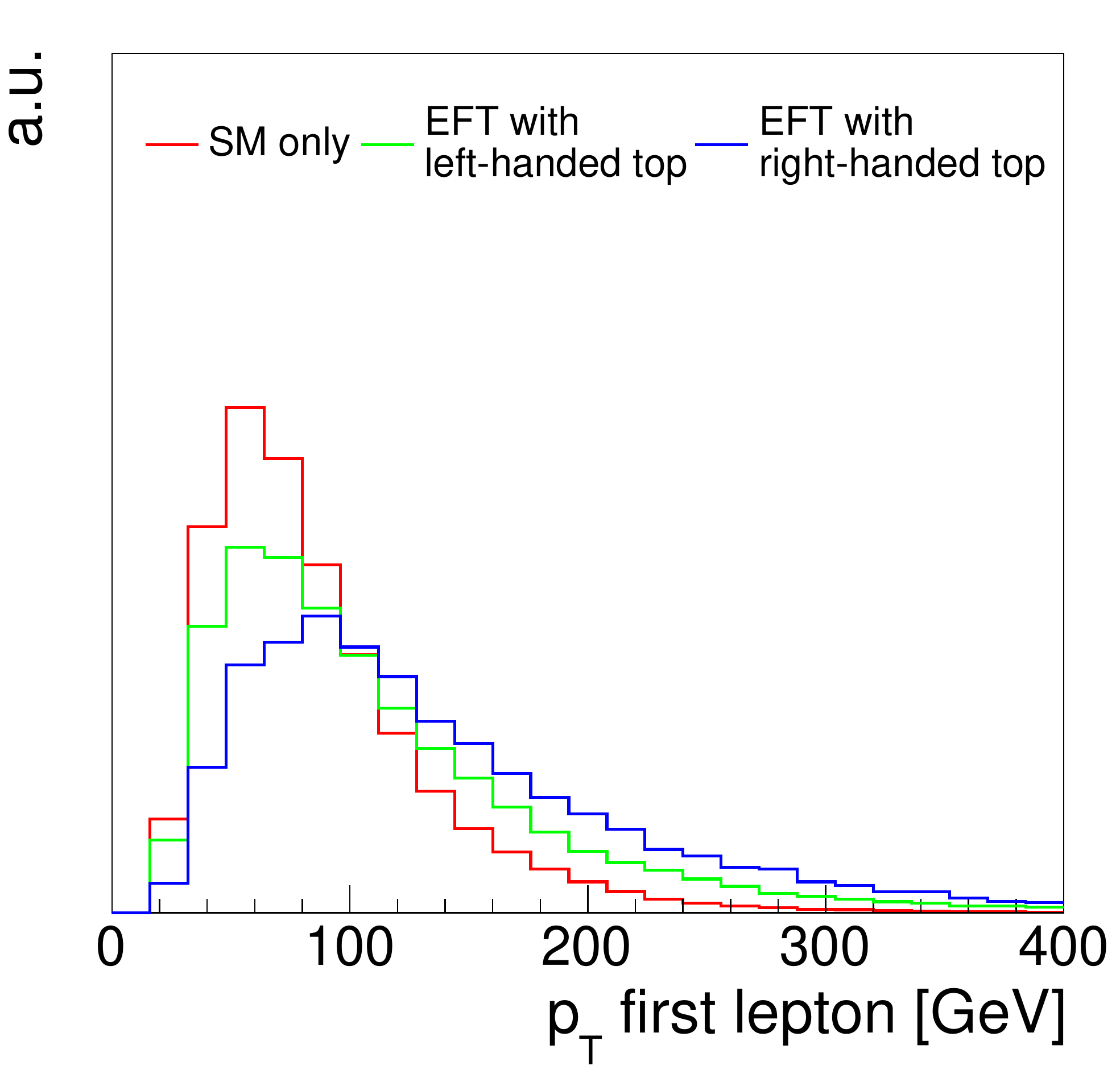}
      \end{minipage}
\end{minipage}\\
\begin{minipage}{\textwidth}
      \centering
      \begin{minipage}{0.3\textwidth}
       \includegraphics[width=\linewidth]{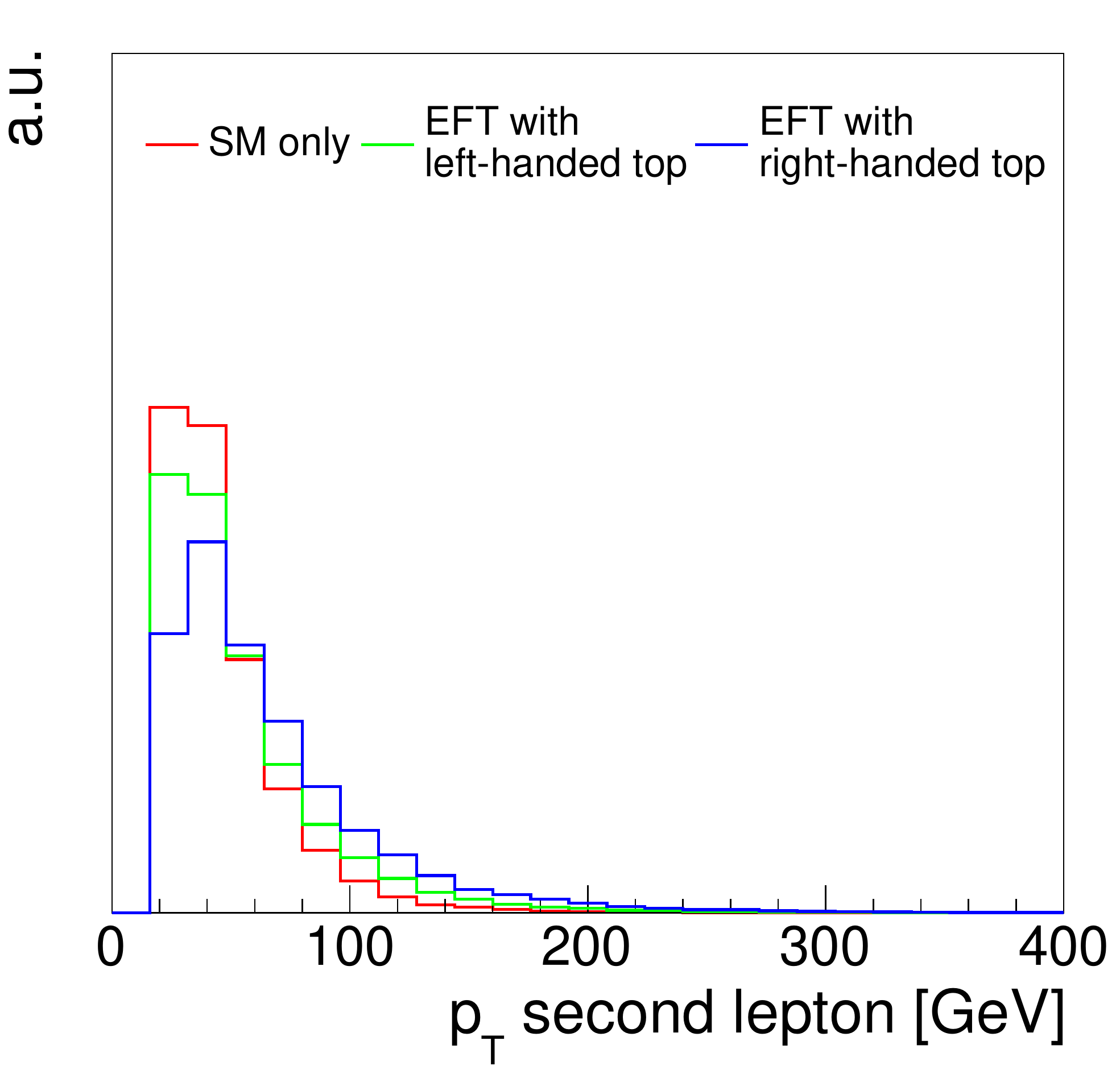}
      \end{minipage}
      \begin{minipage}{0.3\textwidth}
       \includegraphics[width=\textwidth]{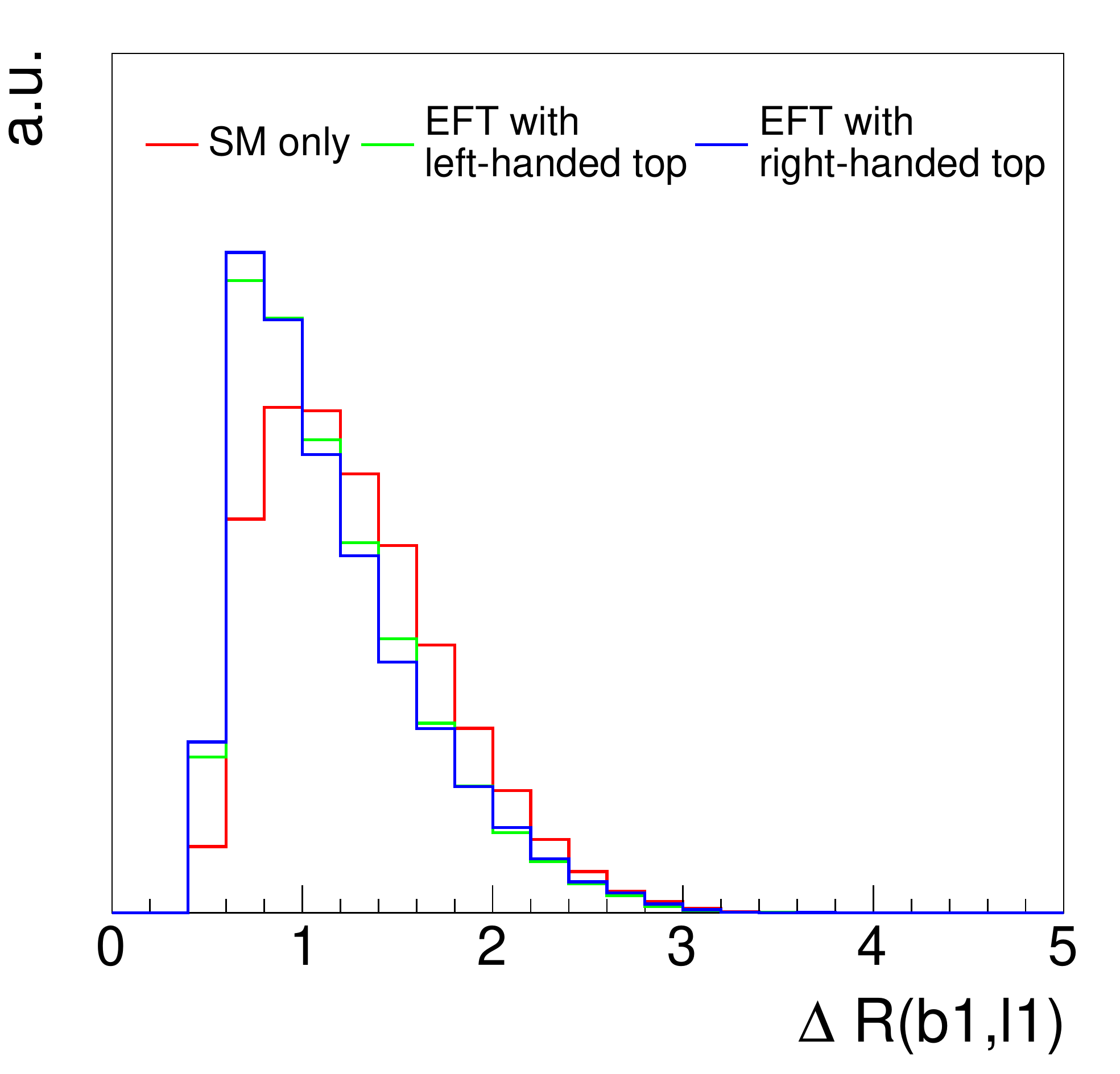}
      \end{minipage}
      \begin{minipage}{0.3\textwidth}
       \includegraphics[width=\textwidth]{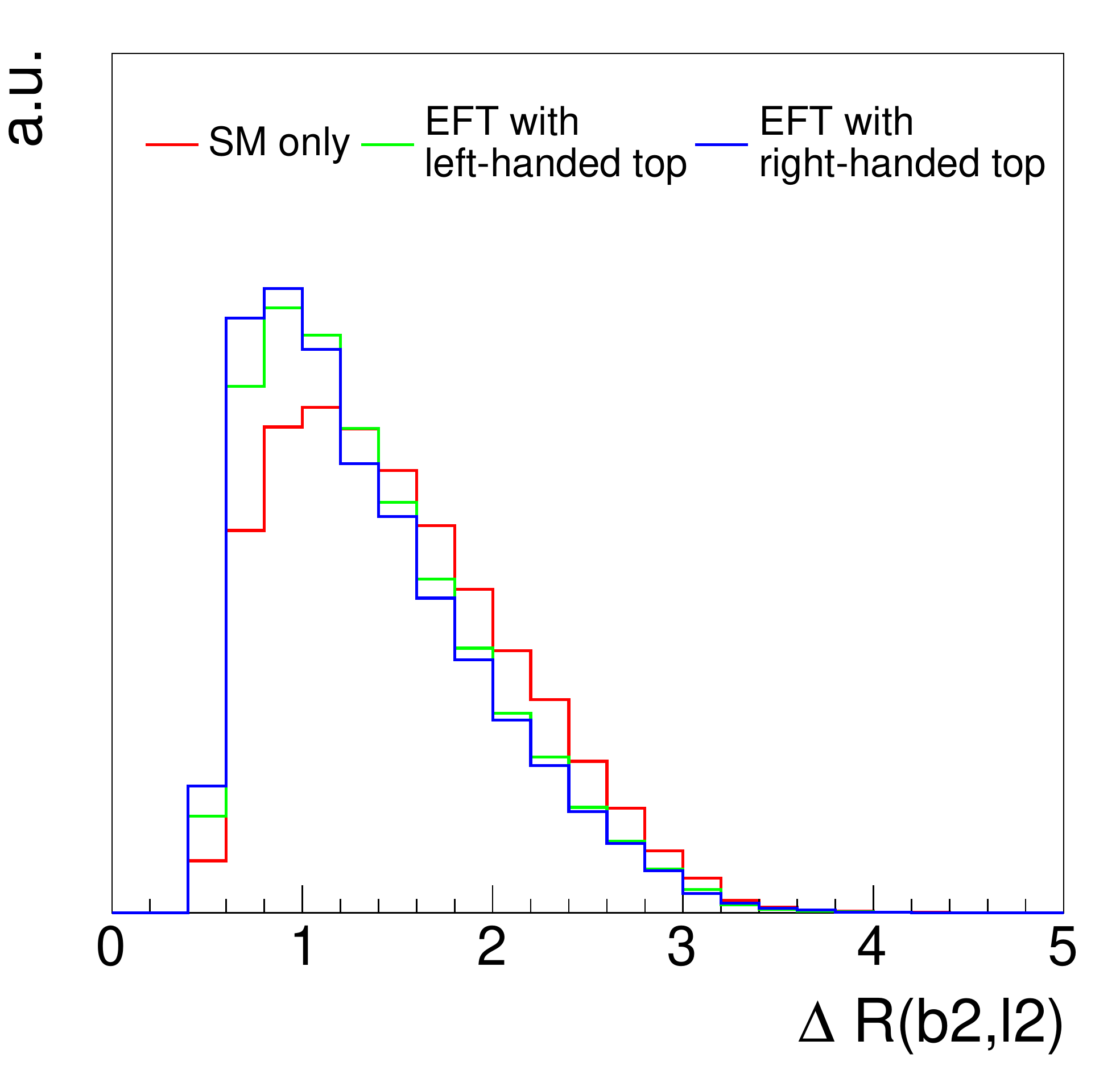}
      \end{minipage}
\end{minipage}\\
\begin{minipage}{\textwidth}
      \centering
      \begin{minipage}{0.3\textwidth}
       \includegraphics[width=\linewidth]{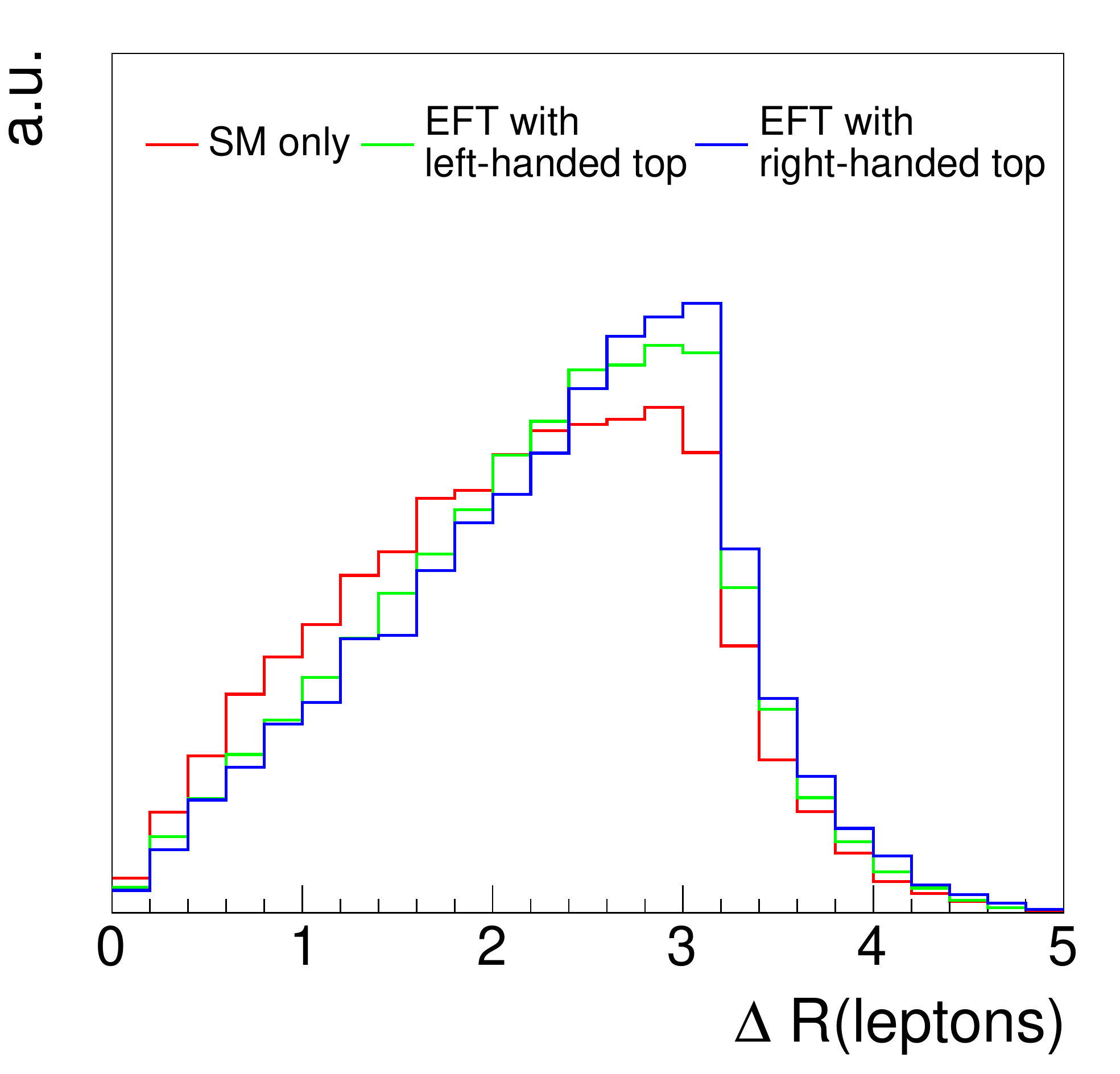}
      \end{minipage}
      \begin{minipage}{0.3\textwidth}
       \includegraphics[width=\textwidth]{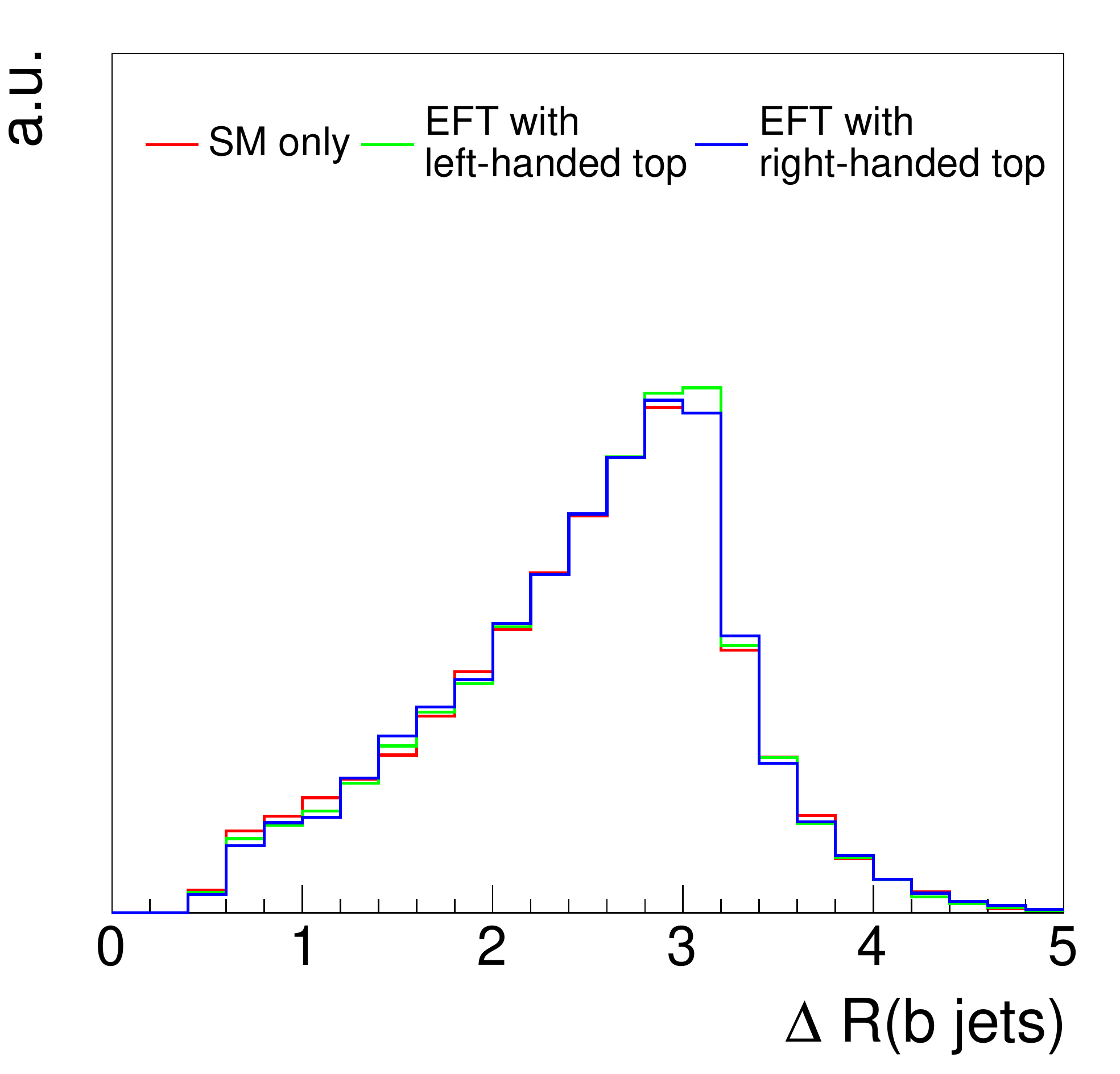}
      \end{minipage}
      \begin{minipage}{0.3\textwidth}
       \includegraphics[width=\textwidth]{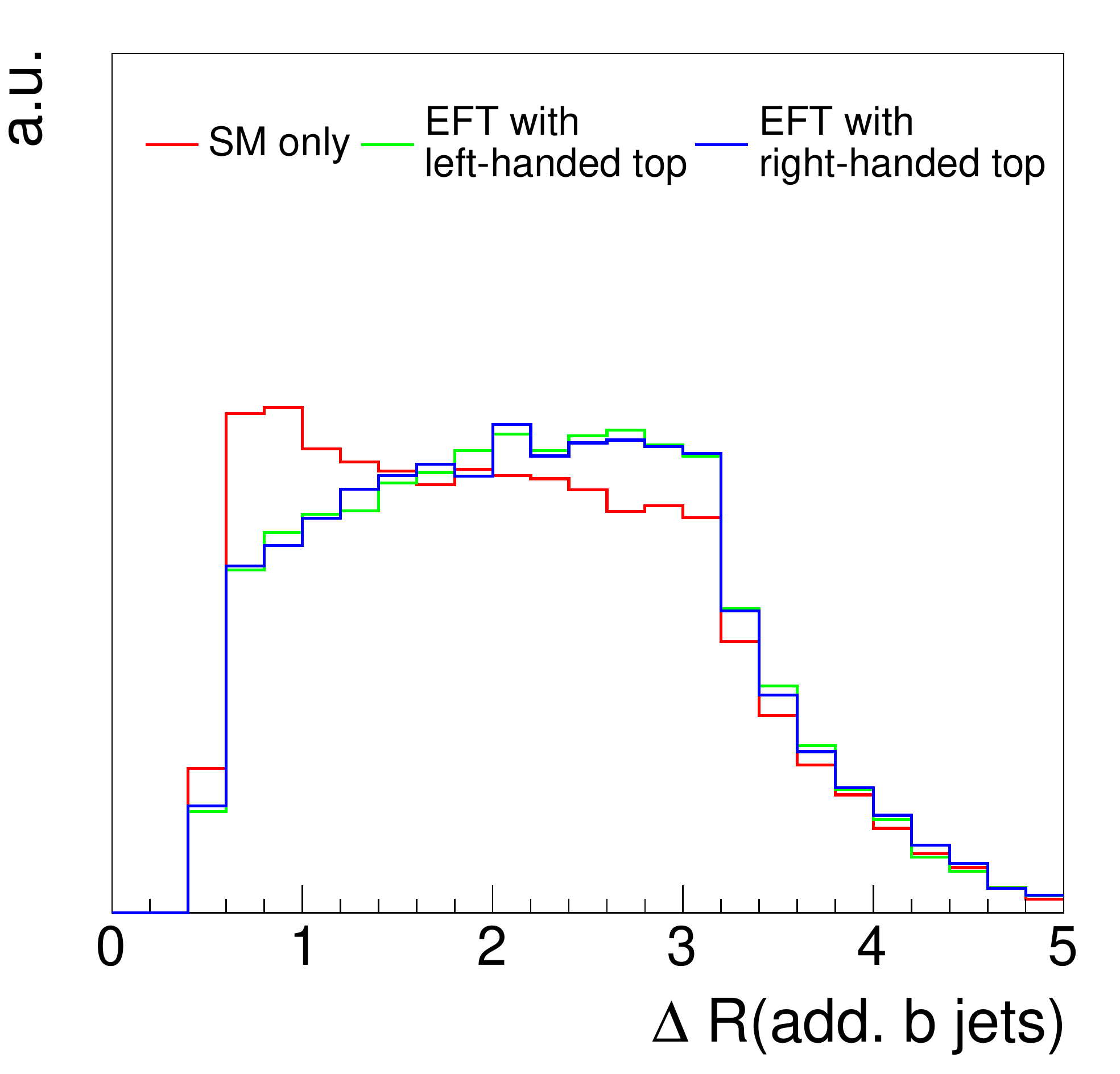}
      \end{minipage}
\end{minipage}\\

